\documentclass[aps,prl,twocolumn,superscriptaddress]{revtex4-2}

\usepackage[colorlinks=true,citecolor=blue]{hyperref}
\usepackage{graphicx}
\usepackage{amsmath}
\usepackage{amssymb}
\usepackage{bbold}
\usepackage{comment}

\DeclareGraphicsExtensions{.png,.jpg,.eps}
\usepackage{xcolor}

\begin{document}

\author{Wojciech Brzezicki}
\affiliation{International Research Centre MagTop, Institute of
  Physics, Polish Academy of Sciences, Aleja Lotnik\'ow 32/46, PL-02668 Warsaw, Poland}
\affiliation{Institute of Theoretical Physics, Jagiellonian University, ulica S. \L{}ojasiewicza 11, PL-30348 Krak\'ow, Poland}

\author{Matti Silveri}
\affiliation{Nano and Molecular Systems Research Unit, University of Oulu, FI-90014 Oulu, Finland}

\author{Marcin P\l{}odzie\'n}
\affiliation{ICFO - Institut de Ciencies Fotoniques, The Barcelona Institute
of Science and Technology, 08860 Castelldefels, Barcelona, Spain}
\affiliation{International Research Centre MagTop, Institute of
  Physics, Polish Academy of Sciences, Aleja Lotnik\'ow 32/46, PL-02668 Warsaw, Poland}

\author{Francesco Massel}
\affiliation{Department of Science and Industry Systems,
University of South-Eastern Norway, PO Box 235, Kongsberg, Norway}

\author{Timo Hyart}
\affiliation{Computational Physics Laboratory, Physics Unit, Faculty of Engineering and Natural Sciences, Tampere University, FI-33014 Tampere, Finland}
\affiliation{Department of Applied Physics, Aalto University, 00076 Aalto, Espoo, Finland}
\affiliation{International Research Centre MagTop, Institute of
  Physics, Polish Academy of Sciences, Aleja Lotnik\'ow 32/46, PL-02668 Warsaw, Poland}

\title{Non-Hermitian topological quantum states in a reservoir-engineered transmon chain}

   
\begin{abstract}

Dissipation in open systems enriches the possible symmetries of the Hamiltonians beyond the Hermitian framework allowing the possibility of novel non-Hermitian topological phases, which exhibit long-living end states that are protected against disorder. So far, non-Hermitian topology has been explored only in settings where probing genuine quantum effects has been challenging.  We theoretically show that a non-Hermitian topological quantum phase can be realized in a reservoir-engineered transmon chain. The spatial modulation of dissipation is obtained by coupling each transmon to a quantum circuit refrigerator allowing in-situ tuning of dissipation strength in a wide range. By solving the many-body Lindblad master equation using a combination of the density matrix renormalization group and third quantization approaches, we show that the topological end modes and the associated phase transition are visible in simple reflection measurements with experimentally realistic parameters. Finally, we  demonstrate that genuine quantum effects are observable in this system via robust and slowly decaying long-range quantum entanglement of the topological end modes, which can be generated passively  starting from a locally excited transmon. 

\end{abstract}

\maketitle


 {\it Introduction.--} Non-Hermitian (NH) phenomena in open systems have motivated   proposals of new families of topological states \cite{El-Ganainy2018, Gong18, Lieu18, Zhou19, Kawabata19, Song2019,Wojtek19, Lieu2020, Ashida2020,bergholtz2020exceptional}, which have been theoretically predicted to be applicable also to fermionic systems 
\cite{Pikulin2012, Pikulin2013, Pikulin14, Ramon1, Ramon2, Bergholtz2019, Lado2021} and exciton-polariton condensates \cite{Comaron20} but so far the paradigmatic experiments probing the NH topology have concentrated on photonic systems and electrical circuits where the quantum effects are not important
\cite{Ozawa, Zeuner2015, Poli2015, Zhan2017, Xiao2017, Weimann2017, Zhao2018, Bandres2018, Parto2018, Helbig2020}. The superconducting circuits, such as arrays of transmon devices \cite{Koch07}, are currently used in the most sophisticated attempts to build a scalable quantum computer \cite{Arute19, Chen21, Krinner22, Zhao21} and to simulate electronic properties \cite{Neill21} and topological phases \cite{Satzinger21, Mi2022}. However, their potential in realizing NH topological quantum phases remains to be explored.

\begin{figure}[h!]
    \centering
    \includegraphics[width=\linewidth]{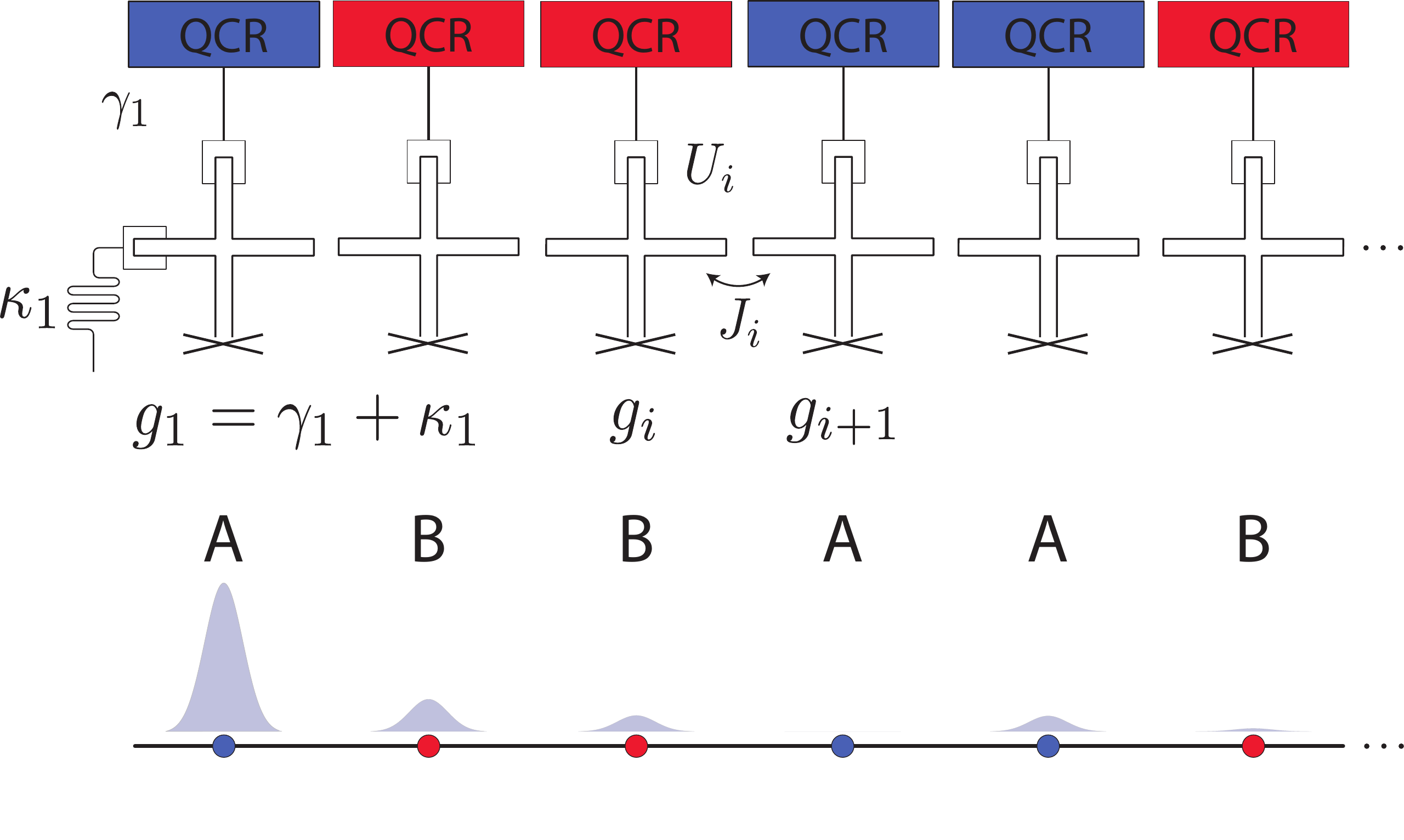}
    \caption{Schematic illustration of the NH Bose-Hubbard transmon chain for the ABBA configuration. (Top) Each transmon is described by the resonance frequency $\omega_i$ and the anharmonicity $U_i$ determining the on-site energy and the Hubbard interaction strength for the bosons. The hopping $J_i$ of the bosons between the lattice sites is determined by the capacitive dipole-dipole interaction between the neighboring transmons. The dissipation strength  $g_i=(\gamma_i + \kappa_i)/2$ at each lattice site is caused by the coupling of the transmon to the measurement circuit $\kappa_i$ and the tunable loss caused by the QCR $\gamma_i$.
    (Bottom) Site population of the topologically protected state for $g_{A}=0.1 J$ and $g_B=3 J$. Given the symmetry of the setup only the 6 leftmost sites of the 12-sites system are depicted.}
    \label{fig:setup}
\end{figure}

In reservoir engineering the idea is to turn the usually detrimental effects of dissipation into a resource. In this Letter, we demonstrate that the flexibility to engineer  dissipation in a controllable manner in  transmon circuits \cite{Silveri18, Silveri19a, Silveri19b, Morstedt21, Orell22} can be utilized for realizing NH topological quantum phases. In our proposal the NH topological phase is created by introducing  a spatial modulation of  dissipation  \cite{Takata2018, Wojtek19} in the one-dimensional Bose-Hubbard transmon chain \cite{Orell19, Mansikkamaki21}, where the dissipation strength in each transmon is controlled by the tunable coupling of the transmon to a quantum circuit refrigerator (QCR) \cite{Silveri18, Silveri19a, Silveri19b, Morstedt21} (see Fig.~\ref{fig:setup}).  In contrast to the earlier realizations of NH topological phases the quantum effects are important in the transmon circuits, and therefore we describe the topological phenomena using the Lindblad master equation approach.
By utilizing the third quantization approach \cite{Prosen_2010} we show that the topology of the Liouvillian superoperator in the non-interacting limit is described by a Chern number \cite{Wojtek19}, which determines the number of topological end modes. We discuss the signatures of the topological end modes and topological phase transition in the reflection measurements, and we utilize density matrix renormalization group (DMRG) approach to show that the effects of the nontrivial topology  can be robustly measured in the presence of a realistic Hubbard interaction caused by the charging energy of the transmons \cite{Orell19, Mansikkamaki21}. Finally, we show that the quantum nature of the NH topological state can be unambiguously demonstrated by utilizing the topologically protected end modes for generation of a long-range entangled state  from a local excitation of a single transmon. Importantly, we obtain robust entanglement between the transmons at the opposite ends of the chain  by just switching on the couplings of the  transmons and QCRs instead of actively controlling the system with a sequence of pulses.

{\it NH topological phase in a  transmon chain.--} 
The Lindblad master equation for a chain of $L$ transmons in the rotating frame \cite{supplementary} can be written as 
\begin{equation}
\frac{d\rho}{dt}={\cal L} \rho, \label{Lindblad-master}
\end{equation}
where at zero temperature the Liouvillian superoperator ${\cal L}$ acting on the density matrix $\rho$ is
\begin{equation}
{\cal L}=-i\left[{\cal H} \square-\square {\cal H} \right]+\sum_{j=1}^L g_{j}\left(2a_{j}\square a_{j}^{\dagger}-a_{j}^{\dagger}a_{j}\square-\square a_{j}^{\dagger}a_{j}\right).
\end{equation}
Here, ${\cal H}$ is the Bose-Hubbard Hamiltonian \cite{Orell19, Mansikkamaki21} 
\begin{equation}
{\cal H}=\vec{a}^{\dagger}\hat{H}\vec{a}+\vec{a}^{\dagger} \vec{f} +\vec{f}^{\dagger}\vec{a}-\sum_{j=1}^{L} \frac{U_j}{2} a_j^\dag a_j (a_j^\dag a_j-1), \label{Ham-full}
\end{equation}
and $\hat{H}$ is the tight-binding Hamiltonian, where the on-site energies $\hat{H}_{i,i}=\omega_i-\omega$ are determined by the driving frequency $\omega$ and the resonance frequencies of the transmons  $\omega_i \approx \sqrt{8 E_{C,i} E_{J,i}}$ ($E_{J,i}$ is the flux-tunable Josephson energy and $E_{C,i}$ is the charging energy of the transmon) \cite{Koch07} and the hoppings $\hat{H}_{i,i+1}=\hat{H}_{i+1,i}=J_i$  originate from the capacitive dipole-dipole interaction between the neighboring transmons. Additionally, the Hubbard-interaction strength $U_i \approx E_{C,i}$ is caused by the anharmonicity  of the transmons \cite{Koch07}, the driving strength $f_j=-i\sqrt{\kappa_j}\alpha_j^{\rm in}$ is determined by the amplitude of the incoming signal $\alpha_j^{\rm in}$ and the coupling of the transmon to the measurement circuit $\kappa_i$, and the dissipation strength 
$g_i=(\gamma_i + \kappa_i)/2$ is mainly controlled by the tunable loss $\gamma_i$ caused by the QCR. We use notations where $\vec{x}$ indicates a column vector,  $\hat{X}$ is a matrix, and $a_i$ are the bosonic annihilation operators. We have set $\hbar=1$ and assumed the zero temperature limit for simplicity. 

The steady state output field amplitudes $\alpha_j^{\rm out}$ and the input field amplitudes $\alpha_j^{\rm in}$ are related as
\begin{equation}
\alpha_j^{\rm out}=\alpha_j^{\rm in}+\sqrt{\kappa_j} {\rm Tr} (a_j \rho_S), \label{input-output}
\end{equation}
where $\rho_S$ is the steady state solution of Eq.~(\ref{Lindblad-master}). In the linear response regime the relationship between the $\vec{\alpha}^{\rm out}$ and  $\vec{\alpha}^{\rm in}$ can be rewritten with the help of a transmission and reflection matrix $\hat{\Gamma}$
\begin{equation}
\vec{\alpha}^{\rm out} = \hat{\Gamma} \vec{\alpha}^{\rm in}.    \label{Gamma-matrix}
\end{equation}
Additionally, we also consider nonlinear responses of the transmons $i$ to a strong driving on one of the transmons $j$ by computing the ratios  $\alpha_i^{\rm out}/\alpha_j^{\rm in}$. These transmission and reflection amplitudes 
are directly measurable in the transmon circuits and allow the detection of the topological end modes and phase transition (see below).

\begin{figure}
    \centering
    \includegraphics[width=1.0\linewidth]{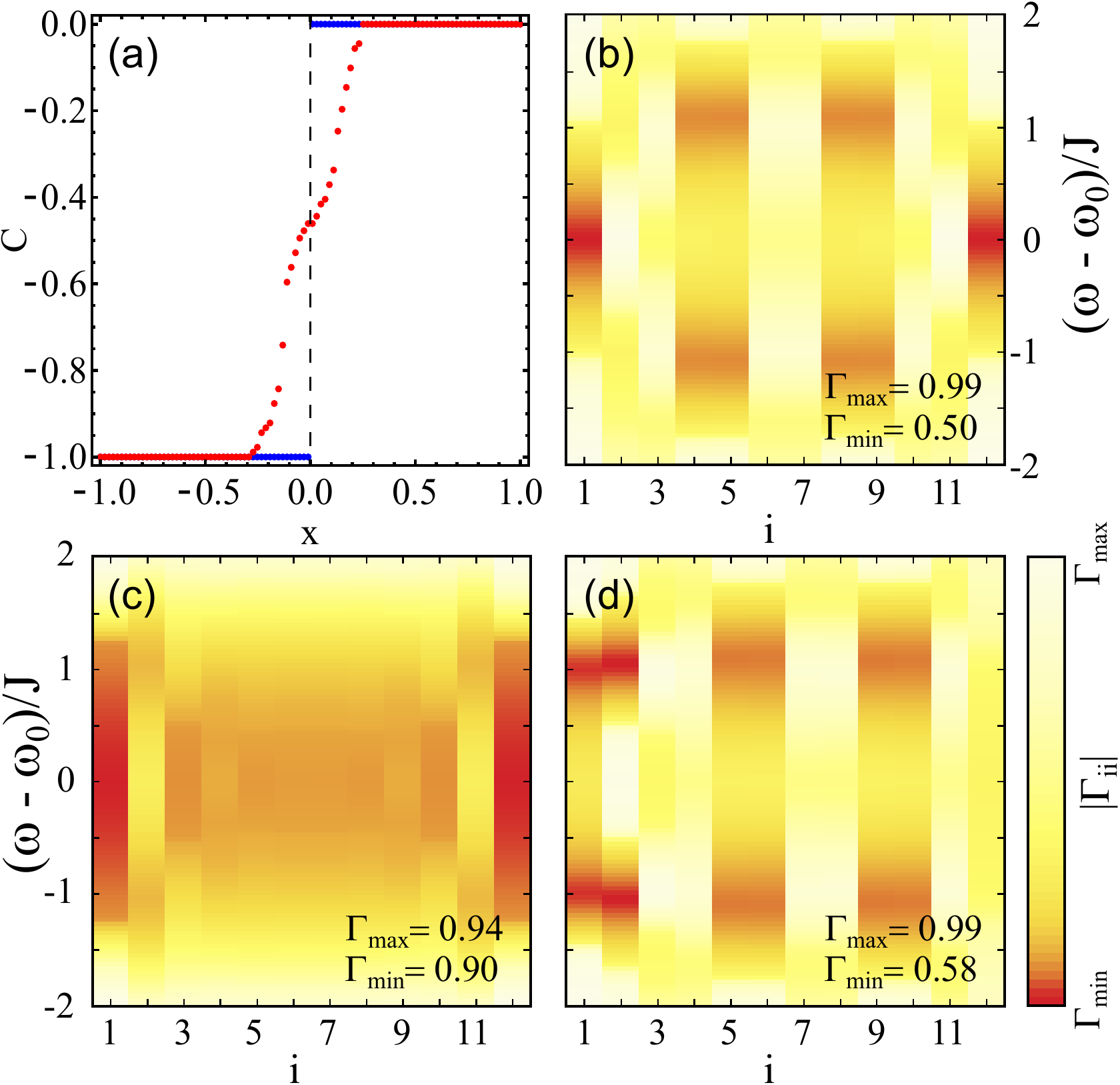}
    \caption{(a) Topological phase diagram as a function of dissipation strengths interpolating between the ABBA ($x=-1$, $C=-1$) and AABB ($x=1$, $C=0$) pattern [Eq.~(\ref{interpolation})]. In the clean limit the topological phase transition takes place at $x=0$ (blue), and the disorder influences the topology only close to $x=0$ (red). We have used a supercell containing $20$ transmons and taken the disorder average of $C$ over $100$ disorder realizations. The disorder in $J_i$ ($g_{i}$) is  sampled from a uniform distribution in interval $[-\delta J, \delta J]$ ($[-\delta g, \delta g]$). We have used $g_A=0.1 J$, $g_B=3.0 J$, $\delta J= 0.1 J$  and $\delta g = 0.29 J$.   (b),(c),(d) Contourmaps of the reflection coefficients $|\Gamma_{ii}(\omega)|$ in the linear response regime [Eq.~(\ref{reflection-transmission-matrix})] as a function of lattice site $i$ and frequency $\omega$ in the nontrivial, gapless and trivial phases. (b) In the nontrivial phase with $x=-1$ the topological end modes show up as a dip in $|\Gamma_{ii}(\omega)|$ at the resonance frequency of the transmon $\omega=\omega_0$ close to the ends of the chain. (c) At the transition ($x=0$) the bulk is gapless leading to features in  $|\Gamma_{ii}(\omega)|$ at most of the lattice sites in a wide range of frequencies $\omega$ around $\omega_0$.  (d) In the trivial phase with $x=1$ the resonant features in the $|\Gamma_{ii}(\omega)|$ are absent at frequencies inside the gap around $\omega =\omega_0$.}
    \label{fig:transmission_reflection}
\end{figure}

The Liouvillian superoperator contains operators acting from left and right on the density matrix, which we denote with superscripts $L$ and $R$. 
The third quantization of the Liouvillian superoperator  is based on definition of new operators 
$\vec{a}_{0}  =  \vec{a}^{L}$,   $(\vec{a}'_{0})^T  =  \vec{a}^{L\dagger}-\vec{a}^{R\dagger}$,  
$(\vec{a}_{1})^T  =  \vec{a}^{R\dagger}$,   $\vec{a}'_{1}  =  \vec{a}^{R}-\vec{a}^{L}$, which satisfy the usual commutation relations of bosonic annihilation and creation operators \cite{Prosen_2010, supplementary}.
Using these definitions the Liouvillian superoperator can be written as 
\begin{eqnarray}
{\cal L}&=&-i (\vec{a}'_{0})^T \hat{H}_{NH} \vec{a}_{0}+i ( \vec{a}'_{1})^T \hat{H}_{NH}^* \vec{a}_{1}-i(\vec{a}'_{0})^T\vec{f}+i \vec{f}^\dag\vec{a}'_{1} \nonumber \\ &&-i\sum_{j} \frac{U_j}{2} \big(2a'_{1,j}a_{1,j}a_{1,j}a_{0,j}+a'_{1,j}a'_{1,j}a_{1,j}a_{1,j} \nonumber \\ && \hspace{1.2cm} -a'_{0,j}a'_{0,j}a_{0,j}a_{0,j}-2a'_{0,j}a_{1,j}a_{0,j}a_{0,j}\big), \label{Liouvillian-third-quantized}
\end{eqnarray}
where the non-Hermitian Hamiltonian is defined as
\begin{equation}
\hat{H}_{NH}=\hat{H}-i{\rm diag}\left(\vec{g}\right).
\end{equation}
 In the linear response regime the interaction effects can be neglected and we obtain \cite{supplementary}
\begin{equation}
\hat{\Gamma}(\omega)=i \sqrt{\hat{K}}  \hat{H}_{NH}^{-1} \sqrt{\hat{K}}+\hat{I}_{L \times L},   \label{reflection-transmission-matrix}
\end{equation}
where $\hat{I}_{L\times L}$ is the identity matrix and 
$\hat{K}={\rm diag}(\vec{\kappa})$. 
In the following, we assume that all the transmons have similar resonance frequencies $\omega_i=\omega_0$ and couplings $J_i=J$. This means that the Hermitian part of the $\hat{H}_{NH}$ describes a trivial tight-binding model with constant on-site energies. On the other hand, we assume that the dissipation $g_i$ is spatially modulated. For simplicity we assume that $\kappa_i=\kappa$, so that the spatial modulation of $g_i$ originates purely from the tunable losses $\gamma_i$ caused by the QCRs. In the presense of an arbitrary dissipation modulation this model always satisfies a NH chiral symmetry ${\cal S}  [\hat{H}_{NH}^\dag-(\omega_0-\omega) \hat{I}_{L \times L}] {\cal S}= -[\hat{H}_{NH}-(\omega_0-\omega) \hat{I}_{L \times L}]$,
where ${\cal S}=\hat{I}_{L/2 \times L/2}\otimes\sigma_{z}$ and the Pauli matrices are denoted as $\sigma_i$ ($i=x,y,z)$. Therefore, the topology of $\hat{H}_{NH}$ is determined by the Chern number $C$  \cite{Wojtek19, supplementary}, which determines the number of topologically protected end modes at the transmon resonance frequency $\omega=\omega_0$. In general, it is possible to construct one-dimensional NH models where $C$ takes all possible integer values \cite{Lado2021} but for simplicity we concentrate here on the simplest models with $|C|=:0,1$ \cite{Wojtek19}. For this purpose we consider a unit cell consisting of $4$ transmons repeating periodically along the chain. 
Based on Ref.~\onlinecite{Wojtek19}, we know that $C=-1$ for the ABBA pattern of dissipation, whereas $C=0$ for the AABB pattern of dissipation. Thus, we can interpolate between the topologically distinct phases by assuming 
\begin{eqnarray}
g_{1,3}(x)&=&[g_A+g_B\pm (g_A-g_B)|x|]/2, \nonumber \\ g_{2,4}(x)&=&[g_A+g_B \pm  (g_A-g_B)x]/2. \label{interpolation}
\end{eqnarray}
The phase diagram as a function of $x$ is shown in Fig.~\ref{fig:transmission_reflection}(a): $C=-1$ for $x<0$ and $C=0$ for $x>0$, so that the phase transition takes place at $x=0$. 

Importantly, in transmons the resonance frequencies $\omega_i$ are flux-tunable and in the state-of-the-art experiments they can be made equal to each other within relative accuracy of $10^{-5}-10^{-4}$~\cite{Ma19}. 
Therefore, the disorder effects in $\omega_i$ can be neglected. On the other hand, we expect that the parameters $J_i$ and $g_i$ will contain significant amount of variations. In Fig.~\ref{fig:transmission_reflection}(a) we demonstrate that the topological phases are robust even in the presence of strong disorder amplitudes $\delta J$ and $\delta g$. These types of disorder can destroy the topology only if they are sufficiently strong to induce a bulk gap closing, and therefore they are important only close to the topological phase transition where the topological gap is small.

{\it Signatures of NH topological phase  in reflection measurements.--} 
The localization length of the end modes in the topologically nontrivial phase depends sensitively on the dissipation parameters. Here we fix them to $g_A=0.1 J$ and $g_B=3J$, so that the topological end modes are strongly localized at the end of the chain and numerical calculations can be performed efficiently using a short chain of length $L=12$. We also set $\kappa=2 g_A$ so that the minimal values of the dissipation originate from the measurement circuits. We use these parameters everywhere in the manuscript unless otherwise stated.

The topological phase diagram shown in Fig.~\ref{fig:transmission_reflection}(a) can be probed by measuring the reflection $|\Gamma_{ii}(\omega)|$  as a function of lattice site $i$ and frequency $\omega$  [see Figs.~\ref{fig:transmission_reflection}(b)-(d)]. In the nontrivial phase  the topological end modes show up as a dip in $|\Gamma_{ii}(\omega)|$ at the resonance frequency of the transmon $\omega=\omega_0$ on lattice sites $i$ close to the ends of the chain [see Fig.~\ref{fig:transmission_reflection}(b)], whereas such kind of features are absent in the trivial phase where the system is gapped around $\omega=\omega_0$ [see Fig.~\ref{fig:transmission_reflection}(d)].  At the transition  the bulk is gapless leading to a broad feature in $|\Gamma_{ii}(\omega)|$ as a function of   $\omega$ at most of the lattice sites [see Fig.~\ref{fig:transmission_reflection}(c)]. 


\begin{figure}
    \centering
    \includegraphics[width=0.95\linewidth]{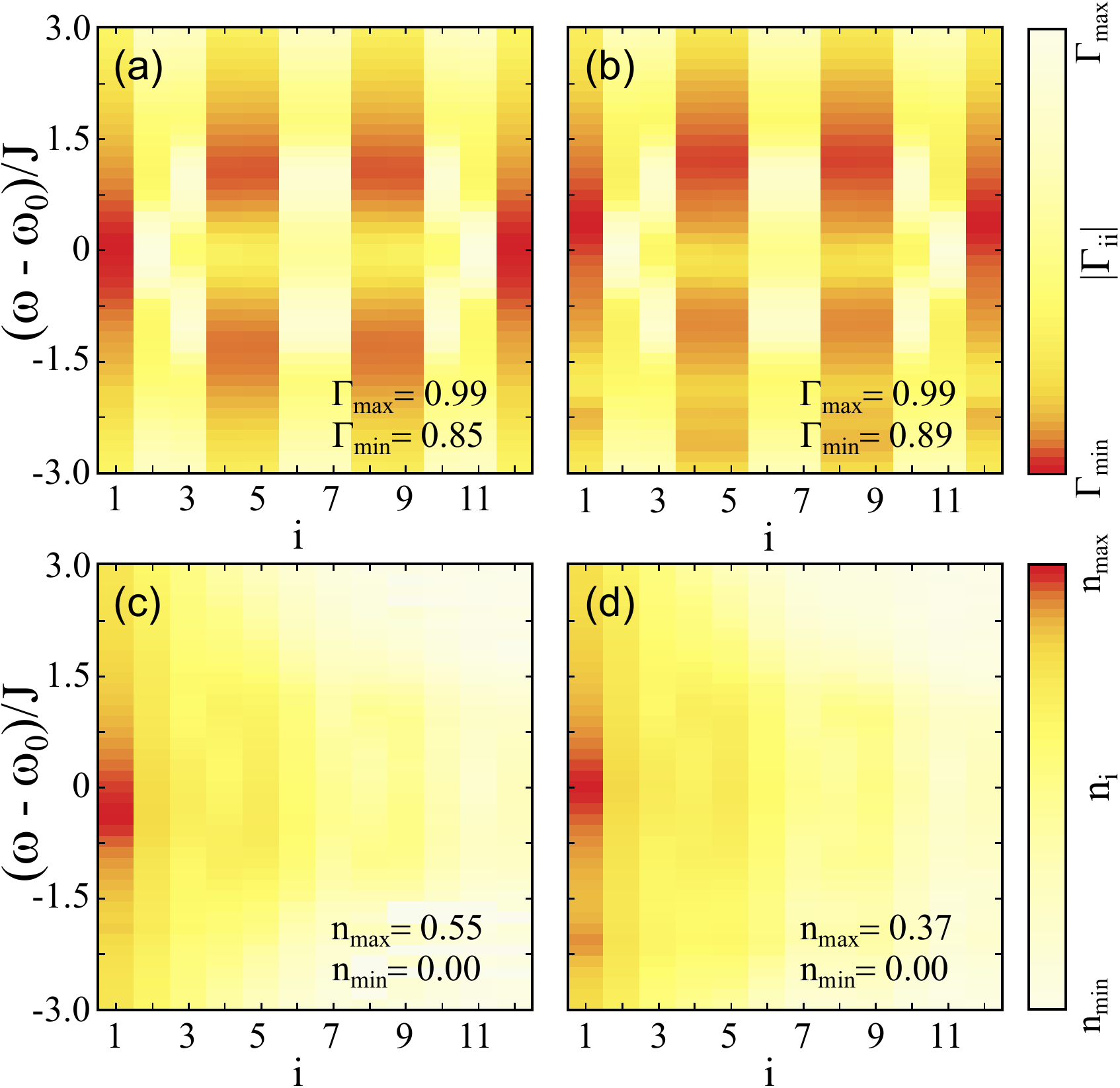}
    \caption{Contourmaps of the reflection coefficients $\Gamma_{ii}(\omega)=\alpha^{\rm out}_i/\alpha^{\rm in}_i$ and density $n_i={\rm Tr}(a_i^\dag a_i \rho_S)$ in a topologically nontrivial chain  [$x=-1$ in Eq.~(\ref{interpolation})] as a function of lattice site $i$ and frequency $\omega$. The driving strength is $|f|=0.4J$ and the interaction strengths are (a),(c) $U=1J$  and (b),(d) $U=4J$. In (a),(b) $\rho_S$ is solved in each case separately when the driving is applied at the different sites $i$. In (c),(d) the driving is always applied  at the site $1$.  
    In addition to the  topological end modes at $\omega \approx \omega_0$ there exists multiboson satellite features at frequencies $\omega \approx \omega_0-n U/2$ ($n \in \mathbb{Z}$).}
    \label{fig:interactions}
\end{figure}

{\it Robustness of the topological states in the presence of interactions.--} 
In the limit of weak driving, where the interactions between the bosons can be neglected, the properties of the steady-state system are completely determined by $\hat{H}_{NH}$, and the expectation values of the normal ordered products of the bosonic annihilation and creation operators separate into products of expectation values \cite{supplementary}.  On the other hand, in the presence of strong driving the steady state of the interacting system is a correlated quantum state with entanglement between the transmons.  We have utilized a generalization of the DMRG approach \cite{supplementary} to describe steady-state density matrix  of  the driven system. This allows us to numerically compute the density profile $n_i={\rm Tr}(a_i^\dag a_i \rho_S)$ and the reflection coefficients $\alpha_i^{\rm out}/\alpha_i^{\rm in}$, and representative results of our numerical calculations are shown in Fig.~\ref{fig:interactions}. Importantly, we find that the resonant feature of the topological end modes at $\omega=\omega_0$ is  very robust in the presence of strong driving and interactions $U_j=U$. Additionally, the topological end modes give rise to satellite features at frequencies $\omega \approx \omega_0-n U/2$ ($n \in \mathbb{Z}_+$) corresponding to multiboson excitations of the interacting system.

{\it Dynamical generation of long-range  entanglement.--} 
While the interactions $U$ are not important for the existence of the topological excitations, they offer interesting new possibilities in the utilization of the topological end modes for the generation of  entangled quantum states.  Namely, in the absence of $U$ the driving initializes the system to a site-wise product of coherent states, and it turns out that in this case there is no entanglement between the transmons developing during the time-evolution of the density matrix described by Eq.~(\ref{Lindblad-master}) \cite{supplementary}. On the other hand, it is well-known that the anharmonicity of the transmons $U\ne 0$ can be utilized for initializing a transmon into a Fock state  \cite{Blais21, Elder20, supplementary}.
\begin{figure}
    \centering
    \includegraphics[width=1.0\linewidth]{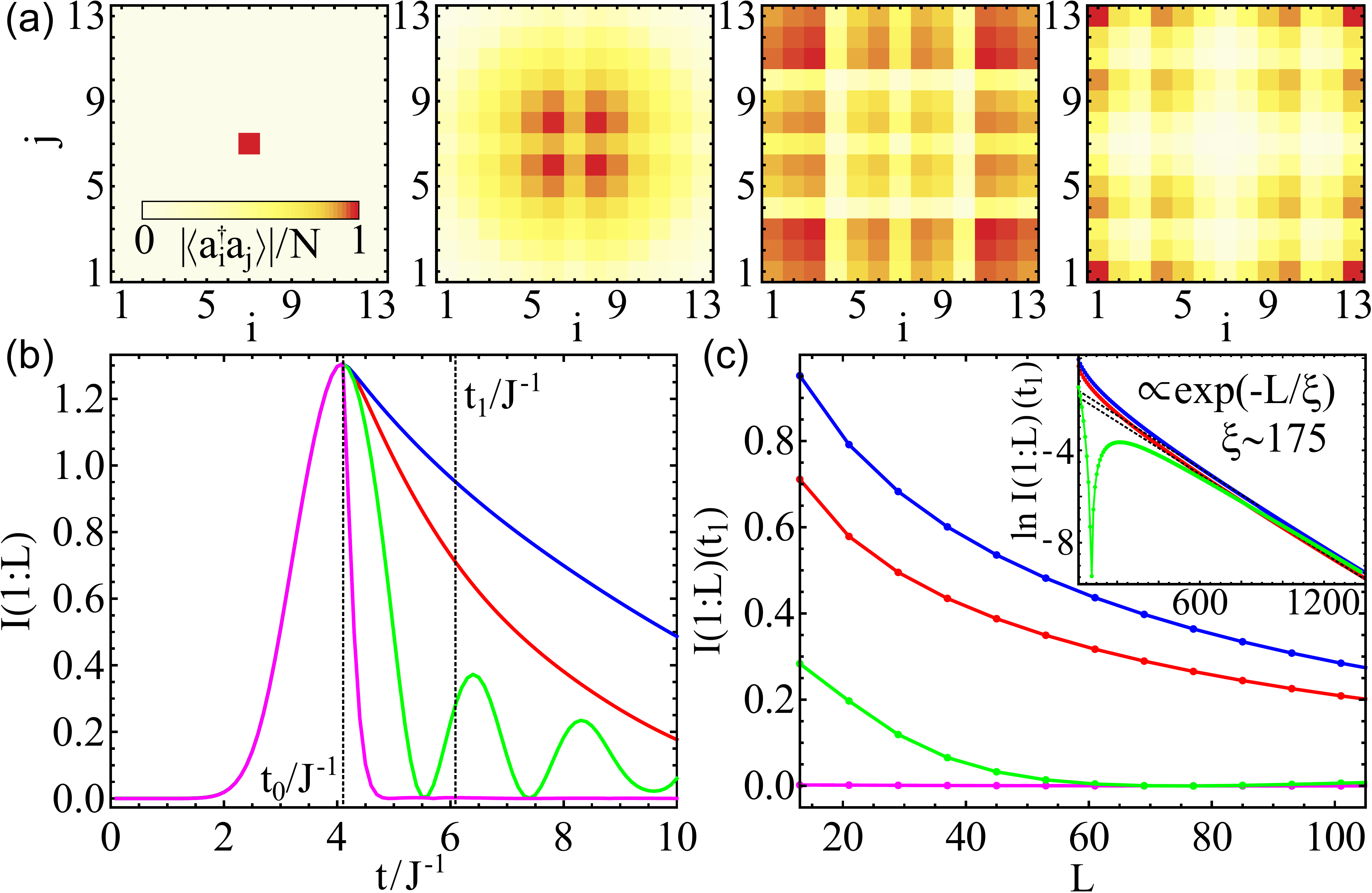}
    \caption{(a)  ${\rm Tr}[a_i^\dag a_j \rho_S(t)]$ vs $i$ and $j$ in the presence of time-dependent dissipation with $x=-1$ in Eq.~(\ref{interpolation}), and $g_{B, {\rm max}}=10 J$   and $t_0\simeq 4.1 J^{-1}$ in Eq.~(\ref{diss-t}). The time instants are $t/J^{-1}=: 0, 1, 3, 10$ from left to right. (b) Mutual information $I(1:L)$ vs $t$  for a reference system with uniform dissipation $g_{B, {\rm max}}=0.01 J$ (green), trivial phase with $x=1$ and $g_{B, {\rm max}}=5 J$ (magenta), and  non-trivial phase  with $x=-1$ and   $g_{B, {\rm max}}/J=:5, 10$ (red, blue). (c) $I(1:L)$ for the same dissipation profiles as a function of $L$ at time $t_1=t_0+2 J^{-1}$, where $I(1:L)(t)$ is maximized at $t=t_0$. Log-plot is shown in the inset. In addition to the dissipation patterns (\ref{interpolation}) we have added one extra site in middle of the chain with $g_{L/2+1}=g_B$, which is excited to a Fock state with $N=5$  
    at $t=0$.}
    \label{fig:dynamics}
\end{figure}
To demonstrate that in this case it is possible to create a long-range entangled state, we consider a protocol
where the system is initilized to a Fock state with $N$ bosons in a single  transmon in the middle of the chain at $t=0$ \cite{supplementary}, and the dissipation strengths are switched on at time $t=t_0$. The dissipation can be controlled fast using the QCRs \cite{Silveri18, Silveri19a, Silveri19b, Morstedt21}, and thus for simplicity we assume that the time-dependence of the dissipation parameters is given by 
\begin{equation}
g_A=0.01 J, \ g_B=\begin{cases}
0.01 J, & t\leq t_0 \\
g_{B,{\rm max}}, & t > t_0
\end{cases}. \label{diss-t}
\end{equation}
We assume that the other transmons have sufficiently small $U$ so that  the interactions can be neglected  during the time-evolution. 
 In Fig.~\ref{fig:dynamics}(a) we show the time-evolution of the expectation values ${\rm Tr}[a_i^\dag a_j \rho_S(t)]$ in the case of a topologically nontrivial dissipation pattern  \cite{supplementary}. It demonstrates that there exists a quasi-stable (slowly decaying) state, where the bosons are dominantly trapped at the end of the chain. We can characterize the entanglement between the end  transmons $1$ and $L$ with the help of time-dependent mutual information 
$I(1:L)(t)$ \cite{supplementary}, and we find that in the case of topologically nontrivial NH phase the generated entanglement is more stable in time than in the reference cases of trivial and uniform chains [Fig.~\ref{fig:dynamics}(b)]. Furthermore, the entanglement decreases only slowly with the increasing length of the chain [Fig.~\ref{fig:dynamics}(c)].

{\it Conclusions.--} To summarize, we have shown that a NH topological quantum phase can be realized in a transmon chain by utilizing a spatial modulation of dissipation obtained  by coupling the transmons to QCRs. The topological end modes and topological phase transition can be detected with the reflection measurements, and the effects of the nontrivial topology  can be robustly measured in the presence of interactions. Moreover, the topologically protected end modes can be utilized for generation of a long-range entangled state  from a local excitation of a single transmon, opening interesting directions for future research in topological initialization of qubits and topological quantum state engineering.

\begin{acknowledgments}
{\it Acknowledgments.--} We acknowledge
the computational resources provided by
the Aalto Science-IT project and the 
financial support from the 
Academy of Finland Projects Nos.~331094 and 316619. The work is supported by the Foundation for Polish Science through the IRA Programme
co-financed by EU within SG OP. W.B. also acknowledges support by Narodowe Centrum Nauki 
(NCN, National Science Centre, Poland) Project No. 2019/34/E/ST3/00404.
M.~P. acknowledges the support of the Polish National Agency for Academic Exchange, the Bekker programme no: PPN/BEK/2020/1/00317, and
Ministerio de Ciencia y Innovation Agencia Estatal de Investigaciones (R\&D project CEX2019-000910-S, AEI/10.13039/501100011033, Plan National FIDEUA PID2019-106901GB-I00, FPI), Fundaci\'{o} Privada Cellex, Fundaci\'{o} Mir-Puig, and from Generalitat de Catalunya (AGAUR Grant No. 2017 SGR 1341, CERCA program).
FM acknowledges financial support from the Research Council of Norway (Grant No. 333937) through participation in the QuantERA ERA-NET Cofund in Quantum Technologies (project MQSens) implemented within the European Union’s Horizon 2020 Programme.

\end{acknowledgments}


\onecolumngrid

\newpage

\setcounter{section}{0}
\renewcommand*{\thesection}{}
\section*{Supplementary material for "Non-Hermitian topological quantum states in a reservoir-engineered transmon chain"}

\section{Driven Bose-Hubbard Hamiltonian in the rotating frame}

In the rotating wave approximation the driven Bose-Hubbard Hamiltonian  can be written as 
\begin{equation}
{\cal H}_{\rm lab}=\sum_{j=1}^{L}  \omega_j a_j^\dag a_j+\sum_{j=1}^{L-1} J_j (a_j^\dag a_{j+1}+a_{j+1}^\dag a_{j}) +\sum_{j=1}^L (a_j^{\dagger} f_j e^{-i \omega t} +a_j f_j^{*} e^{i \omega t})-\sum_{j=1}^{L} \frac{U_j}{2} a_j^\dag a_j (a_j^\dag a_j-1),
\end{equation}
where $\omega$ is the driving frequency. We can now switch to a rotating frame with a time-dependent transformation 
\begin{equation}
{\cal U}(t) = \exp(-i \omega t \sum_{j=1}^L a_j^\dag a_j).
\end{equation}
This way we obtain the driven Bose-Hubbard Hamiltonian in the rotating frame given in the main text
\begin{eqnarray}
{\cal H}&=& {\cal U}(t)^\dag {\cal H}_{\rm lab} {\cal U}(t)-i \ {\cal U}(t)^\dag \frac{d{\cal U}(t)}{dt} \nonumber \\ &=& \sum_{j=1}^{L}  (\omega_j-\omega) a_j^\dag a_j+\sum_{j=1}^{L-1} J_j (a_j^\dag a_{j+1}+a_{j+1}^\dag a_{j}) +\sum_{j=1}^L (a_j^{\dagger} f_j  +a_j f_j^{*})-\sum_{j=1}^{L} \frac{U_j}{2} a_j^\dag a_j (a_j^\dag a_j-1).
\end{eqnarray}

\section{Third quantization of the Liouvillian superoperator}

The Liouvillian superoperator contains operators acting from left and right on the density matrix. Therefore, we use notations
\begin{eqnarray}
A^{L}\rho :=  A\rho, \  A^{R}\rho  :=   \rho A.
\end{eqnarray}
We are interested in calculation of expectation values of observables $O$ of the form
${\rm Tr} (O \rho)$.
Therefore, the above definition allows us also to determine how the right and left operators act on the observables
\begin{eqnarray}
{\rm Tr} (O A^L \rho)&=&{\rm Tr} (O A \rho) \implies O A^L = O A \nonumber \\
{\rm Tr} (O A^R \rho)&=&{\rm Tr} (O \rho A) = {\rm Tr} (A O \rho ) \implies O A^R = A O.
\end{eqnarray}
For two operators we have
\begin{eqnarray}
A^{L}B^{L}\rho  =  AB\rho, \  A^{R}B^{R}\rho  =  \rho BA
\end{eqnarray}

The third quantization is based on new operators $\vec{a}_{\nu}$ and $\vec{a}'_{\nu}$ ($\nu=0,1$) defined as \cite{Prosen_2010}
\begin{eqnarray}
\vec{a}_{0} & := & \vec{a}^{L},  \ \ \  (\vec{a}'_{0})^T  :=  \vec{a}^{L\dagger}-\vec{a}^{R\dagger}, \ \ \ 
(\vec{a}_{1})^T  :=  \vec{a}^{R\dagger}, \ \  \ \vec{a}'_{1}  :=  \vec{a}^{R}-\vec{a}^{L}. \label{trans}
\end{eqnarray}
The inverse transformation is
\begin{eqnarray}
\vec{a}^{L} & := & \vec{a}_{0}, \ \ \  \vec{a}^{L\dagger}  :=  (\vec{a}'_{0})^T+(\vec{a}_{1})^T, \ \ \ 
\vec{a}^{R}  :=  \vec{a}_{0}+\vec{a}'_{1}, \ \ \  \vec{a}^{R\dagger}  :=  (\vec{a}_{1})^T. \label{inverse-trans}
\end{eqnarray}
The calculation of the commutators yields 
\begin{equation}
\left[a_{\nu,j},a'_{\nu',k}\right]\square=\delta_{\nu\nu'}\delta_{jk}\square, \ \left[a_{\nu,j},a_{\nu',k}\right]\square=0=\left[a'_{\nu,j},a'_{\nu',k}\right]\square.
\end{equation}
Therefore, the operators $\vec{a}_{\nu}$ and $\vec{a}'_{\nu}$ act similarly as the bosonic annihilation and creation operators, respectively.
Other useful properties of the operators and the density matrix are
\begin{equation}
1 \vec{a}'_{\nu}=0, \ \  \vec{a}_{\nu} \rho_0=0, \ \ {\rm Tr} \rho_0=1.
\end{equation}
where we have denoted the identity observable with $1$ and the density matrix of the vacuum with   $\rho_0=|\vec{0}\rangle \langle \vec{0}|$.
In particular, the properties described above allow to define dual Fock space for density matrices and observables \cite{Prosen_2010}
\begin{equation}
|\vec{m} \rangle = \prod_{\nu, j} \frac{(a'_{\nu, j})^{m_{\nu, j}}}{\sqrt{m_{\nu,j} !}} \rho_0, \ \ ( \vec{m} | = 1 \prod_{\nu, j} \frac{(a_{\nu, j})^{m_{\nu, j}}}{\sqrt{m_{\nu,j} !}} 
\end{equation}
with bi-orthonormality 
\begin{equation}
{\rm Tr}( \vec{m}' |  \vec{m} \rangle = \delta_{\vec{m}' , \vec{m} }.
\end{equation}
Within this dual Fock space the operators $\vec{a}_{\nu}$ and $\vec{a}'_{\nu}$ have the matrix representations of the bosonic annihilation and creation operators, respectively. Using these definitions the Liouvillian superoperator can be written as 
\begin{eqnarray}
{\cal L}&=&-i (\vec{a}'_{0})^T \hat{H}_{NH} \vec{a}_{0}+i ( \vec{a}'_{1})^T \hat{H}_{NH}^* \vec{a}_{1}-i(\vec{a}'_{0})^T\vec{f}+i \vec{f}^\dag\vec{a}'_{1} \nonumber \\ &&-i \sum_{j} \frac{U_j}{2} \left(2a'_{1,j}a_{1,j}a_{1,j}a_{0,j}+a'_{1,j}a'_{1,j}a_{1,j}a_{1,j}-a'_{0,j}a'_{0,j}a_{0,j}a_{0,j}-2a'_{0,j}a_{1,j}a_{0,j}a_{0,j}\right),
\label{eq:L}
\end{eqnarray}
where the non-Hermitian Hamiltonian is defined as
\begin{equation}
\hat{H}_{NH}=\hat{H}-i{\rm diag}\left(\vec{g}\right).
\end{equation}

\section{Topological invariant in the non-interacting limit $U_i=0$}

In the case of infinite number of $4$-site unit cells, $\omega_i=\omega_0$ and $J_i=J$, the topological invariant for the non-interacting non-Hermitian Hamiltonian  $\hat{H}_{NH}$ can be written as \cite{Wojtek19}
\begin{equation}
C=\frac{1}{2 \pi}\int_{-\infty}^{+\infty}d\eta\int_{0}^{2\pi}dk \ \Omega_{k,\eta},
\end{equation}
where 
\begin{equation*}
\Omega_{k,\eta} = \sum_{{n\leq 2\atop m>2}} {\rm Im}\frac{2\left\langle \psi_{k,\eta}^{n}\right|\!\partial_{k}\hat{H}^{{\rm eff}}\!\left|\psi_{k,\eta}^{m}\right\rangle \!\left\langle \psi_{k,\eta}^{m}\right|\!\partial_{\eta}\hat{H}^{{\rm eff}}\!\left|\psi_{k,\eta}^{n}\right\rangle }{\left(E_{k,\eta}^{(n)}-E_{k,\eta}^{(m)}\right)^{2}}
\end{equation*}
is the Berry curvature corresponding to 2D Hamiltonian 
\begin{equation}
\hat{H}^{{\rm eff}}(k,\eta) =\begin{pmatrix} 
\eta - g_{1} & -iJ & 0 & -iJe^{-ik}\\
iJ & -\eta + g_{2} & iJ & 0\\
0 & -iJ & \eta - g_{3} & -iJ\\
iJe^{ik} & 0 & iJ & -\eta + g_{4}
\end{pmatrix}
\end{equation}
with $\left|\psi_{k,\eta}^{n}\right\rangle$ and $E^{n}_{k, \eta}$ denoting the eigenstates and  eigenenergies of $\hat{H}^{{\rm eff}}(k,\eta)$ (sorted in ascending order of the eigenenergy). 

\section{Transmission and reflection matrix in the linear response regime}

The effects of interactions can be neglected in the linear response regime. By transforming the third-quantized bosonic operators as
\begin{equation}
\vec{a}_{\nu}=\vec{c}_{\nu}+\vec{x}_{\nu},\qquad\vec{a}'_{\nu}=\vec{c}'_{\nu}
\end{equation}
we obtain 
\begin{eqnarray}
{\cal L}&=&-i (\vec{c}'_{0})^T \hat{H}_{NH} \vec{c}_{0}+i ( \vec{c}'_{1})^T \hat{H}_{NH}^* \vec{c}_{1}-i(\vec{c}'_{0})^T\vec{f}+i \vec{f}^\dag\vec{c}'_{1} -i (\vec{c}'_{0})^T \hat{H}_{NH} \vec{x}_{0}+i ( \vec{c}'_{1})^T \hat{H}_{NH}^* \vec{x}_{1}.
\end{eqnarray}
Therefore, we can get rid of the driving terms by requiring that $\vec{x}_{\nu}$ ($\nu=0,1$) satisfy
\begin{eqnarray}
\vec{x}_{0}  =  -\hat{H}_{NH}^{-1}\vec{f}, \ \ \ 
\vec{x}_{1}  =  \vec{x}_{0}^{\star}.
\end{eqnarray}
This way we obtain
\begin{equation}
{\cal L}=-i (\vec{c}'_{0})^T \hat{H}_{NH} \vec{c}_{0}+i ( \vec{c}'_{1})^T \hat{H}_{NH}^* \vec{c}_{1}.
\label{Liouvillian-c}
\end{equation}

The Liouvillian superoperator given by Eq.~(\ref{Liouvillian-c})
can be diagonalized 
\begin{equation}
{\cal L}=\sum_k [ -i E_k b'_{0,k} b_{0,k} + i E_k^* b'_{1,k} b_{1,k} ]
\end{equation}
using a transformation
\begin{equation}
\vec{c}_0=\hat{U} \vec{b}_0, \ (\vec{c}'_0)^T=(\vec{b}_0')^T \hat{U}^{-1}, \  \vec{c}_1=\hat{U}^* \vec{b}_1, \ (\vec{c}'_1)^T= (\vec{b}_1')^T (\hat{U}^{-1})^*, 
\end{equation}
where the matrix $\hat{U}$ diagonalizes the non-Hermitian Hamiltonian
\begin{equation}
\hat{U}^{-1} \hat{H}_{NH} \hat{U}={\rm diag} (\vec{E}).
\end{equation}
Therefore, the non-Hermitian Hamiltonian fully determines the spectrum of the Liouvillian superoperator. 
Here, the operators  $\vec{b}_{\nu}$ and $\vec{b}'_{\nu}$ ($\nu=0,1$) satisfy the bosonic commutation relations 
\begin{equation}
\left[b_{\nu,j},b'_{\nu',k}\right]\square=\delta_{\nu\nu'}\delta_{jk}\square, \ \left[b_{\nu,j},b_{\nu',k}\right]\square=0=\left[b'_{\nu,j},b'_{\nu',k}\right]\square.
\end{equation}
and
\begin{equation}
1 \vec{b}'_{\nu}=0.
\end{equation}
Because all eigenenergies of $H_{NH}$ satisfy ${\rm Im} E_k < 0$, there exists a unique steady-state solution of the density matrix $\rho_S$ and its physical properties are determined by relations
\begin{equation}
b_{\nu, j}\rho_{S}=0, \ \ \   {\rm Tr}\rho_{S}=1. 
\end{equation}

To compute the output fields
\begin{equation}
\alpha_j^{\rm out}=\alpha_j^{\rm in}+\sqrt{\kappa_j} {\rm Tr} (a_j \rho_S),
\end{equation}
we need to calculate the expectation values 
\begin{equation}
{\rm Tr}\left(a_{j}\rho_{S}\right)={\rm Tr}\left(a_{0,j}\rho_{S}\right)={\rm Tr}\left(\left(c_{0,j}+x_{0,j}\right)\rho_{S}\right)=x_{0,j}=-(\hat{H}_{NH}^{-1}\vec{f})_j.  
\end{equation}
Using $f_j=-i\sqrt{\kappa_j}\alpha_j^{\rm in}$, we obtain  
\begin{equation}
\vec{\alpha}^{\rm out}= \hat{\Gamma}(\omega) \vec{\alpha}^{\rm in}, \ \ \ \hat{\Gamma}(\omega)=i \sqrt{\hat{K}}  \hat{H}_{NH}^{-1} \sqrt{\hat{K}}+\hat{I},  
\end{equation}
where $\hat{I}$ is the identity matrix and 
$\hat{K}={\rm diag}(\vec{\kappa})$.

\section{Expectation values of operators in the non-interacting limit}

In the noninteracting limit we can straightforwardly compute the expectation value of an arbitrary normal ordered product of creation and annihilation operators
\begin{eqnarray}
{\rm Tr}\left(a_{i_1}^{\dagger}...a_{i_N}^{\dagger}a_{j_1}...a_{j_M}\rho_{S}\right)&=&{\rm Tr}\left(a_{1, i_1}...a_{1, i_N}a_{0,j_1}...a_{0,j_M}\rho_{S}\right)=
x_{1,i_1}... x_{1,i_N} x_{0,j_1}... x_{0,j_M} \nonumber \\
&=& {\rm Tr}(a_{i_1}^{\dagger}\rho_{S})...{\rm Tr}(a_{i_N}^{\dagger}\rho_{S}){\rm Tr}(a_{j_1}\rho_{S})...{\rm Tr}(a_{j_M}\rho_{S}).
\label{exp-value}
\end{eqnarray}
The expectation value of the product of operators separates into a product of expectation values because in the third quantized formulation the steady state of the system is a product of a coherent state at each lattice site. The expectation values of the other operators can be obtained from this formula by utilizing the commutation relations of the annihilation and creation operators.
In particular, it follows from Eq.~(\ref{exp-value}) that the variance of the density in the steady state $n_j={\rm Tr}(a_j^\dag a_j \rho_S)$ satisfies
\begin{equation}
\Delta n_j^2={\rm Tr}( a_j^\dag a_j a_j^\dag a_j \rho_S ) - [{\rm Tr}( a_j^\dag a_j \rho_S )]^2 = {\rm Tr}( a_j^\dag a_j \rho_S ) =n_j 
\label{density-fluctuations}
\end{equation}
and the covariance of the annihilation operators satisfies
\begin{equation}
{\rm cov}[a_i, a_j]= {\rm Tr} (a_i a_j \rho_{S})-{\rm Tr} (a_i \rho_{S}){\rm Tr} (a_j \rho_{S})=0. \label{covariance}
\end{equation}

\section{Density matrix renormalization group approach  for the interacting problem}

The standard finite-size density renormalization group approach is described in Ref.~\cite{dmrg_lecture}. Here we use this approach, with a few modifications,  to solve the third-quantized Lindblad superoperator of Eq.~(\ref{eq:L}) for its right zero vector - a non-equilibrium stationary state. Firstly, we need to truncate the local Hilbert space of the boson operators $a_{\nu,j}$ to a finite value which becomes a convergence parameter $r$. Thus having fixed $r$ at a certain value we can have no more than $(r-1)$ third-quantized bosons per site. Now, in a finite Hilbert space we can always find a (right) zero vector of $\cal{L}$: vacuum is obviously a left zero state of $\cal{L}$ meaning that $\det{\cal L}^T=\det{\cal L}=0$, so a right zero vector must also exist. The search of it is thus equivalent to finding an eigenvector of $\cal{L}$ with zero eigenvalue.

\begin{figure}
    \centering
    \includegraphics[width=0.95\linewidth]{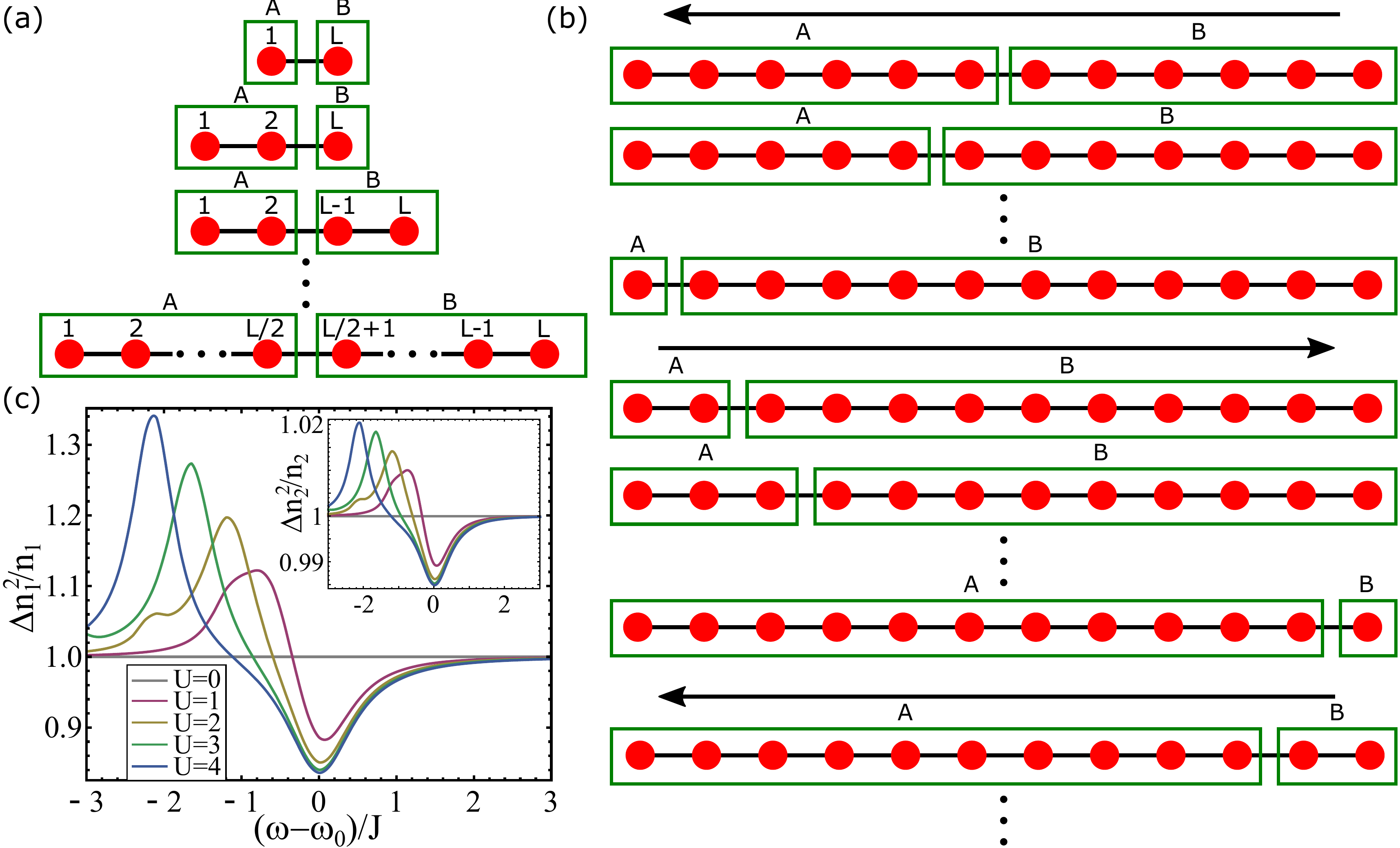}
    \caption{(a,b) Schematics of the DMRG. (a) The building of the system by expanding two blocks, A and B. (b) The sweeping loop for a finite open-ends system.
    (c) The variance of the density $\Delta n_j^2/n_j$  for lattice sites $j=1$ and $j=2$ (inset) as functions of frequency $\omega$ when the system is driven at first lattice site for $f=0.2 J$ and $U/J=:0, 1, 2, 3, 4$.} 
    \label{fig:variance}
\end{figure}

We adapt the general algorithm described in Ref.~\cite{dmrg_lecture}. In the first stage we want to increase the system size starting from two sites, $j=1$ and $j=L$, see Fig. \ref{fig:variance}(a). 
These two sites constitute blocks $A$ and $B$ at the first step of DMRG. Note that we treat $\nu=0,1$ as internal degree of freedom, so each dot in Figs.~\ref{fig:variance}(a,b) represents two bosonic degrees of freedom so that the dimension of the local Hilbert space at a given site $j$ is equal to $r^2$. At a given system size we  search for the eigenvector of $\cal{L}$ whose eigenvalue is closest to zero using Arnoldi algorithm. Then we use singular value decomposition (SVD), as usual in DMRG, and truncate the basis at the cutoff size of $d$, which is another convergence parameter. Now we expand the blocks by adding sites. Here we choose to do it asymmetrically by adding one site to the right in block A. By doing this we increase the dimension of the Hilbert space of block A by $r^2$ and of the whole Hilbert space by $r^2$ as well. If we decide to do it in a more standard way, by growing two blocks symmetrically, then the dimension grows by a factor of $r^4$ which is less handleable. In the next step we add one site to the left of the block B and we continue until we reach the desired system size $L$. All the blocks are stored in memory meaning that one can always recover the Linbladian and any observables for block A or B containing from $1$ up to $L/2$ sites.  

After obtaining the system in desired size $L$ we optimize the stationary state by performing sweeps, see Fig.~\ref{fig:variance}(b). We expand one block and shrink the other each time asking for the eigenstate with  eigenvalue closest to zero and performing SVD followed by the basis truncation. Shrinking of a block means reading a recorded block of the size that we need. Sweeps are done left and right, as shown in Fig.~\ref{fig:variance}(b), until the eigenvalue of our stationary state $| \rho_S \rangle$ is close enough to zero. Here the important difference with respect to the usual Hermitian case is how we calculate observables. From the third quantization we get that the stationary-state average of an operator $O$ is given by, 
\begin{equation}
    {\rm Tr} (O \rho_S) = \langle 0|O| \rho_S \rangle,
\end{equation}
where in bra we have vacuum of the $a_{\nu,j}$ bosons. Thus to calculate any averages in DMRG we need to know current representation of the vacuum state. This is not trivial because from the construction the converged basis is optimized to represent the stationary state only. Nevertheless having ${\cal L}$ in a truncated basis we can always ask for the lowest-magnitude eigenvector of ${\cal L}^T$. If the eigenvalues is as close to zero as for $| \rho_S \rangle$ we can conclude that we have a correct representation of the vacuum state. In our case this approach always works because we drive the system only at one site so that $| \rho_S \rangle$ has large component in the $| 0 \rangle$ direction.

Using DMRG we find that
in the presence of the interactions the simple results for the expectation values (\ref{exp-value}), (\ref{density-fluctuations}) and (\ref{covariance}) are no longer valid. Instead, the steady state of the interacting system is a correlated quantum state with entanglement between the transmons. As an example we show in Fig.~\ref{fig:variance}(c) that the relation (\ref{density-fluctuations}) is not satisfied in the interacting system.
Here we have used the convergence parameters $r=30$ and $d=4$ for $U/J=:0,1,2,3$ and $d=5$ for $U/J=4$. With these parameters we were able to keep the eigenvalue of the stationary state below $10^{-9} J$.

\section{Time evolution of the density matrix} 

Our starting point in the consideration of the dynamics is the  Lindblad master equation in the absence of driving and interactions, but allowing time-dependence of the parameters so that the dissipation can be switched on and off as discussed in the main text. Using the third quantized operators it can be written as 
\begin{equation}
\frac{d}{dt}\rho(t)={\cal L}(t) \rho(t), \ \ \ {\cal L}(t)=-i (\vec{a}'_{0})^T H_{NH}(t)\vec{a}{}_{0}+i (\vec{a}'_{1})^T \hat{H}_{NH}^*(t)\vec{a}{}_{1}. \label{time-dependent-H_NH}
\end{equation}
The solution of Eq.~(\ref{time-dependent-H_NH}) can be written as
\begin{equation}
\rho(t) ={\cal T}\exp\left[\int_0^t dt' {\cal L} (t') \right] \rho(0) = {\cal T} \exp\left[-i \int_0^t dt' (\vec{a}'_{0})^T \hat{H}_{NH}(t') \vec{a}{}_{0}\right] {\cal T} \exp\left[i \int_0^t dt' (\vec{a}'_{1})^T \hat{H}_{NH}^*(t')\vec{a}{}_{1}\right] \rho(0),
\end{equation}
where ${\cal T}$ is the time ordering operator. 
By applying this time-evolution to an initial coherent state in the third quantized operator representation of the density matrix
\begin{equation}
\rho_{\vec{z}}(0)=\rho_{\vec{z}_0, \vec{z}_1}(0) =\exp\left[\vec{z}^T_0 \vec{a}'_{0}+\vec{z}_1^T \vec{a}'_{1}\right]\rho_0 
\end{equation}
we obtain 
\begin{equation}
\rho_{\vec{z}}(t) = \exp\left\{\vec{z}^T_{0} \hat{U}(t)  \vec{a}'_{0}+\vec{z}^T_{1} \hat{U}^*(t) \vec{a}'_{1}\right\}\rho_0,
\end{equation}
where
\begin{equation}
\hat{U}(t)={\cal T} \exp\left[-i \int_0^t dt' \hat{H}_{NH}(t') \right].
\end{equation}
Thus, site-wise product of coherent states keeps this structure under time evolution. Therefore, we can straightforwardly evaluate the time-dependent expectation values of operators once we have an expansion of the initial state $\rho(0)$ in terms of the coherent states $\rho_{\vec{z}}(0)$.  

If the state of the transmons is  initialized into a site-wise product of coherent states in the second quantized formalism, the corresponding density matrix  translates into a site-wise product of coherent states also in the third quantized formalism. Since this structure is kept during the time-evolution there is no entanglement between the transmons developing as a function time.

As discussed in the main text another possibility is to initialize one of the transmons into a Fock state in the second quantized formalism. The Fock state density matrix  $\rho_{n_i}(0)=\frac{1}{n!}(a_{i}^{\dagger})^n\left|0\right\rangle \left\langle 0\right|a_{i}^{n}$ in the third quantized formalism can be expressed as
\begin{equation}
\rho_{n_i}(0)=\frac{1}{n!}\left(a'_{0,i}+a{}_{1,i}\right)^{n}\left(a_{0,i}+a'_{1,i}\right)^{n} \rho_0 =\sum_{k=0}^{n}\frac{(a'_{0,i})^{n-k}a{}_{1,i}^{k}(a'_{1,i})^{n}}{k!(n-k)!} \rho_0 =\sum_{k=0}^{n}\frac{n!}{k!(n-k)!}\left|\left(n-k\right)_{0,i}\left(n-k\right)_{1,i}\right\rangle, 
\end{equation}
and after expressing this with the help of the coherent states we obtain the time-dependent density matrix
\begin{equation}
\rho_{n_i}(t)=\frac{1}{\pi^{2}}\sum_{k=0}^{n}\frac{n!}{k!(n-k)!}\int d^2 z_{0,i} \int d^2 z_{1,i} e^{-|z_{1,i}|^2-|z_{1,i}|^2} \frac{z_{0,i}^{\star(n-k)}z_{1,i}^{\star(n-k)}}{(n-k)!}\rho_{\vec{z}^{(i)}}(t), \ \ \vec{z}^{(i)}_{0,k}=\delta_{i,k} z_{0,i} \ \ \vec{z}^{(i)}_{1,k}=\delta_{i,k} z_{1,i}.
\end{equation}
We are interested on the non-local  entanglement between the transmons $1$ and $L$, and therefore we compute the reduced density matrix by taking the partial trace of $\rho_{n_i}(t)$ over the other transmons
\begin{eqnarray}
\rho^{\rm red}_{n_i}(t) &=& \frac{1}{\pi^{2}}\sum_{k=0}^{n}\frac{n!}{k!(n-k)!}\int d^2 z_{0,i} \int d^2 z_{1,i} e^{-|z_{1,i}|^2-|z_{1,i}|^2} \frac{z_{0,i}^{\star(n-k)}z_{1,i}^{\star(n-k)}}{(n-k)!} e^{z_{0,i} \hat{U}_{i1} a'_{0,1}} e^{z_{1,i} \hat{U}^*_{i1} a'_{1,1}} e^{z_{0,i} \hat{U}_{iL} a'_{0,L}} e^{z_{1,i} \hat{U}^*_{iL} a'_{1,L}} \rho_0  \nonumber \\ 
&=&\sum_{k=0}^{n}\frac{n!}{(n-k)!}\frac{1}{k!^{2}}\left(\hat{U}_{1i}a'_{0,1}+\hat{U}_{Li} a'_{0,L}\right)^{k}\left(\hat{U}^*_{1i}a'_{1,1}+\hat{U}^*_{Li} a'_{1,L}\right)^{k} \rho_0.
\end{eqnarray}
In the second quantized form this density matrix can be written as
\begin{eqnarray}
\rho^{\rm red}_{n_i}(t)&=&\sum_{k=0}^{n}\frac{n!}{(n-k)!}\frac{1}{k!^{2}}\left[\hat{U}_{1i}\left(a_{1}^{L\dagger}-a_{1}^{R\dagger}\right)+\hat{U}_{Li}\left(a_{L}^{L\dagger}-a_{L}^{R\dagger}\right)\right]^{k}\left|0\right\rangle \left\langle 0\right|\left[\hat{U}^*_{1i}a_{1}+\hat{U}^*_{Li}a_{L}\right]^{k} \nonumber \\
&=&\sum_{k=0}^{n}\frac{n!}{(n-k)!}\frac{1}{k!^{2}}\sum_{p=0}^{k}\frac{k!}{p!(k-p)!}\left[\hat{U}_{1i} a_{1}^{\dagger}+\hat{U}_{Li} a_{L}^{\dagger}\right]^{k-p}\left|0\right\rangle \left\langle 0\right|\left[\hat{U}^*_{1i}a_{1}+\hat{U}^*_{Li} a_{L}\right]^{k}\left[-\hat{U}_{1i} a_{1}^{\dagger}-\hat{U}_{Li}a_{L}^{\dagger}\right]^{p} \nonumber \\ 
&=& \sum_{k=0}^{n}\frac{n!}{(n-k)!}\sum_{p=0}^{k}\frac{1}{p!(k-p)!} (-1)^p \left(P_{1,i}+P_{L,i}\right)^{k} | (k-p)_u \rangle \langle (k-p)_u|=\sum_{q=0}^{n}A_{q}\left|q_u\right\rangle \left\langle q_u\right|, \label{red-supp}
\end{eqnarray}
where 
\begin{equation}
A_{q}=\left(1-P_{1,i}-P_{L,i}\right)^{n-q}\left(P_{1,i}+P_{L,i}\right)^{q}\begin{pmatrix}n\\
q
\end{pmatrix}, \ \ |q_u\rangle =\frac{(\hat{U}_{1i} a_{1}^{\dagger}+\hat{U}_{Li} a_{L}^{\dagger})^{q}}{\sqrt{q!\left(P_{1,i}+P_{L,i}\right)^{q}}}\left|0\right\rangle , \ \ P_{1,i}=|\hat{U}_{1i}|^2, \ \ P_{L,i}=|\hat{U}_{Li}|^2. \label{A-q-supp}
\end{equation}

Similarly, we can also compute the reduced density matrix for a single transmon at one end of the chain 
\begin{equation}
\rho^{\rm red, 1}_{n_i}(t)=\sum_{p=0}^{n}B_{1, p}\left|p\right\rangle _{1}\left\langle p\right|_{1}, \ \rho^{\rm red, L}_{n_i}(t)=\sum_{p=0}^{n}B_{L, p}\left|p\right\rangle _{1}\left\langle p\right|_{1}, \  B_{1(L), p}=\left(P_{1(L),i}\right)^{p}\left(1-P_{1(L),i}\right)^{n-p}\begin{pmatrix}n\\
p
\end{pmatrix}. \label{red-red-supp}
\end{equation}

\section{Non-local entanglement}

We can characterize the entanglement between the sites $1$ and $L$ in the reduced density matrix (\ref{red-supp}) using different measures of entanglement. In the main text we  concentrated on the mutual information $I(1:L)$, which is defined as  
\begin{equation}
I(1:L)(t)=S\left(\rho^{\rm red, 1}_{n_i}(t)\right)+S\left(\rho^{\rm red, L}_{n_i}(t)\right)-S\left(\rho^{\rm red}_{n_i}(t)\right),
\end{equation}
where $S(\rho)$ is the 
von-Neumann entropy
\begin{equation}
S(\rho)=-{\rm Tr}\rho\ln\rho.
\end{equation}
If the system is in a simple product state we have $I(1:L)=0$, and therefore the mutual information characterizes correlations between the end transmons.

By using Eqs.~(\ref{red-supp}) and (\ref{red-red-supp}) we obtain 
\begin{equation}
I(1:L)(t)=\sum_{q=0}^{n}A_{q}(t)\ln A_{q}(t)-\sum_{q=0}^{n}B_{1,q}(t)\ln 
B_{1,q}(t)-\sum_{q=0}^{n}B_{L, q}(t)\ln B_{L, q}(t), \label{I1L-supp}
\end{equation}
where $A_{q}(t)$, $B_{1,q}(t)$ and $B_{L,q}(t)$ are given by Eqs.~(\ref{A-q-supp}) and (\ref{red-red-supp}). The formula (\ref{I1L-supp}) has been used to compute the mutual information in the figures shown in the main text. 

\begin{figure}
    \centering
    \includegraphics[width=0.95\linewidth]{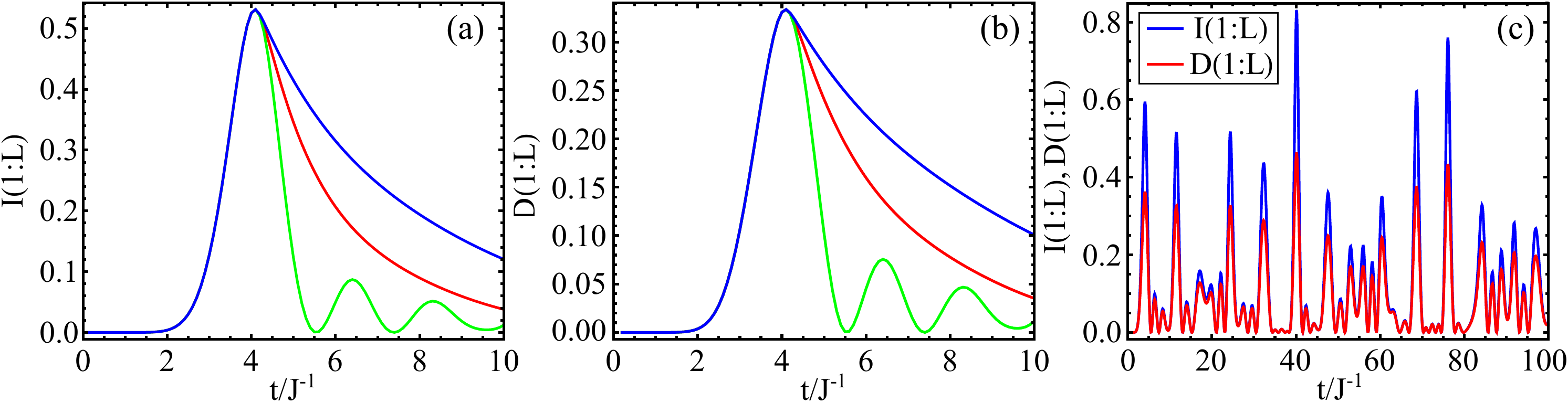}
    \caption{(a) Mutual information $I(1:L)$ vs $t$  for a reference system with $g_{B, {\rm max}}=0.01 J$ (green) and non-trivial phases  with $x=-1$ and   $g_{B, {\rm max}}/J=:5, 10$ (red, blue). (b) Quantum discord $\delta(1:L)$ vs $t$ for the same dissipation patterns. (c) Comparison of mutual information $I(1:L)$ and quantum discord $\delta(1:L)$ for a uniform system with no dissipation, $g_{A,B}=0$. In addition to the usual dissipation patterns we have added one extra site in middle of the chain with $g_{L/2+1}=g_B$, which is excited to a Fock state with $N=1$  at $t=0$.} 
    \label{fig:discord}
\end{figure}

In general the mutual information $I(1:L)$ is not a good measure of quantum entanglement because it can be non-zero also in the case of classically correlated state. However, we have checked that in our case the mutual information originates from the quantum entanglement. For this purpose we have considered the quantum discord \cite{QuantumDiscord}
\begin{equation}
\delta\left(1:L\right)(t)=S\left(\rho^{\rm red, 1}_{n_i}(t)\right)-S\left(\rho^{\rm red}_{n_i}(t)\right)+S\left(\rho^{\rm red, L}_{n_i}(t)|\left\{ \Pi_{1}^{j}\right\}\right)
\end{equation}
where the conditional quantum entropy defined as 
\begin{equation}
S\left(\rho^{\rm red, L}_{n_i}(t)|\left\{ \Pi_{1}^{j}\right\}\right)=\sum_{j=0}^{n}p_{j}S\left(\frac{1}{p_{j}}\Pi_{1}^{j}\rho^{\rm red}_{n_i}(t)\Pi_{1}^{j}\right),
\end{equation}
$p_{j}$ are the  probabilities of the measurements
\begin{equation}
p_{j}={\rm Tr}\left\{ \Pi_{1}^{j}\rho^{\rm red}_{n_i}(t)  \right\}
\end{equation}
and $\Pi_{1}^{j}$ is a complete set of orthogonal projectors at site $1$. Here, we concentrate on the case $n=1$, where the projectors can be written as 
\begin{equation}
\Pi_{1}^{1}=\left(\cos\theta\left|0\right\rangle _{1}+e^{i\phi}\sin\theta\left|1\right\rangle _{1}\right)\left(\cos\theta\left\langle 0\right|_{1}+e^{-i\phi}\sin\theta\left\langle 1\right|_{1}\right), \label{Pi1}
\end{equation}
\begin{equation}
\Pi_{1}^{2}=\left(e^{-i\phi}\sin\theta\left|0\right\rangle _{1}-\cos\theta\left|1\right\rangle _{1}\right)\left(e^{i\phi}\sin\theta\left\langle 0\right|_{1}-\cos\theta\left\langle 1\right|_{1}\right). \label{Pi2}
\end{equation}
In general the discord depends on the basis taken
to make measurement at site $1$, i.e. on the values of $\theta$ and $\phi$ in Eqs.~(\ref{Pi1}) and (\ref{Pi2}), and therefore a meaningful measure of entanglement is 
\[
D=\min_{\left\{ \Pi_{j}^{1}\right\} }\delta\left(1:L\right)
\]
For $n=1$ this means minimization over angles $\theta$ and $\phi$,
which is numerically feasible. We numerically find that 
in our case dependence on the choice of $\left\{ \Pi_{j}^{1}\right\} $
is weak and $D$ shows very similar behavior as $I(1:L)$. In Fig.~\ref{fig:discord} 
we show the comparison between these two quantities in the case of time-dependent
dissipation which was already considered in the main text and in the case of no 
dissipation at all. We note that quantum discord always takes slightly lower or the same
values as mutual information but the shape of the curves is the same.  In the main text we have reported only $I(1:L)$ because of the  
rather high complexity of calculation of $D$ in the case of large number of bosons. 

\section{Initialization and measurement with circuit quantum electrodynamics}

In this section we will briefly review the initialization and measurement of the transmons with circuit quantum electrodynamics  based on Ref.~\cite{Koch07, Blais21}. For this purpose, we consider our transmon chain coupled resonant mode of a cavity
\begin{equation}
{\cal H}_{\rm {cqe}}=\vec{a}^{\dagger}\hat{H}\vec{a}-\sum_{j=1}^{L} \frac{U_j}{2} a_j^\dag a_j (a_j^\dag a_j-1)+\omega_c c^\dag c+ \sum_{j} {\cal G}_j (c^\dag a_j+c a_j^\dag), \label{Ham-full-cavity}
\end{equation}
where
\begin{equation}
\hat{H}=\begin{pmatrix} \omega_1 & J_1\\
J_1 & \omega_2 & J_2\\
 & J_2 & \ddots & \ddots\\
 &  & \ddots &  & J_{L-1}\\
 &  &  & J_{L-1} & \omega_L
\end{pmatrix}
\end{equation}
and ${\cal G}_j$ describes the coupling of the transmon $j$ and the cavity mode. The transmon frequencies $\omega_i$ can be tuned as much as $1$ GHz in as little as $10-20$ ns \cite{Blais21}. Therefore, we can assume that all transmon frequencies  are most of the time tuned very far away from the cavity frequency $\omega_c$ but we can selectively tune specific transmon frequencies  close to the cavity frequency $\omega_c$ so that the coupling between these systems become non-negligible. Notice that even when $\omega_i$ is tuned close to $\omega_c$ we will still stay in the dispersive regime $\omega_i-\omega_c - n U\ll {\cal G}_i$ ($n=0,1, ..., n_{\rm max}$, where the state-dependent cut-off occupation number  $n_{\rm max}$ is defined so that as a good approximation the occupations above $n_{max}$ can be neglected). However, the other transmons are detuned so much more away from $\omega_c$ that their effects can be completely neglected. We also assume that $J_j$ are sufficiently small  so that while transmons are tuned close to $\omega_c$ the couplings to the other transmons can be neglected. Therefore, we assume that during these operations the dissipations $g_i$ are always turned off.

\subsection{Initialization and measurement of the state of a single transmon}

We first consider a single transmon frequency $\omega_i$ tuned close to $\omega_c$. In this case the effective Hamiltonian describing the transmon and the cavity takes a form
\begin{equation}
{\cal H}_{{\rm cqe},i}=\omega_i a_i^\dag a_i-\frac{U_i}{2} a_i^\dag a_i (a_i^\dag a_i-1)+\omega_c c^\dag c+ {\cal G}_i (c^\dag a_i+c a_i^\dag). \label{Ham-full-cavity-i}
\end{equation}

In the dispersive limit one can utilize the Schrieffer-Wolff transformation \cite{Blais21} to show that the Hamiltonian (\ref{Ham-full-cavity-i})
is well approximated by 
\begin{equation}
{\cal H}_{{\rm disp},i}=\omega_i a_i^\dag a_i-\frac{U_i}{2} a_i^\dag a_i (a_i^\dag a_i-1)+\omega_c c^\dag c+  \frac{ {\cal G}_i^2 \  a_i^\dag a_i}{\omega_i-\omega_c-U_i(a_i^\dag a_i-1)} +  \left[\frac{ {\cal G}_i^2 \  a_i^\dag a_i}{\omega_i-\omega_c-U_i(a_i^\dag a_i-1)}-\frac{ {\cal G}_i^2 \  (a_i^\dag a_i+1)}{\omega_i-\omega_c-U_i a_i^\dag a_i}\right] c^\dag c. \label{Ham-disp-i}
\end{equation}
Therefore, the resonator frequency is renormalized so that it is dependent on the Fock state $n$ of the transmon as 
\begin{equation}
\omega_{c}^{\rm eff}(n)=\omega_c+\frac{ {n \cal G}_i^2 }{\omega_i-\omega_c-(n-1) U_i}-\frac{ (n+1) {\cal G}_i^2 }{\omega_i-\omega_c-n U_i}=\omega_c-\frac{  {\cal G}_i^2 (\omega_i-\omega_c +U_i) }{(\omega_i-\omega_c-n U_i)(\omega_i-\omega_c-(n-1) U_i)}.
\end{equation}
In this situation initializing the transmon to a superposition of the Fock states $|n\rangle$ 
\begin{equation}
|\psi_{\rm transmon} \rangle = \sum_{n=0}^{n_{\rm max}} c_n |n \rangle
\end{equation}
and driving the cavity leads to an entangled transmon-resonator state of the form 
\begin{equation}
|\psi \rangle = \sum_{n=0}^{n_{\rm max}} c_n |n, \alpha_n\rangle,
\end{equation}
where $\alpha_n$ describes  the transmon-state $n$ dependent coherent state of the resonator. Assuming that $\omega_{c}^{\rm eff}(n)$ ($n=0,1...,n_{\rm max}$) are sufficiently different from each other the different states of the microwave field $\alpha_n$ can be resolved in heterodyne detection, and this measurement serves as a quantum nondemolition  measurement of $\hat{n} = a^\dag a$. Thus the transmon is projected in the measurement to the Fock state $|n \rangle$ with probability $|c_n|^2$. In the case of two lowest states of the transmons the measurement can be done with above 99$\%$ fidelity in less than $100$ ns \cite{Blais21}. In the case of multiple states of the transmon $|n \rangle$ ($n=0,1...,n_{\rm max}$) we expect the fidelities to be smaller and the required measurement times to be longer.

Alternatively, one can also initialize the transmon to a chosen excited state utilizing the anharmonicity $U_i$. Namely, the frequency difference between states $|n\rangle$ and $|n+1\rangle$ is different for each value of $n$, and therefore one can  sequentially apply $\pi$ pulses at the corresponding resonance frequencies to go from the ground state to a chosen excited state $|0\rangle \to |1\rangle \to .. \to |n\rangle$ \cite{Elder20}. 

\subsection{Two transmon tomography}

After we have dynamically generated the entangled state 
the reduced density matrix of the transmons $1$ and $L$ can be experimentally determined by using quantum state tomography, but it requires that we prepare the transmons $1$ and $L$ to the same state many times by repeating the same  procedure. 
For this purpose we also always need to decouple the transmons $1$ and $L$ from the rest of the chain by tuning the resonance frequencies $\omega_1$ and $\omega_L$ sufficiently far from the resonance frequencies of the other transmons.

The quantum state tomography can then  be performed by applying single transmon gates and  correlating the single transmon measurements of the transmons $1$ and $L$. In the case of two level systems the typical approach  is to measure the probabilities of the qubits being in states $|0 \rangle$ and $|1 \rangle$ e.g.~in $x$-, $y$- and $z$- basis, and then the density matrix is constructed using maximum-likelihood estimation of $\rho$ (see e.g.~\cite{Haffner05, Steffen06}). This method can be easily generalized to our case by noticing that we can formally express the states of the transmon $|n\rangle$ as tensor products of qubit states. The single transmon operations and measurements can be performed for example by driving the transmons with suitable pulses and utilizing the dispersive readout as discussed above.   

Alternatively, the two transmon tomography can be performed using joint dispersive readout \cite{Filipp09}. In this scheme one utilizes the Schrieffer-Wolff transformation for two qubits coupled to the same resonator leading to the Hamiltonian
\begin{eqnarray}
{\cal H}_{{\rm disp},1, L}&=&\omega_1 a_1^\dag a_1-\frac{U_1}{2} a_1^\dag a_1 (a_1^\dag a_1-1)+ \omega_L a_L^\dag a_L-\frac{U_L}{2} a_L^\dag a_L (a_L^\dag a_L-1)+\omega_c c^\dag c \nonumber \\ && +  \frac{ {\cal G}_1^2 \  a_1^\dag a_1}{\omega_1-\omega_c-U_1(a_1^\dag a_1-1)}+ \frac{ {\cal G}_L^2 \  a_L^\dag a_L}{\omega_L-\omega_c-U_L(a_L^\dag a_L-1)}  \\ && +  \left[\frac{ {\cal G}_1^2 \  a_1^\dag a_1}{\omega_1-\omega_c-U_1(a_1^\dag a_1-1)}-\frac{ {\cal G}_1^2 \  (a_1^\dag a_1+1)}{\omega_1-\omega_c-U_1 a_1^\dag a_1}+\frac{ {\cal G}_L^2 \  a_L^\dag a_L}{\omega_L-\omega_c-U_L(a_L^\dag a_L-1)}-\frac{ {\cal G}_L^2 \  (a_L^\dag a_L+1)}{\omega_L-\omega_c-U_L a_L^\dag a_L}\right] c^\dag c. \nonumber \label{Ham-disp-!L}
\end{eqnarray}
in the dispersive limit. Because the renormalized resonator frequency is now dependent on the state of both transmons, the quadratures of the resonator field  correspond to  an operator which comprises also two transmon correlation terms \cite{Filipp09}. Therefore, it is possible to construct the density matrix by performing single transmon operations and the averaged measurements of the transmission amplitudes without the need for single-shot readout of individual transmons \cite{Filipp09}. 

\bibliography{bibliography}

\begin{thebibliography}{54}%
\makeatletter
\providecommand \@ifxundefined [1]{%
 \@ifx{#1\undefined}
}%
\providecommand \@ifnum [1]{%
 \ifnum #1\expandafter \@firstoftwo
 \else \expandafter \@secondoftwo
 \fi
}%
\providecommand \@ifx [1]{%
 \ifx #1\expandafter \@firstoftwo
 \else \expandafter \@secondoftwo
 \fi
}%
\providecommand \natexlab [1]{#1}%
\providecommand \enquote  [1]{``#1''}%
\providecommand \bibnamefont  [1]{#1}%
\providecommand \bibfnamefont [1]{#1}%
\providecommand \citenamefont [1]{#1}%
\providecommand \href@noop [0]{\@secondoftwo}%
\providecommand \href [0]{\begingroup \@sanitize@url \@href}%
\providecommand \@href[1]{\@@startlink{#1}\@@href}%
\providecommand \@@href[1]{\endgroup#1\@@endlink}%
\providecommand \@sanitize@url [0]{\catcode `\\12\catcode `\$12\catcode
  `\&12\catcode `\#12\catcode `\^12\catcode `\_12\catcode `\%12\relax}%
\providecommand \@@startlink[1]{}%
\providecommand \@@endlink[0]{}%
\providecommand \url  [0]{\begingroup\@sanitize@url \@url }%
\providecommand \@url [1]{\endgroup\@href {#1}{\urlprefix }}%
\providecommand \urlprefix  [0]{URL }%
\providecommand \Eprint [0]{\href }%
\providecommand \doibase [0]{https://doi.org/}%
\providecommand \selectlanguage [0]{\@gobble}%
\providecommand \bibinfo  [0]{\@secondoftwo}%
\providecommand \bibfield  [0]{\@secondoftwo}%
\providecommand \translation [1]{[#1]}%
\providecommand \BibitemOpen [0]{}%
\providecommand \bibitemStop [0]{}%
\providecommand \bibitemNoStop [0]{.\EOS\space}%
\providecommand \EOS [0]{\spacefactor3000\relax}%
\providecommand \BibitemShut  [1]{\csname bibitem#1\endcsname}%
\let\auto@bib@innerbib\@empty
\bibitem [{\citenamefont {El-Ganainy}\ \emph {et~al.}(2018)\citenamefont
  {El-Ganainy}, \citenamefont {Makris}, \citenamefont {Khajavikhan},
  \citenamefont {Musslimani}, \citenamefont {Rotter},\ and\ \citenamefont
  {Christodoulides}}]{El-Ganainy2018}%
  \BibitemOpen
  \bibfield  {author} {\bibinfo {author} {\bibfnamefont {R.}~\bibnamefont
  {El-Ganainy}}, \bibinfo {author} {\bibfnamefont {K.~G.}\ \bibnamefont
  {Makris}}, \bibinfo {author} {\bibfnamefont {M.}~\bibnamefont {Khajavikhan}},
  \bibinfo {author} {\bibfnamefont {Z.~H.}\ \bibnamefont {Musslimani}},
  \bibinfo {author} {\bibfnamefont {S.}~\bibnamefont {Rotter}},\ and\ \bibinfo
  {author} {\bibfnamefont {D.~N.}\ \bibnamefont {Christodoulides}},\ }\bibfield
   {title} {\bibinfo {title} {{Non-Hermitian physics and PT symmetry}},\ }\href
  {https://doi.org/10.1038/nphys4323} {\bibfield  {journal} {\bibinfo
  {journal} {Nature Physics}\ }\textbf {\bibinfo {volume} {14}},\ \bibinfo
  {pages} {11} (\bibinfo {year} {2018})}\BibitemShut {NoStop}%
\bibitem [{\citenamefont {Gong}\ \emph {et~al.}(2018)\citenamefont {Gong},
  \citenamefont {Ashida}, \citenamefont {Kawabata}, \citenamefont {Takasan},
  \citenamefont {Higashikawa},\ and\ \citenamefont {Ueda}}]{Gong18}%
  \BibitemOpen
  \bibfield  {author} {\bibinfo {author} {\bibfnamefont {Z.}~\bibnamefont
  {Gong}}, \bibinfo {author} {\bibfnamefont {Y.}~\bibnamefont {Ashida}},
  \bibinfo {author} {\bibfnamefont {K.}~\bibnamefont {Kawabata}}, \bibinfo
  {author} {\bibfnamefont {K.}~\bibnamefont {Takasan}}, \bibinfo {author}
  {\bibfnamefont {S.}~\bibnamefont {Higashikawa}},\ and\ \bibinfo {author}
  {\bibfnamefont {M.}~\bibnamefont {Ueda}},\ }\bibfield  {title} {\bibinfo
  {title} {{Topological Phases of Non-Hermitian Systems}},\ }\href
  {https://doi.org/10.1103/PhysRevX.8.031079} {\bibfield  {journal} {\bibinfo
  {journal} {Phys. Rev. X}\ }\textbf {\bibinfo {volume} {8}},\ \bibinfo {pages}
  {031079} (\bibinfo {year} {2018})}\BibitemShut {NoStop}%
\bibitem [{\citenamefont {Lieu}(2018)}]{Lieu18}%
  \BibitemOpen
  \bibfield  {author} {\bibinfo {author} {\bibfnamefont {S.}~\bibnamefont
  {Lieu}},\ }\bibfield  {title} {\bibinfo {title} {{Topological symmetry
  classes for non-Hermitian models and connections to the bosonic
  Bogoliubov--de Gennes equation}},\ }\href
  {https://doi.org/10.1103/PhysRevB.98.115135} {\bibfield  {journal} {\bibinfo
  {journal} {Phys. Rev. B}\ }\textbf {\bibinfo {volume} {98}},\ \bibinfo
  {pages} {115135} (\bibinfo {year} {2018})}\BibitemShut {NoStop}%
\bibitem [{\citenamefont {Zhou}\ and\ \citenamefont {Lee}(2019)}]{Zhou19}%
  \BibitemOpen
  \bibfield  {author} {\bibinfo {author} {\bibfnamefont {H.}~\bibnamefont
  {Zhou}}\ and\ \bibinfo {author} {\bibfnamefont {J.~Y.}\ \bibnamefont {Lee}},\
  }\bibfield  {title} {\bibinfo {title} {{Periodic table for topological bands
  with non-Hermitian symmetries}},\ }\href
  {https://doi.org/10.1103/PhysRevB.99.235112} {\bibfield  {journal} {\bibinfo
  {journal} {Phys. Rev. B}\ }\textbf {\bibinfo {volume} {99}},\ \bibinfo
  {pages} {235112} (\bibinfo {year} {2019})}\BibitemShut {NoStop}%
\bibitem [{\citenamefont {Kawabata}\ \emph {et~al.}(2019)\citenamefont
  {Kawabata}, \citenamefont {Shiozaki}, \citenamefont {Ueda},\ and\
  \citenamefont {Sato}}]{Kawabata19}%
  \BibitemOpen
  \bibfield  {author} {\bibinfo {author} {\bibfnamefont {K.}~\bibnamefont
  {Kawabata}}, \bibinfo {author} {\bibfnamefont {K.}~\bibnamefont {Shiozaki}},
  \bibinfo {author} {\bibfnamefont {M.}~\bibnamefont {Ueda}},\ and\ \bibinfo
  {author} {\bibfnamefont {M.}~\bibnamefont {Sato}},\ }\bibfield  {title}
  {\bibinfo {title} {{Symmetry and Topology in Non-Hermitian Physics}},\ }\href
  {https://doi.org/10.1103/PhysRevX.9.041015} {\bibfield  {journal} {\bibinfo
  {journal} {Phys. Rev. X}\ }\textbf {\bibinfo {volume} {9}},\ \bibinfo {pages}
  {041015} (\bibinfo {year} {2019})}\BibitemShut {NoStop}%
\bibitem [{\citenamefont {Song}\ \emph {et~al.}(2019)\citenamefont {Song},
  \citenamefont {Yao},\ and\ \citenamefont {Wang}}]{Song2019}%
  \BibitemOpen
  \bibfield  {author} {\bibinfo {author} {\bibfnamefont {F.}~\bibnamefont
  {Song}}, \bibinfo {author} {\bibfnamefont {S.}~\bibnamefont {Yao}},\ and\
  \bibinfo {author} {\bibfnamefont {Z.}~\bibnamefont {Wang}},\ }\bibfield
  {title} {\bibinfo {title} {{Non-Hermitian Skin Effect and Chiral Damping in
  Open Quantum Systems}},\ }\href
  {https://doi.org/10.1103/PhysRevLett.123.170401} {\bibfield  {journal}
  {\bibinfo  {journal} {Phys. Rev. Lett.}\ }\textbf {\bibinfo {volume} {123}},\
  \bibinfo {pages} {170401} (\bibinfo {year} {2019})}\BibitemShut {NoStop}%
\bibitem [{\citenamefont {Brzezicki}\ and\ \citenamefont
  {Hyart}(2019)}]{Wojtek19}%
  \BibitemOpen
  \bibfield  {author} {\bibinfo {author} {\bibfnamefont {W.}~\bibnamefont
  {Brzezicki}}\ and\ \bibinfo {author} {\bibfnamefont {T.}~\bibnamefont
  {Hyart}},\ }\bibfield  {title} {\bibinfo {title} {{Hidden Chern number in
  one-dimensional non-Hermitian chiral-symmetric systems}},\ }\href
  {https://doi.org/10.1103/PhysRevB.100.161105} {\bibfield  {journal} {\bibinfo
   {journal} {Phys. Rev. B}\ }\textbf {\bibinfo {volume} {100}},\ \bibinfo
  {pages} {161105} (\bibinfo {year} {2019})}\BibitemShut {NoStop}%
\bibitem [{\citenamefont {Lieu}\ \emph {et~al.}(2020)\citenamefont {Lieu},
  \citenamefont {McGinley},\ and\ \citenamefont {Cooper}}]{Lieu2020}%
  \BibitemOpen
  \bibfield  {author} {\bibinfo {author} {\bibfnamefont {S.}~\bibnamefont
  {Lieu}}, \bibinfo {author} {\bibfnamefont {M.}~\bibnamefont {McGinley}},\
  and\ \bibinfo {author} {\bibfnamefont {N.~R.}\ \bibnamefont {Cooper}},\
  }\bibfield  {title} {\bibinfo {title} {{Tenfold Way for Quadratic
  Lindbladians}},\ }\href {https://doi.org/10.1103/PhysRevLett.124.040401}
  {\bibfield  {journal} {\bibinfo  {journal} {Phys. Rev. Lett.}\ }\textbf
  {\bibinfo {volume} {124}},\ \bibinfo {pages} {040401} (\bibinfo {year}
  {2020})}\BibitemShut {NoStop}%
\bibitem [{\citenamefont {Ashida}\ \emph {et~al.}(2020)\citenamefont {Ashida},
  \citenamefont {Gong},\ and\ \citenamefont {Ueda}}]{Ashida2020}%
  \BibitemOpen
  \bibfield  {author} {\bibinfo {author} {\bibfnamefont {Y.}~\bibnamefont
  {Ashida}}, \bibinfo {author} {\bibfnamefont {Z.}~\bibnamefont {Gong}},\ and\
  \bibinfo {author} {\bibfnamefont {M.}~\bibnamefont {Ueda}},\ }\bibfield
  {title} {\bibinfo {title} {{Non-Hermitian physics}},\ }\href
  {https://doi.org/10.1080/00018732.2021.1876991} {\bibfield  {journal}
  {\bibinfo  {journal} {Advances in Physics}\ }\textbf {\bibinfo {volume}
  {69}},\ \bibinfo {pages} {249} (\bibinfo {year} {2020})}\BibitemShut
  {NoStop}%
\bibitem [{\citenamefont {Bergholtz}\ \emph {et~al.}(2021)\citenamefont
  {Bergholtz}, \citenamefont {Budich},\ and\ \citenamefont
  {Kunst}}]{bergholtz2020exceptional}%
  \BibitemOpen
  \bibfield  {author} {\bibinfo {author} {\bibfnamefont {E.~J.}\ \bibnamefont
  {Bergholtz}}, \bibinfo {author} {\bibfnamefont {J.~C.}\ \bibnamefont
  {Budich}},\ and\ \bibinfo {author} {\bibfnamefont {F.~K.}\ \bibnamefont
  {Kunst}},\ }\bibfield  {title} {\bibinfo {title} {{Exceptional topology of
  non-Hermitian systems}},\ }\href
  {https://doi.org/10.1103/RevModPhys.93.015005} {\bibfield  {journal}
  {\bibinfo  {journal} {Rev. Mod. Phys.}\ }\textbf {\bibinfo {volume} {93}},\
  \bibinfo {pages} {015005} (\bibinfo {year} {2021})}\BibitemShut {NoStop}%
\bibitem [{\citenamefont {Pikulin}\ and\ \citenamefont
  {Nazarov}(2012)}]{Pikulin2012}%
  \BibitemOpen
  \bibfield  {author} {\bibinfo {author} {\bibfnamefont {D.~I.}\ \bibnamefont
  {Pikulin}}\ and\ \bibinfo {author} {\bibfnamefont {Y.~V.}\ \bibnamefont
  {Nazarov}},\ }\bibfield  {title} {\bibinfo {title} {{Topological properties
  of superconducting junctions}},\ }\href
  {https://doi.org/10.1134/S0021364011210090} {\bibfield  {journal} {\bibinfo
  {journal} {JETP Letters}\ }\textbf {\bibinfo {volume} {94}},\ \bibinfo
  {pages} {693} (\bibinfo {year} {2012})},\ \Eprint
  {https://arxiv.org/abs/1103.0780} {1103.0780} \BibitemShut {NoStop}%
\bibitem [{\citenamefont {Pikulin}\ and\ \citenamefont
  {Nazarov}(2013)}]{Pikulin2013}%
  \BibitemOpen
  \bibfield  {author} {\bibinfo {author} {\bibfnamefont {D.~I.}\ \bibnamefont
  {Pikulin}}\ and\ \bibinfo {author} {\bibfnamefont {Y.~V.}\ \bibnamefont
  {Nazarov}},\ }\bibfield  {title} {\bibinfo {title} {{Two types of topological
  transitions in finite Majorana wires}},\ }\href
  {https://doi.org/10.1103/PhysRevB.87.235421} {\bibfield  {journal} {\bibinfo
  {journal} {Phys. Rev. B}\ }\textbf {\bibinfo {volume} {87}},\ \bibinfo
  {pages} {235421} (\bibinfo {year} {2013})}\BibitemShut {NoStop}%
\bibitem [{\citenamefont {Mi}\ \emph {et~al.}(2014)\citenamefont {Mi},
  \citenamefont {Pikulin}, \citenamefont {Marciani},\ and\ \citenamefont
  {Beenakker}}]{Pikulin14}%
  \BibitemOpen
  \bibfield  {author} {\bibinfo {author} {\bibfnamefont {S.}~\bibnamefont
  {Mi}}, \bibinfo {author} {\bibfnamefont {D.~I.}\ \bibnamefont {Pikulin}},
  \bibinfo {author} {\bibfnamefont {M.}~\bibnamefont {Marciani}},\ and\
  \bibinfo {author} {\bibfnamefont {C.~W.~J.}\ \bibnamefont {Beenakker}},\
  }\bibfield  {title} {\bibinfo {title} {{X-shaped and Y-shaped Andreev
  resonance profiles in a superconducting quantum dot}},\ }\href
  {https://doi.org/10.1134/S1063776114120176} {\bibfield  {journal} {\bibinfo
  {journal} {Journal of Experimental and Theoretical Physics}\ }\textbf
  {\bibinfo {volume} {119}},\ \bibinfo {pages} {1018} (\bibinfo {year}
  {2014})}\BibitemShut {NoStop}%
\bibitem [{\citenamefont {San-Jose}\ \emph {et~al.}(2016)\citenamefont
  {San-Jose}, \citenamefont {Cayao}, \citenamefont {Prada},\ and\ \citenamefont
  {Aguado}}]{Ramon1}%
  \BibitemOpen
  \bibfield  {author} {\bibinfo {author} {\bibfnamefont {P.}~\bibnamefont
  {San-Jose}}, \bibinfo {author} {\bibfnamefont {J.}~\bibnamefont {Cayao}},
  \bibinfo {author} {\bibfnamefont {E.}~\bibnamefont {Prada}},\ and\ \bibinfo
  {author} {\bibfnamefont {R.}~\bibnamefont {Aguado}},\ }\bibfield  {title}
  {\bibinfo {title} {{Majorana bound states from exceptional points in
  non-topological superconductors}},\ }\href
  {https://doi.org/10.1038/srep21427} {\bibfield  {journal} {\bibinfo
  {journal} {Scientific Reports}\ }\textbf {\bibinfo {volume} {6}},\ \bibinfo
  {pages} {21427} (\bibinfo {year} {2016})}\BibitemShut {NoStop}%
\bibitem [{\citenamefont {Avila}\ \emph {et~al.}(2019)\citenamefont {Avila},
  \citenamefont {Pe{\~n}aranda}, \citenamefont {Prada}, \citenamefont
  {San-Jose},\ and\ \citenamefont {Aguado}}]{Ramon2}%
  \BibitemOpen
  \bibfield  {author} {\bibinfo {author} {\bibfnamefont {J.}~\bibnamefont
  {Avila}}, \bibinfo {author} {\bibfnamefont {F.}~\bibnamefont
  {Pe{\~n}aranda}}, \bibinfo {author} {\bibfnamefont {E.}~\bibnamefont
  {Prada}}, \bibinfo {author} {\bibfnamefont {P.}~\bibnamefont {San-Jose}},\
  and\ \bibinfo {author} {\bibfnamefont {R.}~\bibnamefont {Aguado}},\
  }\bibfield  {title} {\bibinfo {title} {{Non-hermitian topology as a unifying
  framework for the Andreev versus Majorana states controversy}},\ }\href
  {https://doi.org/10.1038/s42005-019-0231-8} {\bibfield  {journal} {\bibinfo
  {journal} {Communications Physics}\ }\textbf {\bibinfo {volume} {2}},\
  \bibinfo {pages} {133} (\bibinfo {year} {2019})}\BibitemShut {NoStop}%
\bibitem [{\citenamefont {Bergholtz}\ and\ \citenamefont
  {Budich}(2019)}]{Bergholtz2019}%
  \BibitemOpen
  \bibfield  {author} {\bibinfo {author} {\bibfnamefont {E.~J.}\ \bibnamefont
  {Bergholtz}}\ and\ \bibinfo {author} {\bibfnamefont {J.~C.}\ \bibnamefont
  {Budich}},\ }\bibfield  {title} {\bibinfo {title} {{Non-Hermitian Weyl
  physics in topological insulator ferromagnet junctions}},\ }\href
  {https://doi.org/10.1103/PhysRevResearch.1.012003} {\bibfield  {journal}
  {\bibinfo  {journal} {Phys. Rev. Research}\ }\textbf {\bibinfo {volume}
  {1}},\ \bibinfo {pages} {012003} (\bibinfo {year} {2019})}\BibitemShut
  {NoStop}%
\bibitem [{\citenamefont {Hyart}\ and\ \citenamefont {Lado}(2022)}]{Lado2021}%
  \BibitemOpen
  \bibfield  {author} {\bibinfo {author} {\bibfnamefont {T.}~\bibnamefont
  {Hyart}}\ and\ \bibinfo {author} {\bibfnamefont {J.~L.}\ \bibnamefont
  {Lado}},\ }\bibfield  {title} {\bibinfo {title} {{Non-Hermitian many-body
  topological excitations in interacting quantum dots}},\ }\href
  {https://doi.org/10.1103/PhysRevResearch.4.L012006} {\bibfield  {journal}
  {\bibinfo  {journal} {Phys. Rev. Research}\ }\textbf {\bibinfo {volume}
  {4}},\ \bibinfo {pages} {L012006} (\bibinfo {year} {2022})}\BibitemShut
  {NoStop}%
\bibitem [{\citenamefont {Comaron}\ \emph {et~al.}(2020)\citenamefont
  {Comaron}, \citenamefont {Shahnazaryan}, \citenamefont {Brzezicki},
  \citenamefont {Hyart},\ and\ \citenamefont {Matuszewski}}]{Comaron20}%
  \BibitemOpen
  \bibfield  {author} {\bibinfo {author} {\bibfnamefont {P.}~\bibnamefont
  {Comaron}}, \bibinfo {author} {\bibfnamefont {V.}~\bibnamefont
  {Shahnazaryan}}, \bibinfo {author} {\bibfnamefont {W.}~\bibnamefont
  {Brzezicki}}, \bibinfo {author} {\bibfnamefont {T.}~\bibnamefont {Hyart}},\
  and\ \bibinfo {author} {\bibfnamefont {M.}~\bibnamefont {Matuszewski}},\
  }\bibfield  {title} {\bibinfo {title} {{Non-Hermitian topological end-mode
  lasing in polariton systems}},\ }\href
  {https://doi.org/10.1103/PhysRevResearch.2.022051} {\bibfield  {journal}
  {\bibinfo  {journal} {Phys. Rev. Research}\ }\textbf {\bibinfo {volume}
  {2}},\ \bibinfo {pages} {022051} (\bibinfo {year} {2020})}\BibitemShut
  {NoStop}%
\bibitem [{\citenamefont {Ozawa}\ \emph {et~al.}(2019)\citenamefont {Ozawa},
  \citenamefont {Price}, \citenamefont {Amo}, \citenamefont {Goldman},
  \citenamefont {Hafezi}, \citenamefont {Lu}, \citenamefont {Rechtsman},
  \citenamefont {Schuster}, \citenamefont {Simon}, \citenamefont {Zilberberg},\
  and\ \citenamefont {Carusotto}}]{Ozawa}%
  \BibitemOpen
  \bibfield  {author} {\bibinfo {author} {\bibfnamefont {T.}~\bibnamefont
  {Ozawa}}, \bibinfo {author} {\bibfnamefont {H.~M.}\ \bibnamefont {Price}},
  \bibinfo {author} {\bibfnamefont {A.}~\bibnamefont {Amo}}, \bibinfo {author}
  {\bibfnamefont {N.}~\bibnamefont {Goldman}}, \bibinfo {author} {\bibfnamefont
  {M.}~\bibnamefont {Hafezi}}, \bibinfo {author} {\bibfnamefont
  {L.}~\bibnamefont {Lu}}, \bibinfo {author} {\bibfnamefont {M.~C.}\
  \bibnamefont {Rechtsman}}, \bibinfo {author} {\bibfnamefont {D.}~\bibnamefont
  {Schuster}}, \bibinfo {author} {\bibfnamefont {J.}~\bibnamefont {Simon}},
  \bibinfo {author} {\bibfnamefont {O.}~\bibnamefont {Zilberberg}},\ and\
  \bibinfo {author} {\bibfnamefont {I.}~\bibnamefont {Carusotto}},\ }\bibfield
  {title} {\bibinfo {title} {{Topological photonics}},\ }\href
  {https://doi.org/10.1103/RevModPhys.91.015006} {\bibfield  {journal}
  {\bibinfo  {journal} {Rev. Mod. Phys.}\ }\textbf {\bibinfo {volume} {91}},\
  \bibinfo {pages} {015006} (\bibinfo {year} {2019})}\BibitemShut {NoStop}%
\bibitem [{\citenamefont {Zeuner}\ \emph {et~al.}(2015)\citenamefont {Zeuner},
  \citenamefont {Rechtsman}, \citenamefont {Plotnik}, \citenamefont {Lumer},
  \citenamefont {Nolte}, \citenamefont {Rudner}, \citenamefont {Segev},\ and\
  \citenamefont {Szameit}}]{Zeuner2015}%
  \BibitemOpen
  \bibfield  {author} {\bibinfo {author} {\bibfnamefont {J.~M.}\ \bibnamefont
  {Zeuner}}, \bibinfo {author} {\bibfnamefont {M.~C.}\ \bibnamefont
  {Rechtsman}}, \bibinfo {author} {\bibfnamefont {Y.}~\bibnamefont {Plotnik}},
  \bibinfo {author} {\bibfnamefont {Y.}~\bibnamefont {Lumer}}, \bibinfo
  {author} {\bibfnamefont {S.}~\bibnamefont {Nolte}}, \bibinfo {author}
  {\bibfnamefont {M.~S.}\ \bibnamefont {Rudner}}, \bibinfo {author}
  {\bibfnamefont {M.}~\bibnamefont {Segev}},\ and\ \bibinfo {author}
  {\bibfnamefont {A.}~\bibnamefont {Szameit}},\ }\bibfield  {title} {\bibinfo
  {title} {{Observation of a Topological Transition in the Bulk of a
  Non-Hermitian System}},\ }\href
  {https://doi.org/10.1103/PhysRevLett.115.040402} {\bibfield  {journal}
  {\bibinfo  {journal} {Phys. Rev. Lett.}\ }\textbf {\bibinfo {volume} {115}},\
  \bibinfo {pages} {040402} (\bibinfo {year} {2015})}\BibitemShut {NoStop}%
\bibitem [{\citenamefont {Poli}\ \emph {et~al.}(2015)\citenamefont {Poli},
  \citenamefont {Bellec}, \citenamefont {Kuhl}, \citenamefont {Mortessagne},\
  and\ \citenamefont {Schomerus}}]{Poli2015}%
  \BibitemOpen
  \bibfield  {author} {\bibinfo {author} {\bibfnamefont {C.}~\bibnamefont
  {Poli}}, \bibinfo {author} {\bibfnamefont {M.}~\bibnamefont {Bellec}},
  \bibinfo {author} {\bibfnamefont {U.}~\bibnamefont {Kuhl}}, \bibinfo {author}
  {\bibfnamefont {F.}~\bibnamefont {Mortessagne}},\ and\ \bibinfo {author}
  {\bibfnamefont {H.}~\bibnamefont {Schomerus}},\ }\bibfield  {title} {\bibinfo
  {title} {{Selective enhancement of topologically induced interface states in
  a dielectric resonator chain}},\ }\href {https://doi.org/10.1038/ncomms7710}
  {\bibfield  {journal} {\bibinfo  {journal} {Nature Communications}\ }\textbf
  {\bibinfo {volume} {6}},\ \bibinfo {pages} {6710} (\bibinfo {year}
  {2015})}\BibitemShut {NoStop}%
\bibitem [{\citenamefont {Zhan}\ \emph {et~al.}(2017)\citenamefont {Zhan},
  \citenamefont {Xiao}, \citenamefont {Bian}, \citenamefont {Wang},
  \citenamefont {Qiu}, \citenamefont {Sanders}, \citenamefont {Yi},\ and\
  \citenamefont {Xue}}]{Zhan2017}%
  \BibitemOpen
  \bibfield  {author} {\bibinfo {author} {\bibfnamefont {X.}~\bibnamefont
  {Zhan}}, \bibinfo {author} {\bibfnamefont {L.}~\bibnamefont {Xiao}}, \bibinfo
  {author} {\bibfnamefont {Z.}~\bibnamefont {Bian}}, \bibinfo {author}
  {\bibfnamefont {K.}~\bibnamefont {Wang}}, \bibinfo {author} {\bibfnamefont
  {X.}~\bibnamefont {Qiu}}, \bibinfo {author} {\bibfnamefont {B.~C.}\
  \bibnamefont {Sanders}}, \bibinfo {author} {\bibfnamefont {W.}~\bibnamefont
  {Yi}},\ and\ \bibinfo {author} {\bibfnamefont {P.}~\bibnamefont {Xue}},\
  }\bibfield  {title} {\bibinfo {title} {{Detecting Topological Invariants in
  Nonunitary Discrete-Time Quantum Walks}},\ }\href
  {https://doi.org/10.1103/PhysRevLett.119.130501} {\bibfield  {journal}
  {\bibinfo  {journal} {Phys. Rev. Lett.}\ }\textbf {\bibinfo {volume} {119}},\
  \bibinfo {pages} {130501} (\bibinfo {year} {2017})}\BibitemShut {NoStop}%
\bibitem [{\citenamefont {Xiao}\ \emph {et~al.}(2017)\citenamefont {Xiao},
  \citenamefont {Zhan}, \citenamefont {Bian}, \citenamefont {Wang},
  \citenamefont {Zhang}, \citenamefont {Wang}, \citenamefont {Li},
  \citenamefont {Mochizuki}, \citenamefont {Kim}, \citenamefont {Kawakami},
  \citenamefont {Yi}, \citenamefont {Obuse}, \citenamefont {Sanders},\ and\
  \citenamefont {Xue}}]{Xiao2017}%
  \BibitemOpen
  \bibfield  {author} {\bibinfo {author} {\bibfnamefont {L.}~\bibnamefont
  {Xiao}}, \bibinfo {author} {\bibfnamefont {X.}~\bibnamefont {Zhan}}, \bibinfo
  {author} {\bibfnamefont {Z.~H.}\ \bibnamefont {Bian}}, \bibinfo {author}
  {\bibfnamefont {K.~K.}\ \bibnamefont {Wang}}, \bibinfo {author}
  {\bibfnamefont {X.}~\bibnamefont {Zhang}}, \bibinfo {author} {\bibfnamefont
  {X.~P.}\ \bibnamefont {Wang}}, \bibinfo {author} {\bibfnamefont
  {J.}~\bibnamefont {Li}}, \bibinfo {author} {\bibfnamefont {K.}~\bibnamefont
  {Mochizuki}}, \bibinfo {author} {\bibfnamefont {D.}~\bibnamefont {Kim}},
  \bibinfo {author} {\bibfnamefont {N.}~\bibnamefont {Kawakami}}, \bibinfo
  {author} {\bibfnamefont {W.}~\bibnamefont {Yi}}, \bibinfo {author}
  {\bibfnamefont {H.}~\bibnamefont {Obuse}}, \bibinfo {author} {\bibfnamefont
  {B.~C.}\ \bibnamefont {Sanders}},\ and\ \bibinfo {author} {\bibfnamefont
  {P.}~\bibnamefont {Xue}},\ }\bibfield  {title} {\bibinfo {title}
  {{Observation of topological edge states in parity-time-symmetric quantum
  walks}},\ }\href {https://doi.org/10.1038/nphys4204} {\bibfield  {journal}
  {\bibinfo  {journal} {Nature Physics}\ }\textbf {\bibinfo {volume} {13}},\
  \bibinfo {pages} {1117} (\bibinfo {year} {2017})}\BibitemShut {NoStop}%
\bibitem [{\citenamefont {Weimann}\ \emph {et~al.}(2017)\citenamefont
  {Weimann}, \citenamefont {Kremer}, \citenamefont {Plotnik}, \citenamefont
  {Lumer}, \citenamefont {Nolte}, \citenamefont {Makris}, \citenamefont
  {Segev}, \citenamefont {Rechtsman},\ and\ \citenamefont
  {Szameit}}]{Weimann2017}%
  \BibitemOpen
  \bibfield  {author} {\bibinfo {author} {\bibfnamefont {S.}~\bibnamefont
  {Weimann}}, \bibinfo {author} {\bibfnamefont {M.}~\bibnamefont {Kremer}},
  \bibinfo {author} {\bibfnamefont {Y.}~\bibnamefont {Plotnik}}, \bibinfo
  {author} {\bibfnamefont {Y.}~\bibnamefont {Lumer}}, \bibinfo {author}
  {\bibfnamefont {S.}~\bibnamefont {Nolte}}, \bibinfo {author} {\bibfnamefont
  {K.~G.}\ \bibnamefont {Makris}}, \bibinfo {author} {\bibfnamefont
  {M.}~\bibnamefont {Segev}}, \bibinfo {author} {\bibfnamefont {M.~C.}\
  \bibnamefont {Rechtsman}},\ and\ \bibinfo {author} {\bibfnamefont
  {A.}~\bibnamefont {Szameit}},\ }\bibfield  {title} {\bibinfo {title}
  {{Topologically protected bound states in photonic parity–time-symmetric
  crystals}},\ }\href {https://doi.org/10.1038/nmat4811} {\bibfield  {journal}
  {\bibinfo  {journal} {Nature Materials}\ }\textbf {\bibinfo {volume} {16}},\
  \bibinfo {pages} {433} (\bibinfo {year} {2017})}\BibitemShut {NoStop}%
\bibitem [{\citenamefont {Zhao}\ \emph {et~al.}(2018)\citenamefont {Zhao},
  \citenamefont {Miao}, \citenamefont {Teimourpour}, \citenamefont {Malzard},
  \citenamefont {El-Ganainy}, \citenamefont {Schomerus},\ and\ \citenamefont
  {Feng}}]{Zhao2018}%
  \BibitemOpen
  \bibfield  {author} {\bibinfo {author} {\bibfnamefont {H.}~\bibnamefont
  {Zhao}}, \bibinfo {author} {\bibfnamefont {P.}~\bibnamefont {Miao}}, \bibinfo
  {author} {\bibfnamefont {M.~H.}\ \bibnamefont {Teimourpour}}, \bibinfo
  {author} {\bibfnamefont {S.}~\bibnamefont {Malzard}}, \bibinfo {author}
  {\bibfnamefont {R.}~\bibnamefont {El-Ganainy}}, \bibinfo {author}
  {\bibfnamefont {H.}~\bibnamefont {Schomerus}},\ and\ \bibinfo {author}
  {\bibfnamefont {L.}~\bibnamefont {Feng}},\ }\bibfield  {title} {\bibinfo
  {title} {{Topological hybrid silicon microlasers}},\ }\href
  {https://doi.org/10.1038/s41467-018-03434-2} {\bibfield  {journal} {\bibinfo
  {journal} {Nature Communications}\ }\textbf {\bibinfo {volume} {9}},\
  \bibinfo {pages} {981} (\bibinfo {year} {2018})}\BibitemShut {NoStop}%
\bibitem [{\citenamefont {Bandres}\ \emph {et~al.}(2018)\citenamefont
  {Bandres}, \citenamefont {Wittek}, \citenamefont {Harari}, \citenamefont
  {Parto}, \citenamefont {Ren}, \citenamefont {Segev}, \citenamefont
  {Christodoulides},\ and\ \citenamefont {Khajavikhan}}]{Bandres2018}%
  \BibitemOpen
  \bibfield  {author} {\bibinfo {author} {\bibfnamefont {M.~A.}\ \bibnamefont
  {Bandres}}, \bibinfo {author} {\bibfnamefont {S.}~\bibnamefont {Wittek}},
  \bibinfo {author} {\bibfnamefont {G.}~\bibnamefont {Harari}}, \bibinfo
  {author} {\bibfnamefont {M.}~\bibnamefont {Parto}}, \bibinfo {author}
  {\bibfnamefont {J.}~\bibnamefont {Ren}}, \bibinfo {author} {\bibfnamefont
  {M.}~\bibnamefont {Segev}}, \bibinfo {author} {\bibfnamefont {D.~N.}\
  \bibnamefont {Christodoulides}},\ and\ \bibinfo {author} {\bibfnamefont
  {M.}~\bibnamefont {Khajavikhan}},\ }\bibfield  {title} {\bibinfo {title}
  {{Topological insulator laser: Experiments}},\ }\bibfield  {journal}
  {\bibinfo  {journal} {Science}\ }\textbf {\bibinfo {volume} {359}},\ \href
  {https://doi.org/10.1126/science.aar4005} {10.1126/science.aar4005} (\bibinfo
  {year} {2018})\BibitemShut {NoStop}%
\bibitem [{\citenamefont {Parto}\ \emph {et~al.}(2018)\citenamefont {Parto},
  \citenamefont {Wittek}, \citenamefont {Hodaei}, \citenamefont {Harari},
  \citenamefont {Bandres}, \citenamefont {Ren}, \citenamefont {Rechtsman},
  \citenamefont {Segev}, \citenamefont {Christodoulides},\ and\ \citenamefont
  {Khajavikhan}}]{Parto2018}%
  \BibitemOpen
  \bibfield  {author} {\bibinfo {author} {\bibfnamefont {M.}~\bibnamefont
  {Parto}}, \bibinfo {author} {\bibfnamefont {S.}~\bibnamefont {Wittek}},
  \bibinfo {author} {\bibfnamefont {H.}~\bibnamefont {Hodaei}}, \bibinfo
  {author} {\bibfnamefont {G.}~\bibnamefont {Harari}}, \bibinfo {author}
  {\bibfnamefont {M.~A.}\ \bibnamefont {Bandres}}, \bibinfo {author}
  {\bibfnamefont {J.}~\bibnamefont {Ren}}, \bibinfo {author} {\bibfnamefont
  {M.~C.}\ \bibnamefont {Rechtsman}}, \bibinfo {author} {\bibfnamefont
  {M.}~\bibnamefont {Segev}}, \bibinfo {author} {\bibfnamefont {D.~N.}\
  \bibnamefont {Christodoulides}},\ and\ \bibinfo {author} {\bibfnamefont
  {M.}~\bibnamefont {Khajavikhan}},\ }\bibfield  {title} {\bibinfo {title}
  {{Edge-Mode Lasing in 1D Topological Active Arrays}},\ }\href
  {https://doi.org/10.1103/PhysRevLett.120.113901} {\bibfield  {journal}
  {\bibinfo  {journal} {Phys. Rev. Lett.}\ }\textbf {\bibinfo {volume} {120}},\
  \bibinfo {pages} {113901} (\bibinfo {year} {2018})}\BibitemShut {NoStop}%
\bibitem [{\citenamefont {Helbig}\ \emph {et~al.}(2020)\citenamefont {Helbig},
  \citenamefont {Hofmann}, \citenamefont {Imhof}, \citenamefont {Abdelghany},
  \citenamefont {Kiessling}, \citenamefont {Molenkamp}, \citenamefont {Lee},
  \citenamefont {Szameit}, \citenamefont {Greiter},\ and\ \citenamefont
  {Thomale}}]{Helbig2020}%
  \BibitemOpen
  \bibfield  {author} {\bibinfo {author} {\bibfnamefont {T.}~\bibnamefont
  {Helbig}}, \bibinfo {author} {\bibfnamefont {T.}~\bibnamefont {Hofmann}},
  \bibinfo {author} {\bibfnamefont {S.}~\bibnamefont {Imhof}}, \bibinfo
  {author} {\bibfnamefont {M.}~\bibnamefont {Abdelghany}}, \bibinfo {author}
  {\bibfnamefont {T.}~\bibnamefont {Kiessling}}, \bibinfo {author}
  {\bibfnamefont {L.~W.}\ \bibnamefont {Molenkamp}}, \bibinfo {author}
  {\bibfnamefont {C.~H.}\ \bibnamefont {Lee}}, \bibinfo {author} {\bibfnamefont
  {A.}~\bibnamefont {Szameit}}, \bibinfo {author} {\bibfnamefont
  {M.}~\bibnamefont {Greiter}},\ and\ \bibinfo {author} {\bibfnamefont
  {R.}~\bibnamefont {Thomale}},\ }\bibfield  {title} {\bibinfo {title}
  {{Generalized bulk–boundary correspondence in non-Hermitian topolectrical
  circuits}},\ }\href {https://doi.org/10.1038/s41567-020-0922-9} {\bibfield
  {journal} {\bibinfo  {journal} {Nature Physics}\ }\textbf {\bibinfo {volume}
  {16}},\ \bibinfo {pages} {747} (\bibinfo {year} {2020})}\BibitemShut
  {NoStop}%
\bibitem [{\citenamefont {Koch}\ \emph {et~al.}(2007)\citenamefont {Koch},
  \citenamefont {Yu}, \citenamefont {Gambetta}, \citenamefont {Houck},
  \citenamefont {Schuster}, \citenamefont {Majer}, \citenamefont {Blais},
  \citenamefont {Devoret}, \citenamefont {Girvin},\ and\ \citenamefont
  {Schoelkopf}}]{Koch07}%
  \BibitemOpen
  \bibfield  {author} {\bibinfo {author} {\bibfnamefont {J.}~\bibnamefont
  {Koch}}, \bibinfo {author} {\bibfnamefont {T.~M.}\ \bibnamefont {Yu}},
  \bibinfo {author} {\bibfnamefont {J.}~\bibnamefont {Gambetta}}, \bibinfo
  {author} {\bibfnamefont {A.~A.}\ \bibnamefont {Houck}}, \bibinfo {author}
  {\bibfnamefont {D.~I.}\ \bibnamefont {Schuster}}, \bibinfo {author}
  {\bibfnamefont {J.}~\bibnamefont {Majer}}, \bibinfo {author} {\bibfnamefont
  {A.}~\bibnamefont {Blais}}, \bibinfo {author} {\bibfnamefont {M.~H.}\
  \bibnamefont {Devoret}}, \bibinfo {author} {\bibfnamefont {S.~M.}\
  \bibnamefont {Girvin}},\ and\ \bibinfo {author} {\bibfnamefont {R.~J.}\
  \bibnamefont {Schoelkopf}},\ }\bibfield  {title} {\bibinfo {title}
  {{Charge-insensitive qubit design derived from the Cooper pair box}},\ }\href
  {https://doi.org/10.1103/PhysRevA.76.042319} {\bibfield  {journal} {\bibinfo
  {journal} {Phys. Rev. A}\ }\textbf {\bibinfo {volume} {76}},\ \bibinfo
  {pages} {042319} (\bibinfo {year} {2007})}\BibitemShut {NoStop}%
\bibitem [{\citenamefont {Arute}\ \emph {et~al.}(2019)\citenamefont {Arute},
  \citenamefont {Arya}, \citenamefont {Babbush}, \citenamefont {Bacon},
  \citenamefont {Bardin}, \citenamefont {Barends}, \citenamefont {Biswas},
  \citenamefont {Boixo}, \citenamefont {Brandao}, \citenamefont {Buell},
  \citenamefont {Burkett}, \citenamefont {Chen}, \citenamefont {Chen},
  \citenamefont {Chiaro}, \citenamefont {Collins}, \citenamefont {Courtney},
  \citenamefont {Dunsworth}, \citenamefont {Farhi}, \citenamefont {Foxen},
  \citenamefont {Fowler}, \citenamefont {Gidney}, \citenamefont {Giustina},
  \citenamefont {Graff}, \citenamefont {Guerin}, \citenamefont {Habegger},
  \citenamefont {Harrigan}, \citenamefont {Hartmann}, \citenamefont {Ho},
  \citenamefont {Hoffmann}, \citenamefont {Huang}, \citenamefont {Humble},
  \citenamefont {Isakov}, \citenamefont {Jeffrey}, \citenamefont {Jiang},
  \citenamefont {Kafri}, \citenamefont {Kechedzhi}, \citenamefont {Kelly},
  \citenamefont {Klimov}, \citenamefont {Knysh}, \citenamefont {Korotkov},
  \citenamefont {Kostritsa}, \citenamefont {Landhuis}, \citenamefont
  {Lindmark}, \citenamefont {Lucero}, \citenamefont {Lyakh}, \citenamefont
  {Mandr{\`a}}, \citenamefont {McClean}, \citenamefont {McEwen}, \citenamefont
  {Megrant}, \citenamefont {Mi}, \citenamefont {Michielsen}, \citenamefont
  {Mohseni}, \citenamefont {Mutus}, \citenamefont {Naaman}, \citenamefont
  {Neeley}, \citenamefont {Neill}, \citenamefont {Niu}, \citenamefont {Ostby},
  \citenamefont {Petukhov}, \citenamefont {Platt}, \citenamefont {Quintana},
  \citenamefont {Rieffel}, \citenamefont {Roushan}, \citenamefont {Rubin},
  \citenamefont {Sank}, \citenamefont {Satzinger}, \citenamefont {Smelyanskiy},
  \citenamefont {Sung}, \citenamefont {Trevithick}, \citenamefont
  {Vainsencher}, \citenamefont {Villalonga}, \citenamefont {White},
  \citenamefont {Yao}, \citenamefont {Yeh}, \citenamefont {Zalcman},
  \citenamefont {Neven},\ and\ \citenamefont {Martinis}}]{Arute19}%
  \BibitemOpen
  \bibfield  {author} {\bibinfo {author} {\bibfnamefont {F.}~\bibnamefont
  {Arute}}, \bibinfo {author} {\bibfnamefont {K.}~\bibnamefont {Arya}},
  \bibinfo {author} {\bibfnamefont {R.}~\bibnamefont {Babbush}}, \bibinfo
  {author} {\bibfnamefont {D.}~\bibnamefont {Bacon}}, \bibinfo {author}
  {\bibfnamefont {J.~C.}\ \bibnamefont {Bardin}}, \bibinfo {author}
  {\bibfnamefont {R.}~\bibnamefont {Barends}}, \bibinfo {author} {\bibfnamefont
  {R.}~\bibnamefont {Biswas}}, \bibinfo {author} {\bibfnamefont
  {S.}~\bibnamefont {Boixo}}, \bibinfo {author} {\bibfnamefont {F.~G. S.~L.}\
  \bibnamefont {Brandao}}, \bibinfo {author} {\bibfnamefont {D.~A.}\
  \bibnamefont {Buell}}, \bibinfo {author} {\bibfnamefont {B.}~\bibnamefont
  {Burkett}}, \bibinfo {author} {\bibfnamefont {Y.}~\bibnamefont {Chen}},
  \bibinfo {author} {\bibfnamefont {Z.}~\bibnamefont {Chen}}, \bibinfo {author}
  {\bibfnamefont {B.}~\bibnamefont {Chiaro}}, \bibinfo {author} {\bibfnamefont
  {R.}~\bibnamefont {Collins}}, \bibinfo {author} {\bibfnamefont
  {W.}~\bibnamefont {Courtney}}, \bibinfo {author} {\bibfnamefont
  {A.}~\bibnamefont {Dunsworth}}, \bibinfo {author} {\bibfnamefont
  {E.}~\bibnamefont {Farhi}}, \bibinfo {author} {\bibfnamefont
  {B.}~\bibnamefont {Foxen}}, \bibinfo {author} {\bibfnamefont
  {A.}~\bibnamefont {Fowler}}, \bibinfo {author} {\bibfnamefont
  {C.}~\bibnamefont {Gidney}}, \bibinfo {author} {\bibfnamefont
  {M.}~\bibnamefont {Giustina}}, \bibinfo {author} {\bibfnamefont
  {R.}~\bibnamefont {Graff}}, \bibinfo {author} {\bibfnamefont
  {K.}~\bibnamefont {Guerin}}, \bibinfo {author} {\bibfnamefont
  {S.}~\bibnamefont {Habegger}}, \bibinfo {author} {\bibfnamefont {M.~P.}\
  \bibnamefont {Harrigan}}, \bibinfo {author} {\bibfnamefont {M.~J.}\
  \bibnamefont {Hartmann}}, \bibinfo {author} {\bibfnamefont {A.}~\bibnamefont
  {Ho}}, \bibinfo {author} {\bibfnamefont {M.}~\bibnamefont {Hoffmann}},
  \bibinfo {author} {\bibfnamefont {T.}~\bibnamefont {Huang}}, \bibinfo
  {author} {\bibfnamefont {T.~S.}\ \bibnamefont {Humble}}, \bibinfo {author}
  {\bibfnamefont {S.~V.}\ \bibnamefont {Isakov}}, \bibinfo {author}
  {\bibfnamefont {E.}~\bibnamefont {Jeffrey}}, \bibinfo {author} {\bibfnamefont
  {Z.}~\bibnamefont {Jiang}}, \bibinfo {author} {\bibfnamefont
  {D.}~\bibnamefont {Kafri}}, \bibinfo {author} {\bibfnamefont
  {K.}~\bibnamefont {Kechedzhi}}, \bibinfo {author} {\bibfnamefont
  {J.}~\bibnamefont {Kelly}}, \bibinfo {author} {\bibfnamefont {P.~V.}\
  \bibnamefont {Klimov}}, \bibinfo {author} {\bibfnamefont {S.}~\bibnamefont
  {Knysh}}, \bibinfo {author} {\bibfnamefont {A.}~\bibnamefont {Korotkov}},
  \bibinfo {author} {\bibfnamefont {F.}~\bibnamefont {Kostritsa}}, \bibinfo
  {author} {\bibfnamefont {D.}~\bibnamefont {Landhuis}}, \bibinfo {author}
  {\bibfnamefont {M.}~\bibnamefont {Lindmark}}, \bibinfo {author}
  {\bibfnamefont {E.}~\bibnamefont {Lucero}}, \bibinfo {author} {\bibfnamefont
  {D.}~\bibnamefont {Lyakh}}, \bibinfo {author} {\bibfnamefont
  {S.}~\bibnamefont {Mandr{\`a}}}, \bibinfo {author} {\bibfnamefont {J.~R.}\
  \bibnamefont {McClean}}, \bibinfo {author} {\bibfnamefont {M.}~\bibnamefont
  {McEwen}}, \bibinfo {author} {\bibfnamefont {A.}~\bibnamefont {Megrant}},
  \bibinfo {author} {\bibfnamefont {X.}~\bibnamefont {Mi}}, \bibinfo {author}
  {\bibfnamefont {K.}~\bibnamefont {Michielsen}}, \bibinfo {author}
  {\bibfnamefont {M.}~\bibnamefont {Mohseni}}, \bibinfo {author} {\bibfnamefont
  {J.}~\bibnamefont {Mutus}}, \bibinfo {author} {\bibfnamefont
  {O.}~\bibnamefont {Naaman}}, \bibinfo {author} {\bibfnamefont
  {M.}~\bibnamefont {Neeley}}, \bibinfo {author} {\bibfnamefont
  {C.}~\bibnamefont {Neill}}, \bibinfo {author} {\bibfnamefont {M.~Y.}\
  \bibnamefont {Niu}}, \bibinfo {author} {\bibfnamefont {E.}~\bibnamefont
  {Ostby}}, \bibinfo {author} {\bibfnamefont {A.}~\bibnamefont {Petukhov}},
  \bibinfo {author} {\bibfnamefont {J.~C.}\ \bibnamefont {Platt}}, \bibinfo
  {author} {\bibfnamefont {C.}~\bibnamefont {Quintana}}, \bibinfo {author}
  {\bibfnamefont {E.~G.}\ \bibnamefont {Rieffel}}, \bibinfo {author}
  {\bibfnamefont {P.}~\bibnamefont {Roushan}}, \bibinfo {author} {\bibfnamefont
  {N.~C.}\ \bibnamefont {Rubin}}, \bibinfo {author} {\bibfnamefont
  {D.}~\bibnamefont {Sank}}, \bibinfo {author} {\bibfnamefont {K.~J.}\
  \bibnamefont {Satzinger}}, \bibinfo {author} {\bibfnamefont {V.}~\bibnamefont
  {Smelyanskiy}}, \bibinfo {author} {\bibfnamefont {K.~J.}\ \bibnamefont
  {Sung}}, \bibinfo {author} {\bibfnamefont {M.~D.}\ \bibnamefont
  {Trevithick}}, \bibinfo {author} {\bibfnamefont {A.}~\bibnamefont
  {Vainsencher}}, \bibinfo {author} {\bibfnamefont {B.}~\bibnamefont
  {Villalonga}}, \bibinfo {author} {\bibfnamefont {T.}~\bibnamefont {White}},
  \bibinfo {author} {\bibfnamefont {Z.~J.}\ \bibnamefont {Yao}}, \bibinfo
  {author} {\bibfnamefont {P.}~\bibnamefont {Yeh}}, \bibinfo {author}
  {\bibfnamefont {A.}~\bibnamefont {Zalcman}}, \bibinfo {author} {\bibfnamefont
  {H.}~\bibnamefont {Neven}},\ and\ \bibinfo {author} {\bibfnamefont {J.~M.}\
  \bibnamefont {Martinis}},\ }\bibfield  {title} {\bibinfo {title} {Quantum
  supremacy using a programmable superconducting processor},\ }\href
  {https://doi.org/10.1038/s41586-019-1666-5} {\bibfield  {journal} {\bibinfo
  {journal} {Nature}\ }\textbf {\bibinfo {volume} {574}},\ \bibinfo {pages}
  {505} (\bibinfo {year} {2019})}\BibitemShut {NoStop}%
\bibitem [{\citenamefont {Chen}\ \emph {et~al.}(2021)\citenamefont {Chen},
  \citenamefont {Satzinger}, \citenamefont {Atalaya}, \citenamefont {Korotkov},
  \citenamefont {Dunsworth}, \citenamefont {Sank}, \citenamefont {Quintana},
  \citenamefont {McEwen}, \citenamefont {Barends}, \citenamefont {Klimov},
  \citenamefont {Hong}, \citenamefont {Jones}, \citenamefont {Petukhov},
  \citenamefont {Kafri}, \citenamefont {Demura}, \citenamefont {Burkett},
  \citenamefont {Gidney}, \citenamefont {Fowler}, \citenamefont {Paler},
  \citenamefont {Putterman}, \citenamefont {Aleiner}, \citenamefont {Arute},
  \citenamefont {Arya}, \citenamefont {Babbush}, \citenamefont {Bardin},
  \citenamefont {Bengtsson}, \citenamefont {Bourassa}, \citenamefont
  {Broughton}, \citenamefont {Buckley}, \citenamefont {Buell}, \citenamefont
  {Bushnell}, \citenamefont {Chiaro}, \citenamefont {Collins}, \citenamefont
  {Courtney}, \citenamefont {Derk}, \citenamefont {Eppens}, \citenamefont
  {Erickson}, \citenamefont {Farhi}, \citenamefont {Foxen}, \citenamefont
  {Giustina}, \citenamefont {Greene}, \citenamefont {Gross}, \citenamefont
  {Harrigan}, \citenamefont {Harrington}, \citenamefont {Hilton}, \citenamefont
  {Ho}, \citenamefont {Huang}, \citenamefont {Huggins}, \citenamefont {Ioffe},
  \citenamefont {Isakov}, \citenamefont {Jeffrey}, \citenamefont {Jiang},
  \citenamefont {Kechedzhi}, \citenamefont {Kim}, \citenamefont {Kitaev},
  \citenamefont {Kostritsa}, \citenamefont {Landhuis}, \citenamefont {Laptev},
  \citenamefont {Lucero}, \citenamefont {Martin}, \citenamefont {McClean},
  \citenamefont {McCourt}, \citenamefont {Mi}, \citenamefont {Miao},
  \citenamefont {Mohseni}, \citenamefont {Montazeri}, \citenamefont
  {Mruczkiewicz}, \citenamefont {Mutus}, \citenamefont {Naaman}, \citenamefont
  {Neeley}, \citenamefont {Neill}, \citenamefont {Newman}, \citenamefont {Niu},
  \citenamefont {O'Brien}, \citenamefont {Opremcak}, \citenamefont {Ostby},
  \citenamefont {Pat{\'o}}, \citenamefont {Redd}, \citenamefont {Roushan},
  \citenamefont {Rubin}, \citenamefont {Shvarts}, \citenamefont {Strain},
  \citenamefont {Szalay}, \citenamefont {Trevithick}, \citenamefont
  {Villalonga}, \citenamefont {White}, \citenamefont {Yao}, \citenamefont
  {Yeh}, \citenamefont {Yoo}, \citenamefont {Zalcman}, \citenamefont {Neven},
  \citenamefont {Boixo}, \citenamefont {Smelyanskiy}, \citenamefont {Chen},
  \citenamefont {Megrant}, \citenamefont {Kelly},\ and\ \citenamefont
  {AI}}]{Chen21}%
  \BibitemOpen
  \bibfield  {author} {\bibinfo {author} {\bibfnamefont {Z.}~\bibnamefont
  {Chen}}, \bibinfo {author} {\bibfnamefont {K.~J.}\ \bibnamefont {Satzinger}},
  \bibinfo {author} {\bibfnamefont {J.}~\bibnamefont {Atalaya}}, \bibinfo
  {author} {\bibfnamefont {A.~N.}\ \bibnamefont {Korotkov}}, \bibinfo {author}
  {\bibfnamefont {A.}~\bibnamefont {Dunsworth}}, \bibinfo {author}
  {\bibfnamefont {D.}~\bibnamefont {Sank}}, \bibinfo {author} {\bibfnamefont
  {C.}~\bibnamefont {Quintana}}, \bibinfo {author} {\bibfnamefont
  {M.}~\bibnamefont {McEwen}}, \bibinfo {author} {\bibfnamefont
  {R.}~\bibnamefont {Barends}}, \bibinfo {author} {\bibfnamefont {P.~V.}\
  \bibnamefont {Klimov}}, \bibinfo {author} {\bibfnamefont {S.}~\bibnamefont
  {Hong}}, \bibinfo {author} {\bibfnamefont {C.}~\bibnamefont {Jones}},
  \bibinfo {author} {\bibfnamefont {A.}~\bibnamefont {Petukhov}}, \bibinfo
  {author} {\bibfnamefont {D.}~\bibnamefont {Kafri}}, \bibinfo {author}
  {\bibfnamefont {S.}~\bibnamefont {Demura}}, \bibinfo {author} {\bibfnamefont
  {B.}~\bibnamefont {Burkett}}, \bibinfo {author} {\bibfnamefont
  {C.}~\bibnamefont {Gidney}}, \bibinfo {author} {\bibfnamefont {A.~G.}\
  \bibnamefont {Fowler}}, \bibinfo {author} {\bibfnamefont {A.}~\bibnamefont
  {Paler}}, \bibinfo {author} {\bibfnamefont {H.}~\bibnamefont {Putterman}},
  \bibinfo {author} {\bibfnamefont {I.}~\bibnamefont {Aleiner}}, \bibinfo
  {author} {\bibfnamefont {F.}~\bibnamefont {Arute}}, \bibinfo {author}
  {\bibfnamefont {K.}~\bibnamefont {Arya}}, \bibinfo {author} {\bibfnamefont
  {R.}~\bibnamefont {Babbush}}, \bibinfo {author} {\bibfnamefont {J.~C.}\
  \bibnamefont {Bardin}}, \bibinfo {author} {\bibfnamefont {A.}~\bibnamefont
  {Bengtsson}}, \bibinfo {author} {\bibfnamefont {A.}~\bibnamefont {Bourassa}},
  \bibinfo {author} {\bibfnamefont {M.}~\bibnamefont {Broughton}}, \bibinfo
  {author} {\bibfnamefont {B.~B.}\ \bibnamefont {Buckley}}, \bibinfo {author}
  {\bibfnamefont {D.~A.}\ \bibnamefont {Buell}}, \bibinfo {author}
  {\bibfnamefont {N.}~\bibnamefont {Bushnell}}, \bibinfo {author}
  {\bibfnamefont {B.}~\bibnamefont {Chiaro}}, \bibinfo {author} {\bibfnamefont
  {R.}~\bibnamefont {Collins}}, \bibinfo {author} {\bibfnamefont
  {W.}~\bibnamefont {Courtney}}, \bibinfo {author} {\bibfnamefont {A.~R.}\
  \bibnamefont {Derk}}, \bibinfo {author} {\bibfnamefont {D.}~\bibnamefont
  {Eppens}}, \bibinfo {author} {\bibfnamefont {C.}~\bibnamefont {Erickson}},
  \bibinfo {author} {\bibfnamefont {E.}~\bibnamefont {Farhi}}, \bibinfo
  {author} {\bibfnamefont {B.}~\bibnamefont {Foxen}}, \bibinfo {author}
  {\bibfnamefont {M.}~\bibnamefont {Giustina}}, \bibinfo {author}
  {\bibfnamefont {A.}~\bibnamefont {Greene}}, \bibinfo {author} {\bibfnamefont
  {J.~A.}\ \bibnamefont {Gross}}, \bibinfo {author} {\bibfnamefont {M.~P.}\
  \bibnamefont {Harrigan}}, \bibinfo {author} {\bibfnamefont {S.~D.}\
  \bibnamefont {Harrington}}, \bibinfo {author} {\bibfnamefont
  {J.}~\bibnamefont {Hilton}}, \bibinfo {author} {\bibfnamefont
  {A.}~\bibnamefont {Ho}}, \bibinfo {author} {\bibfnamefont {T.}~\bibnamefont
  {Huang}}, \bibinfo {author} {\bibfnamefont {W.~J.}\ \bibnamefont {Huggins}},
  \bibinfo {author} {\bibfnamefont {L.~B.}\ \bibnamefont {Ioffe}}, \bibinfo
  {author} {\bibfnamefont {S.~V.}\ \bibnamefont {Isakov}}, \bibinfo {author}
  {\bibfnamefont {E.}~\bibnamefont {Jeffrey}}, \bibinfo {author} {\bibfnamefont
  {Z.}~\bibnamefont {Jiang}}, \bibinfo {author} {\bibfnamefont
  {K.}~\bibnamefont {Kechedzhi}}, \bibinfo {author} {\bibfnamefont
  {S.}~\bibnamefont {Kim}}, \bibinfo {author} {\bibfnamefont {A.}~\bibnamefont
  {Kitaev}}, \bibinfo {author} {\bibfnamefont {F.}~\bibnamefont {Kostritsa}},
  \bibinfo {author} {\bibfnamefont {D.}~\bibnamefont {Landhuis}}, \bibinfo
  {author} {\bibfnamefont {P.}~\bibnamefont {Laptev}}, \bibinfo {author}
  {\bibfnamefont {E.}~\bibnamefont {Lucero}}, \bibinfo {author} {\bibfnamefont
  {O.}~\bibnamefont {Martin}}, \bibinfo {author} {\bibfnamefont {J.~R.}\
  \bibnamefont {McClean}}, \bibinfo {author} {\bibfnamefont {T.}~\bibnamefont
  {McCourt}}, \bibinfo {author} {\bibfnamefont {X.}~\bibnamefont {Mi}},
  \bibinfo {author} {\bibfnamefont {K.~C.}\ \bibnamefont {Miao}}, \bibinfo
  {author} {\bibfnamefont {M.}~\bibnamefont {Mohseni}}, \bibinfo {author}
  {\bibfnamefont {S.}~\bibnamefont {Montazeri}}, \bibinfo {author}
  {\bibfnamefont {W.}~\bibnamefont {Mruczkiewicz}}, \bibinfo {author}
  {\bibfnamefont {J.}~\bibnamefont {Mutus}}, \bibinfo {author} {\bibfnamefont
  {O.}~\bibnamefont {Naaman}}, \bibinfo {author} {\bibfnamefont
  {M.}~\bibnamefont {Neeley}}, \bibinfo {author} {\bibfnamefont
  {C.}~\bibnamefont {Neill}}, \bibinfo {author} {\bibfnamefont
  {M.}~\bibnamefont {Newman}}, \bibinfo {author} {\bibfnamefont {M.~Y.}\
  \bibnamefont {Niu}}, \bibinfo {author} {\bibfnamefont {T.~E.}\ \bibnamefont
  {O'Brien}}, \bibinfo {author} {\bibfnamefont {A.}~\bibnamefont {Opremcak}},
  \bibinfo {author} {\bibfnamefont {E.}~\bibnamefont {Ostby}}, \bibinfo
  {author} {\bibfnamefont {B.}~\bibnamefont {Pat{\'o}}}, \bibinfo {author}
  {\bibfnamefont {N.}~\bibnamefont {Redd}}, \bibinfo {author} {\bibfnamefont
  {P.}~\bibnamefont {Roushan}}, \bibinfo {author} {\bibfnamefont {N.~C.}\
  \bibnamefont {Rubin}}, \bibinfo {author} {\bibfnamefont {V.}~\bibnamefont
  {Shvarts}}, \bibinfo {author} {\bibfnamefont {D.}~\bibnamefont {Strain}},
  \bibinfo {author} {\bibfnamefont {M.}~\bibnamefont {Szalay}}, \bibinfo
  {author} {\bibfnamefont {M.~D.}\ \bibnamefont {Trevithick}}, \bibinfo
  {author} {\bibfnamefont {B.}~\bibnamefont {Villalonga}}, \bibinfo {author}
  {\bibfnamefont {T.}~\bibnamefont {White}}, \bibinfo {author} {\bibfnamefont
  {Z.~J.}\ \bibnamefont {Yao}}, \bibinfo {author} {\bibfnamefont
  {P.}~\bibnamefont {Yeh}}, \bibinfo {author} {\bibfnamefont {J.}~\bibnamefont
  {Yoo}}, \bibinfo {author} {\bibfnamefont {A.}~\bibnamefont {Zalcman}},
  \bibinfo {author} {\bibfnamefont {H.}~\bibnamefont {Neven}}, \bibinfo
  {author} {\bibfnamefont {S.}~\bibnamefont {Boixo}}, \bibinfo {author}
  {\bibfnamefont {V.}~\bibnamefont {Smelyanskiy}}, \bibinfo {author}
  {\bibfnamefont {Y.}~\bibnamefont {Chen}}, \bibinfo {author} {\bibfnamefont
  {A.}~\bibnamefont {Megrant}}, \bibinfo {author} {\bibfnamefont
  {J.}~\bibnamefont {Kelly}},\ and\ \bibinfo {author} {\bibfnamefont {G.~Q.}\
  \bibnamefont {AI}},\ }\bibfield  {title} {\bibinfo {title} {Exponential
  suppression of bit or phase errors with cyclic error correction},\ }\href
  {https://doi.org/10.1038/s41586-021-03588-y} {\bibfield  {journal} {\bibinfo
  {journal} {Nature}\ }\textbf {\bibinfo {volume} {595}},\ \bibinfo {pages}
  {383} (\bibinfo {year} {2021})}\BibitemShut {NoStop}%
\bibitem [{\citenamefont {Krinner}\ \emph {et~al.}(2022)\citenamefont
  {Krinner}, \citenamefont {Lacroix}, \citenamefont {Remm}, \citenamefont
  {Di~Paolo}, \citenamefont {Genois}, \citenamefont {Leroux}, \citenamefont
  {Hellings}, \citenamefont {Lazar}, \citenamefont {Swiadek}, \citenamefont
  {Herrmann}, \citenamefont {Norris}, \citenamefont {Andersen}, \citenamefont
  {M{\"u}ller}, \citenamefont {Blais}, \citenamefont {Eichler},\ and\
  \citenamefont {Wallraff}}]{Krinner22}%
  \BibitemOpen
  \bibfield  {author} {\bibinfo {author} {\bibfnamefont {S.}~\bibnamefont
  {Krinner}}, \bibinfo {author} {\bibfnamefont {N.}~\bibnamefont {Lacroix}},
  \bibinfo {author} {\bibfnamefont {A.}~\bibnamefont {Remm}}, \bibinfo {author}
  {\bibfnamefont {A.}~\bibnamefont {Di~Paolo}}, \bibinfo {author}
  {\bibfnamefont {E.}~\bibnamefont {Genois}}, \bibinfo {author} {\bibfnamefont
  {C.}~\bibnamefont {Leroux}}, \bibinfo {author} {\bibfnamefont
  {C.}~\bibnamefont {Hellings}}, \bibinfo {author} {\bibfnamefont
  {S.}~\bibnamefont {Lazar}}, \bibinfo {author} {\bibfnamefont
  {F.}~\bibnamefont {Swiadek}}, \bibinfo {author} {\bibfnamefont
  {J.}~\bibnamefont {Herrmann}}, \bibinfo {author} {\bibfnamefont {G.~J.}\
  \bibnamefont {Norris}}, \bibinfo {author} {\bibfnamefont {C.~K.}\
  \bibnamefont {Andersen}}, \bibinfo {author} {\bibfnamefont {M.}~\bibnamefont
  {M{\"u}ller}}, \bibinfo {author} {\bibfnamefont {A.}~\bibnamefont {Blais}},
  \bibinfo {author} {\bibfnamefont {C.}~\bibnamefont {Eichler}},\ and\ \bibinfo
  {author} {\bibfnamefont {A.}~\bibnamefont {Wallraff}},\ }\bibfield  {title}
  {\bibinfo {title} {Realizing repeated quantum error correction in a
  distance-three surface code},\ }\href
  {https://doi.org/10.1038/s41586-022-04566-8} {\bibfield  {journal} {\bibinfo
  {journal} {Nature}\ }\textbf {\bibinfo {volume} {605}},\ \bibinfo {pages}
  {669} (\bibinfo {year} {2022})}\BibitemShut {NoStop}%
\bibitem [{\citenamefont {Zhao}\ \emph {et~al.}(2022)\citenamefont {Zhao},
  \citenamefont {Ye}, \citenamefont {Huang}, \citenamefont {Zhang},
  \citenamefont {Wu}, \citenamefont {Guan}, \citenamefont {Zhu}, \citenamefont
  {Wei}, \citenamefont {He}, \citenamefont {Cao}, \citenamefont {Chen},
  \citenamefont {Chung}, \citenamefont {Deng}, \citenamefont {Fan},
  \citenamefont {Gong}, \citenamefont {Guo}, \citenamefont {Guo}, \citenamefont
  {Han}, \citenamefont {Li}, \citenamefont {Li}, \citenamefont {Li},
  \citenamefont {Liang}, \citenamefont {Lin}, \citenamefont {Qian},
  \citenamefont {Rong}, \citenamefont {Su}, \citenamefont {Sun}, \citenamefont
  {Wang}, \citenamefont {Wu}, \citenamefont {Xu}, \citenamefont {Ying},
  \citenamefont {Yu}, \citenamefont {Zha}, \citenamefont {Zhang}, \citenamefont
  {Huo}, \citenamefont {Lu}, \citenamefont {Peng}, \citenamefont {Zhu},\ and\
  \citenamefont {Pan}}]{Zhao21}%
  \BibitemOpen
  \bibfield  {author} {\bibinfo {author} {\bibfnamefont {Y.}~\bibnamefont
  {Zhao}}, \bibinfo {author} {\bibfnamefont {Y.}~\bibnamefont {Ye}}, \bibinfo
  {author} {\bibfnamefont {H.-L.}\ \bibnamefont {Huang}}, \bibinfo {author}
  {\bibfnamefont {Y.}~\bibnamefont {Zhang}}, \bibinfo {author} {\bibfnamefont
  {D.}~\bibnamefont {Wu}}, \bibinfo {author} {\bibfnamefont {H.}~\bibnamefont
  {Guan}}, \bibinfo {author} {\bibfnamefont {Q.}~\bibnamefont {Zhu}}, \bibinfo
  {author} {\bibfnamefont {Z.}~\bibnamefont {Wei}}, \bibinfo {author}
  {\bibfnamefont {T.}~\bibnamefont {He}}, \bibinfo {author} {\bibfnamefont
  {S.}~\bibnamefont {Cao}}, \bibinfo {author} {\bibfnamefont {F.}~\bibnamefont
  {Chen}}, \bibinfo {author} {\bibfnamefont {T.-H.}\ \bibnamefont {Chung}},
  \bibinfo {author} {\bibfnamefont {H.}~\bibnamefont {Deng}}, \bibinfo {author}
  {\bibfnamefont {D.}~\bibnamefont {Fan}}, \bibinfo {author} {\bibfnamefont
  {M.}~\bibnamefont {Gong}}, \bibinfo {author} {\bibfnamefont {C.}~\bibnamefont
  {Guo}}, \bibinfo {author} {\bibfnamefont {S.}~\bibnamefont {Guo}}, \bibinfo
  {author} {\bibfnamefont {L.}~\bibnamefont {Han}}, \bibinfo {author}
  {\bibfnamefont {N.}~\bibnamefont {Li}}, \bibinfo {author} {\bibfnamefont
  {S.}~\bibnamefont {Li}}, \bibinfo {author} {\bibfnamefont {Y.}~\bibnamefont
  {Li}}, \bibinfo {author} {\bibfnamefont {F.}~\bibnamefont {Liang}}, \bibinfo
  {author} {\bibfnamefont {J.}~\bibnamefont {Lin}}, \bibinfo {author}
  {\bibfnamefont {H.}~\bibnamefont {Qian}}, \bibinfo {author} {\bibfnamefont
  {H.}~\bibnamefont {Rong}}, \bibinfo {author} {\bibfnamefont {H.}~\bibnamefont
  {Su}}, \bibinfo {author} {\bibfnamefont {L.}~\bibnamefont {Sun}}, \bibinfo
  {author} {\bibfnamefont {S.}~\bibnamefont {Wang}}, \bibinfo {author}
  {\bibfnamefont {Y.}~\bibnamefont {Wu}}, \bibinfo {author} {\bibfnamefont
  {Y.}~\bibnamefont {Xu}}, \bibinfo {author} {\bibfnamefont {C.}~\bibnamefont
  {Ying}}, \bibinfo {author} {\bibfnamefont {J.}~\bibnamefont {Yu}}, \bibinfo
  {author} {\bibfnamefont {C.}~\bibnamefont {Zha}}, \bibinfo {author}
  {\bibfnamefont {K.}~\bibnamefont {Zhang}}, \bibinfo {author} {\bibfnamefont
  {Y.-H.}\ \bibnamefont {Huo}}, \bibinfo {author} {\bibfnamefont {C.-Y.}\
  \bibnamefont {Lu}}, \bibinfo {author} {\bibfnamefont {C.-Z.}\ \bibnamefont
  {Peng}}, \bibinfo {author} {\bibfnamefont {X.}~\bibnamefont {Zhu}},\ and\
  \bibinfo {author} {\bibfnamefont {J.-W.}\ \bibnamefont {Pan}},\ }\bibfield
  {title} {\bibinfo {title} {Realization of an error-correcting surface code
  with superconducting qubits},\ }\href
  {https://doi.org/10.1103/PhysRevLett.129.030501} {\bibfield  {journal}
  {\bibinfo  {journal} {Phys. Rev. Lett.}\ }\textbf {\bibinfo {volume} {129}},\
  \bibinfo {pages} {030501} (\bibinfo {year} {2022})}\BibitemShut {NoStop}%
\bibitem [{\citenamefont {Neill}\ \emph {et~al.}(2021)\citenamefont {Neill},
  \citenamefont {McCourt}, \citenamefont {Mi}, \citenamefont {Jiang},
  \citenamefont {Niu}, \citenamefont {Mruczkiewicz}, \citenamefont {Aleiner},
  \citenamefont {Arute}, \citenamefont {Arya}, \citenamefont {Atalaya},
  \citenamefont {Babbush}, \citenamefont {Bardin}, \citenamefont {Barends},
  \citenamefont {Bengtsson}, \citenamefont {Bourassa}, \citenamefont
  {Broughton}, \citenamefont {Buckley}, \citenamefont {Buell}, \citenamefont
  {Burkett}, \citenamefont {Bushnell}, \citenamefont {Campero}, \citenamefont
  {Chen}, \citenamefont {Chiaro}, \citenamefont {Collins}, \citenamefont
  {Courtney}, \citenamefont {Demura}, \citenamefont {Derk}, \citenamefont
  {Dunsworth}, \citenamefont {Eppens}, \citenamefont {Erickson}, \citenamefont
  {Farhi}, \citenamefont {Fowler}, \citenamefont {Foxen}, \citenamefont
  {Gidney}, \citenamefont {Giustina}, \citenamefont {Gross}, \citenamefont
  {Harrigan}, \citenamefont {Harrington}, \citenamefont {Hilton}, \citenamefont
  {Ho}, \citenamefont {Hong}, \citenamefont {Huang}, \citenamefont {Huggins},
  \citenamefont {Isakov}, \citenamefont {Jacob-Mitos}, \citenamefont {Jeffrey},
  \citenamefont {Jones}, \citenamefont {Kafri}, \citenamefont {Kechedzhi},
  \citenamefont {Kelly}, \citenamefont {Kim}, \citenamefont {Klimov},
  \citenamefont {Korotkov}, \citenamefont {Kostritsa}, \citenamefont
  {Landhuis}, \citenamefont {Laptev}, \citenamefont {Lucero}, \citenamefont
  {Martin}, \citenamefont {McClean}, \citenamefont {McEwen}, \citenamefont
  {Megrant}, \citenamefont {Miao}, \citenamefont {Mohseni}, \citenamefont
  {Mutus}, \citenamefont {Naaman}, \citenamefont {Neeley}, \citenamefont
  {Newman}, \citenamefont {O'Brien}, \citenamefont {Opremcak}, \citenamefont
  {Ostby}, \citenamefont {Pat{\'o}}, \citenamefont {Petukhov}, \citenamefont
  {Quintana}, \citenamefont {Redd}, \citenamefont {Rubin}, \citenamefont
  {Sank}, \citenamefont {Satzinger}, \citenamefont {Shvarts}, \citenamefont
  {Strain}, \citenamefont {Szalay}, \citenamefont {Trevithick}, \citenamefont
  {Villalonga}, \citenamefont {White}, \citenamefont {Yao}, \citenamefont
  {Yeh}, \citenamefont {Zalcman}, \citenamefont {Neven}, \citenamefont {Boixo},
  \citenamefont {Ioffe}, \citenamefont {Roushan}, \citenamefont {Chen},\ and\
  \citenamefont {Smelyanskiy}}]{Neill21}%
  \BibitemOpen
  \bibfield  {author} {\bibinfo {author} {\bibfnamefont {C.}~\bibnamefont
  {Neill}}, \bibinfo {author} {\bibfnamefont {T.}~\bibnamefont {McCourt}},
  \bibinfo {author} {\bibfnamefont {X.}~\bibnamefont {Mi}}, \bibinfo {author}
  {\bibfnamefont {Z.}~\bibnamefont {Jiang}}, \bibinfo {author} {\bibfnamefont
  {M.~Y.}\ \bibnamefont {Niu}}, \bibinfo {author} {\bibfnamefont
  {W.}~\bibnamefont {Mruczkiewicz}}, \bibinfo {author} {\bibfnamefont
  {I.}~\bibnamefont {Aleiner}}, \bibinfo {author} {\bibfnamefont
  {F.}~\bibnamefont {Arute}}, \bibinfo {author} {\bibfnamefont
  {K.}~\bibnamefont {Arya}}, \bibinfo {author} {\bibfnamefont {J.}~\bibnamefont
  {Atalaya}}, \bibinfo {author} {\bibfnamefont {R.}~\bibnamefont {Babbush}},
  \bibinfo {author} {\bibfnamefont {J.~C.}\ \bibnamefont {Bardin}}, \bibinfo
  {author} {\bibfnamefont {R.}~\bibnamefont {Barends}}, \bibinfo {author}
  {\bibfnamefont {A.}~\bibnamefont {Bengtsson}}, \bibinfo {author}
  {\bibfnamefont {A.}~\bibnamefont {Bourassa}}, \bibinfo {author}
  {\bibfnamefont {M.}~\bibnamefont {Broughton}}, \bibinfo {author}
  {\bibfnamefont {B.~B.}\ \bibnamefont {Buckley}}, \bibinfo {author}
  {\bibfnamefont {D.~A.}\ \bibnamefont {Buell}}, \bibinfo {author}
  {\bibfnamefont {B.}~\bibnamefont {Burkett}}, \bibinfo {author} {\bibfnamefont
  {N.}~\bibnamefont {Bushnell}}, \bibinfo {author} {\bibfnamefont
  {J.}~\bibnamefont {Campero}}, \bibinfo {author} {\bibfnamefont
  {Z.}~\bibnamefont {Chen}}, \bibinfo {author} {\bibfnamefont {B.}~\bibnamefont
  {Chiaro}}, \bibinfo {author} {\bibfnamefont {R.}~\bibnamefont {Collins}},
  \bibinfo {author} {\bibfnamefont {W.}~\bibnamefont {Courtney}}, \bibinfo
  {author} {\bibfnamefont {S.}~\bibnamefont {Demura}}, \bibinfo {author}
  {\bibfnamefont {A.~R.}\ \bibnamefont {Derk}}, \bibinfo {author}
  {\bibfnamefont {A.}~\bibnamefont {Dunsworth}}, \bibinfo {author}
  {\bibfnamefont {D.}~\bibnamefont {Eppens}}, \bibinfo {author} {\bibfnamefont
  {C.}~\bibnamefont {Erickson}}, \bibinfo {author} {\bibfnamefont
  {E.}~\bibnamefont {Farhi}}, \bibinfo {author} {\bibfnamefont {A.~G.}\
  \bibnamefont {Fowler}}, \bibinfo {author} {\bibfnamefont {B.}~\bibnamefont
  {Foxen}}, \bibinfo {author} {\bibfnamefont {C.}~\bibnamefont {Gidney}},
  \bibinfo {author} {\bibfnamefont {M.}~\bibnamefont {Giustina}}, \bibinfo
  {author} {\bibfnamefont {J.~A.}\ \bibnamefont {Gross}}, \bibinfo {author}
  {\bibfnamefont {M.~P.}\ \bibnamefont {Harrigan}}, \bibinfo {author}
  {\bibfnamefont {S.~D.}\ \bibnamefont {Harrington}}, \bibinfo {author}
  {\bibfnamefont {J.}~\bibnamefont {Hilton}}, \bibinfo {author} {\bibfnamefont
  {A.}~\bibnamefont {Ho}}, \bibinfo {author} {\bibfnamefont {S.}~\bibnamefont
  {Hong}}, \bibinfo {author} {\bibfnamefont {T.}~\bibnamefont {Huang}},
  \bibinfo {author} {\bibfnamefont {W.~J.}\ \bibnamefont {Huggins}}, \bibinfo
  {author} {\bibfnamefont {S.~V.}\ \bibnamefont {Isakov}}, \bibinfo {author}
  {\bibfnamefont {M.}~\bibnamefont {Jacob-Mitos}}, \bibinfo {author}
  {\bibfnamefont {E.}~\bibnamefont {Jeffrey}}, \bibinfo {author} {\bibfnamefont
  {C.}~\bibnamefont {Jones}}, \bibinfo {author} {\bibfnamefont
  {D.}~\bibnamefont {Kafri}}, \bibinfo {author} {\bibfnamefont
  {K.}~\bibnamefont {Kechedzhi}}, \bibinfo {author} {\bibfnamefont
  {J.}~\bibnamefont {Kelly}}, \bibinfo {author} {\bibfnamefont
  {S.}~\bibnamefont {Kim}}, \bibinfo {author} {\bibfnamefont {P.~V.}\
  \bibnamefont {Klimov}}, \bibinfo {author} {\bibfnamefont {A.~N.}\
  \bibnamefont {Korotkov}}, \bibinfo {author} {\bibfnamefont {F.}~\bibnamefont
  {Kostritsa}}, \bibinfo {author} {\bibfnamefont {D.}~\bibnamefont {Landhuis}},
  \bibinfo {author} {\bibfnamefont {P.}~\bibnamefont {Laptev}}, \bibinfo
  {author} {\bibfnamefont {E.}~\bibnamefont {Lucero}}, \bibinfo {author}
  {\bibfnamefont {O.}~\bibnamefont {Martin}}, \bibinfo {author} {\bibfnamefont
  {J.~R.}\ \bibnamefont {McClean}}, \bibinfo {author} {\bibfnamefont
  {M.}~\bibnamefont {McEwen}}, \bibinfo {author} {\bibfnamefont
  {A.}~\bibnamefont {Megrant}}, \bibinfo {author} {\bibfnamefont {K.~C.}\
  \bibnamefont {Miao}}, \bibinfo {author} {\bibfnamefont {M.}~\bibnamefont
  {Mohseni}}, \bibinfo {author} {\bibfnamefont {J.}~\bibnamefont {Mutus}},
  \bibinfo {author} {\bibfnamefont {O.}~\bibnamefont {Naaman}}, \bibinfo
  {author} {\bibfnamefont {M.}~\bibnamefont {Neeley}}, \bibinfo {author}
  {\bibfnamefont {M.}~\bibnamefont {Newman}}, \bibinfo {author} {\bibfnamefont
  {T.~E.}\ \bibnamefont {O'Brien}}, \bibinfo {author} {\bibfnamefont
  {A.}~\bibnamefont {Opremcak}}, \bibinfo {author} {\bibfnamefont
  {E.}~\bibnamefont {Ostby}}, \bibinfo {author} {\bibfnamefont
  {B.}~\bibnamefont {Pat{\'o}}}, \bibinfo {author} {\bibfnamefont
  {A.}~\bibnamefont {Petukhov}}, \bibinfo {author} {\bibfnamefont
  {C.}~\bibnamefont {Quintana}}, \bibinfo {author} {\bibfnamefont
  {N.}~\bibnamefont {Redd}}, \bibinfo {author} {\bibfnamefont {N.~C.}\
  \bibnamefont {Rubin}}, \bibinfo {author} {\bibfnamefont {D.}~\bibnamefont
  {Sank}}, \bibinfo {author} {\bibfnamefont {K.~J.}\ \bibnamefont {Satzinger}},
  \bibinfo {author} {\bibfnamefont {V.}~\bibnamefont {Shvarts}}, \bibinfo
  {author} {\bibfnamefont {D.}~\bibnamefont {Strain}}, \bibinfo {author}
  {\bibfnamefont {M.}~\bibnamefont {Szalay}}, \bibinfo {author} {\bibfnamefont
  {M.~D.}\ \bibnamefont {Trevithick}}, \bibinfo {author} {\bibfnamefont
  {B.}~\bibnamefont {Villalonga}}, \bibinfo {author} {\bibfnamefont {T.~C.}\
  \bibnamefont {White}}, \bibinfo {author} {\bibfnamefont {Z.}~\bibnamefont
  {Yao}}, \bibinfo {author} {\bibfnamefont {P.}~\bibnamefont {Yeh}}, \bibinfo
  {author} {\bibfnamefont {A.}~\bibnamefont {Zalcman}}, \bibinfo {author}
  {\bibfnamefont {H.}~\bibnamefont {Neven}}, \bibinfo {author} {\bibfnamefont
  {S.}~\bibnamefont {Boixo}}, \bibinfo {author} {\bibfnamefont {L.~B.}\
  \bibnamefont {Ioffe}}, \bibinfo {author} {\bibfnamefont {P.}~\bibnamefont
  {Roushan}}, \bibinfo {author} {\bibfnamefont {Y.}~\bibnamefont {Chen}},\ and\
  \bibinfo {author} {\bibfnamefont {V.}~\bibnamefont {Smelyanskiy}},\
  }\bibfield  {title} {\bibinfo {title} {Accurately computing the electronic
  properties of a quantum ring},\ }\href
  {https://doi.org/10.1038/s41586-021-03576-2} {\bibfield  {journal} {\bibinfo
  {journal} {Nature}\ }\textbf {\bibinfo {volume} {594}},\ \bibinfo {pages}
  {508} (\bibinfo {year} {2021})}\BibitemShut {NoStop}%
\bibitem [{\citenamefont {Satzinger}\ \emph {et~al.}(2021)\citenamefont
  {Satzinger}, \citenamefont {Liu}, \citenamefont {Smith}, \citenamefont
  {Knapp}, \citenamefont {Newman}, \citenamefont {Jones}, \citenamefont {Chen},
  \citenamefont {Quintana}, \citenamefont {Mi}, \citenamefont {Dunsworth},
  \citenamefont {Gidney}, \citenamefont {Aleiner}, \citenamefont {Arute},
  \citenamefont {Arya}, \citenamefont {Atalaya}, \citenamefont {Babbush},
  \citenamefont {Bardin}, \citenamefont {Barends}, \citenamefont {Basso},
  \citenamefont {Bengtsson}, \citenamefont {Bilmes}, \citenamefont {Broughton},
  \citenamefont {Buckley}, \citenamefont {Buell}, \citenamefont {Burkett},
  \citenamefont {Bushnell}, \citenamefont {Chiaro}, \citenamefont {Collins},
  \citenamefont {Courtney}, \citenamefont {Demura}, \citenamefont {Derk},
  \citenamefont {Eppens}, \citenamefont {Erickson}, \citenamefont {Faoro},
  \citenamefont {Farhi}, \citenamefont {Fowler}, \citenamefont {Foxen},
  \citenamefont {Giustina}, \citenamefont {Greene}, \citenamefont {Gross},
  \citenamefont {Harrigan}, \citenamefont {Harrington}, \citenamefont {Hilton},
  \citenamefont {Hong}, \citenamefont {Huang}, \citenamefont {Huggins},
  \citenamefont {Ioffe}, \citenamefont {Isakov}, \citenamefont {Jeffrey},
  \citenamefont {Jiang}, \citenamefont {Kafri}, \citenamefont {Kechedzhi},
  \citenamefont {Khattar}, \citenamefont {Kim}, \citenamefont {Klimov},
  \citenamefont {Korotkov}, \citenamefont {Kostritsa}, \citenamefont
  {Landhuis}, \citenamefont {Laptev}, \citenamefont {Locharla}, \citenamefont
  {Lucero}, \citenamefont {Martin}, \citenamefont {McClean}, \citenamefont
  {McEwen}, \citenamefont {Miao}, \citenamefont {Mohseni}, \citenamefont
  {Montazeri}, \citenamefont {Mruczkiewicz}, \citenamefont {Mutus},
  \citenamefont {Naaman}, \citenamefont {Neeley}, \citenamefont {Neill},
  \citenamefont {Niu}, \citenamefont {O'Brien}, \citenamefont {Opremcak},
  \citenamefont {Pat{\'o}}, \citenamefont {Petukhov}, \citenamefont {Rubin},
  \citenamefont {Sank}, \citenamefont {Shvarts}, \citenamefont {Strain},
  \citenamefont {Szalay}, \citenamefont {Villalonga}, \citenamefont {White},
  \citenamefont {Yao}, \citenamefont {Yeh}, \citenamefont {Yoo}, \citenamefont
  {Zalcman}, \citenamefont {Neven}, \citenamefont {Boixo}, \citenamefont
  {Megrant}, \citenamefont {Chen}, \citenamefont {Kelly}, \citenamefont
  {Smelyanskiy}, \citenamefont {Kitaev}, \citenamefont {Knap}, \citenamefont
  {Pollmann},\ and\ \citenamefont {Roushan}}]{Satzinger21}%
  \BibitemOpen
  \bibfield  {author} {\bibinfo {author} {\bibfnamefont {K.~J.}\ \bibnamefont
  {Satzinger}}, \bibinfo {author} {\bibfnamefont {Y.-J.}\ \bibnamefont {Liu}},
  \bibinfo {author} {\bibfnamefont {A.}~\bibnamefont {Smith}}, \bibinfo
  {author} {\bibfnamefont {C.}~\bibnamefont {Knapp}}, \bibinfo {author}
  {\bibfnamefont {M.}~\bibnamefont {Newman}}, \bibinfo {author} {\bibfnamefont
  {C.}~\bibnamefont {Jones}}, \bibinfo {author} {\bibfnamefont
  {Z.}~\bibnamefont {Chen}}, \bibinfo {author} {\bibfnamefont {C.}~\bibnamefont
  {Quintana}}, \bibinfo {author} {\bibfnamefont {X.}~\bibnamefont {Mi}},
  \bibinfo {author} {\bibfnamefont {A.}~\bibnamefont {Dunsworth}}, \bibinfo
  {author} {\bibfnamefont {C.}~\bibnamefont {Gidney}}, \bibinfo {author}
  {\bibfnamefont {I.}~\bibnamefont {Aleiner}}, \bibinfo {author} {\bibfnamefont
  {F.}~\bibnamefont {Arute}}, \bibinfo {author} {\bibfnamefont
  {K.}~\bibnamefont {Arya}}, \bibinfo {author} {\bibfnamefont {J.}~\bibnamefont
  {Atalaya}}, \bibinfo {author} {\bibfnamefont {R.}~\bibnamefont {Babbush}},
  \bibinfo {author} {\bibfnamefont {J.~C.}\ \bibnamefont {Bardin}}, \bibinfo
  {author} {\bibfnamefont {R.}~\bibnamefont {Barends}}, \bibinfo {author}
  {\bibfnamefont {J.}~\bibnamefont {Basso}}, \bibinfo {author} {\bibfnamefont
  {A.}~\bibnamefont {Bengtsson}}, \bibinfo {author} {\bibfnamefont
  {A.}~\bibnamefont {Bilmes}}, \bibinfo {author} {\bibfnamefont
  {M.}~\bibnamefont {Broughton}}, \bibinfo {author} {\bibfnamefont {B.~B.}\
  \bibnamefont {Buckley}}, \bibinfo {author} {\bibfnamefont {D.~A.}\
  \bibnamefont {Buell}}, \bibinfo {author} {\bibfnamefont {B.}~\bibnamefont
  {Burkett}}, \bibinfo {author} {\bibfnamefont {N.}~\bibnamefont {Bushnell}},
  \bibinfo {author} {\bibfnamefont {B.}~\bibnamefont {Chiaro}}, \bibinfo
  {author} {\bibfnamefont {R.}~\bibnamefont {Collins}}, \bibinfo {author}
  {\bibfnamefont {W.}~\bibnamefont {Courtney}}, \bibinfo {author}
  {\bibfnamefont {S.}~\bibnamefont {Demura}}, \bibinfo {author} {\bibfnamefont
  {A.~R.}\ \bibnamefont {Derk}}, \bibinfo {author} {\bibfnamefont
  {D.}~\bibnamefont {Eppens}}, \bibinfo {author} {\bibfnamefont
  {C.}~\bibnamefont {Erickson}}, \bibinfo {author} {\bibfnamefont
  {L.}~\bibnamefont {Faoro}}, \bibinfo {author} {\bibfnamefont
  {E.}~\bibnamefont {Farhi}}, \bibinfo {author} {\bibfnamefont {A.~G.}\
  \bibnamefont {Fowler}}, \bibinfo {author} {\bibfnamefont {B.}~\bibnamefont
  {Foxen}}, \bibinfo {author} {\bibfnamefont {M.}~\bibnamefont {Giustina}},
  \bibinfo {author} {\bibfnamefont {A.}~\bibnamefont {Greene}}, \bibinfo
  {author} {\bibfnamefont {J.~A.}\ \bibnamefont {Gross}}, \bibinfo {author}
  {\bibfnamefont {M.~P.}\ \bibnamefont {Harrigan}}, \bibinfo {author}
  {\bibfnamefont {S.~D.}\ \bibnamefont {Harrington}}, \bibinfo {author}
  {\bibfnamefont {J.}~\bibnamefont {Hilton}}, \bibinfo {author} {\bibfnamefont
  {S.}~\bibnamefont {Hong}}, \bibinfo {author} {\bibfnamefont {T.}~\bibnamefont
  {Huang}}, \bibinfo {author} {\bibfnamefont {W.~J.}\ \bibnamefont {Huggins}},
  \bibinfo {author} {\bibfnamefont {L.~B.}\ \bibnamefont {Ioffe}}, \bibinfo
  {author} {\bibfnamefont {S.~V.}\ \bibnamefont {Isakov}}, \bibinfo {author}
  {\bibfnamefont {E.}~\bibnamefont {Jeffrey}}, \bibinfo {author} {\bibfnamefont
  {Z.}~\bibnamefont {Jiang}}, \bibinfo {author} {\bibfnamefont
  {D.}~\bibnamefont {Kafri}}, \bibinfo {author} {\bibfnamefont
  {K.}~\bibnamefont {Kechedzhi}}, \bibinfo {author} {\bibfnamefont
  {T.}~\bibnamefont {Khattar}}, \bibinfo {author} {\bibfnamefont
  {S.}~\bibnamefont {Kim}}, \bibinfo {author} {\bibfnamefont {P.~V.}\
  \bibnamefont {Klimov}}, \bibinfo {author} {\bibfnamefont {A.~N.}\
  \bibnamefont {Korotkov}}, \bibinfo {author} {\bibfnamefont {F.}~\bibnamefont
  {Kostritsa}}, \bibinfo {author} {\bibfnamefont {D.}~\bibnamefont {Landhuis}},
  \bibinfo {author} {\bibfnamefont {P.}~\bibnamefont {Laptev}}, \bibinfo
  {author} {\bibfnamefont {A.}~\bibnamefont {Locharla}}, \bibinfo {author}
  {\bibfnamefont {E.}~\bibnamefont {Lucero}}, \bibinfo {author} {\bibfnamefont
  {O.}~\bibnamefont {Martin}}, \bibinfo {author} {\bibfnamefont {J.~R.}\
  \bibnamefont {McClean}}, \bibinfo {author} {\bibfnamefont {M.}~\bibnamefont
  {McEwen}}, \bibinfo {author} {\bibfnamefont {K.~C.}\ \bibnamefont {Miao}},
  \bibinfo {author} {\bibfnamefont {M.}~\bibnamefont {Mohseni}}, \bibinfo
  {author} {\bibfnamefont {S.}~\bibnamefont {Montazeri}}, \bibinfo {author}
  {\bibfnamefont {W.}~\bibnamefont {Mruczkiewicz}}, \bibinfo {author}
  {\bibfnamefont {J.}~\bibnamefont {Mutus}}, \bibinfo {author} {\bibfnamefont
  {O.}~\bibnamefont {Naaman}}, \bibinfo {author} {\bibfnamefont
  {M.}~\bibnamefont {Neeley}}, \bibinfo {author} {\bibfnamefont
  {C.}~\bibnamefont {Neill}}, \bibinfo {author} {\bibfnamefont {M.~Y.}\
  \bibnamefont {Niu}}, \bibinfo {author} {\bibfnamefont {T.~E.}\ \bibnamefont
  {O'Brien}}, \bibinfo {author} {\bibfnamefont {A.}~\bibnamefont {Opremcak}},
  \bibinfo {author} {\bibfnamefont {B.}~\bibnamefont {Pat{\'o}}}, \bibinfo
  {author} {\bibfnamefont {A.}~\bibnamefont {Petukhov}}, \bibinfo {author}
  {\bibfnamefont {N.~C.}\ \bibnamefont {Rubin}}, \bibinfo {author}
  {\bibfnamefont {D.}~\bibnamefont {Sank}}, \bibinfo {author} {\bibfnamefont
  {V.}~\bibnamefont {Shvarts}}, \bibinfo {author} {\bibfnamefont
  {D.}~\bibnamefont {Strain}}, \bibinfo {author} {\bibfnamefont
  {M.}~\bibnamefont {Szalay}}, \bibinfo {author} {\bibfnamefont
  {B.}~\bibnamefont {Villalonga}}, \bibinfo {author} {\bibfnamefont {T.~C.}\
  \bibnamefont {White}}, \bibinfo {author} {\bibfnamefont {Z.}~\bibnamefont
  {Yao}}, \bibinfo {author} {\bibfnamefont {P.}~\bibnamefont {Yeh}}, \bibinfo
  {author} {\bibfnamefont {J.}~\bibnamefont {Yoo}}, \bibinfo {author}
  {\bibfnamefont {A.}~\bibnamefont {Zalcman}}, \bibinfo {author} {\bibfnamefont
  {H.}~\bibnamefont {Neven}}, \bibinfo {author} {\bibfnamefont
  {S.}~\bibnamefont {Boixo}}, \bibinfo {author} {\bibfnamefont
  {A.}~\bibnamefont {Megrant}}, \bibinfo {author} {\bibfnamefont
  {Y.}~\bibnamefont {Chen}}, \bibinfo {author} {\bibfnamefont {J.}~\bibnamefont
  {Kelly}}, \bibinfo {author} {\bibfnamefont {V.}~\bibnamefont {Smelyanskiy}},
  \bibinfo {author} {\bibfnamefont {A.}~\bibnamefont {Kitaev}}, \bibinfo
  {author} {\bibfnamefont {M.}~\bibnamefont {Knap}}, \bibinfo {author}
  {\bibfnamefont {F.}~\bibnamefont {Pollmann}},\ and\ \bibinfo {author}
  {\bibfnamefont {P.}~\bibnamefont {Roushan}},\ }\bibfield  {title} {\bibinfo
  {title} {Realizing topologically ordered states on a quantum processor},\
  }\href {https://doi.org/10.1126/science.abi8378} {\bibfield  {journal}
  {\bibinfo  {journal} {Science}\ }\textbf {\bibinfo {volume} {374}},\ \bibinfo
  {pages} {1237} (\bibinfo {year} {2021})}\BibitemShut {NoStop}%
\bibitem [{\citenamefont {Mi}\ \emph {et~al.}(2022)\citenamefont {Mi},
  \citenamefont {Sonner}, \citenamefont {Niu}, \citenamefont {Lee},
  \citenamefont {Foxen}, \citenamefont {Acharya}, \citenamefont {Aleiner},
  \citenamefont {Andersen}, \citenamefont {Arute}, \citenamefont {Arya},
  \citenamefont {Asfaw}, \citenamefont {Atalaya}, \citenamefont {Babbush},
  \citenamefont {Bacon}, \citenamefont {Bardin}, \citenamefont {Basso},
  \citenamefont {Bengtsson}, \citenamefont {Bortoli}, \citenamefont {Bourassa},
  \citenamefont {Brill}, \citenamefont {Broughton}, \citenamefont {Buckley},
  \citenamefont {Buell}, \citenamefont {Burkett}, \citenamefont {Bushnell},
  \citenamefont {Chen}, \citenamefont {Chiaro}, \citenamefont {Collins},
  \citenamefont {Conner}, \citenamefont {Courtney}, \citenamefont {Crook},
  \citenamefont {Debroy}, \citenamefont {Demura}, \citenamefont {Dunsworth},
  \citenamefont {Eppens}, \citenamefont {Erickson}, \citenamefont {Faoro},
  \citenamefont {Farhi}, \citenamefont {Fatemi}, \citenamefont {Flores},
  \citenamefont {Forati}, \citenamefont {Fowler}, \citenamefont {Giang},
  \citenamefont {Gidney}, \citenamefont {Gilboa}, \citenamefont {Giustina},
  \citenamefont {Dau}, \citenamefont {Gross}, \citenamefont {Habegger},
  \citenamefont {Harrigan}, \citenamefont {Hilton}, \citenamefont {Hoffmann},
  \citenamefont {Hong}, \citenamefont {Huang}, \citenamefont {Huff},
  \citenamefont {Huggins}, \citenamefont {Ioffe}, \citenamefont {Isakov},
  \citenamefont {Iveland}, \citenamefont {Jeffrey}, \citenamefont {Jiang},
  \citenamefont {Jones}, \citenamefont {Kafri}, \citenamefont {Kechedzhi},
  \citenamefont {Khattar}, \citenamefont {Kim}, \citenamefont {Kitaev},
  \citenamefont {Klimov}, \citenamefont {Klots}, \citenamefont {Korotkov},
  \citenamefont {Kostritsa}, \citenamefont {Kreikebaum}, \citenamefont
  {Landhuis}, \citenamefont {Laptev}, \citenamefont {Lau}, \citenamefont {Lee},
  \citenamefont {Laws}, \citenamefont {Liu}, \citenamefont {Locharla},
  \citenamefont {Lucero}, \citenamefont {Martin}, \citenamefont {McClean},
  \citenamefont {McEwen}, \citenamefont {Costa}, \citenamefont {Miao},
  \citenamefont {Mohseni}, \citenamefont {Montazeri}, \citenamefont {Morvan},
  \citenamefont {Mount}, \citenamefont {Mruczkiewicz}, \citenamefont {Naaman},
  \citenamefont {Neeley}, \citenamefont {Neill}, \citenamefont {Newman},
  \citenamefont {O'Brien}, \citenamefont {Opremcak}, \citenamefont {Petukhov},
  \citenamefont {Potter}, \citenamefont {Quintana}, \citenamefont {Rubin},
  \citenamefont {Saei}, \citenamefont {Sank}, \citenamefont {Sankaragomathi},
  \citenamefont {Satzinger}, \citenamefont {Schuster}, \citenamefont {Shearn},
  \citenamefont {Shvarts}, \citenamefont {Strain}, \citenamefont {Su},
  \citenamefont {Szalay}, \citenamefont {Vidal}, \citenamefont {Villalonga},
  \citenamefont {Vollgraff-Heidweiller}, \citenamefont {White}, \citenamefont
  {Yao}, \citenamefont {Yeh}, \citenamefont {Yoo}, \citenamefont {Zalcman},
  \citenamefont {Zhang}, \citenamefont {Zhu}, \citenamefont {Neven},
  \citenamefont {Boixo}, \citenamefont {Megrant}, \citenamefont {Chen},
  \citenamefont {Kelly}, \citenamefont {Smelyanskiy}, \citenamefont {Abanin},\
  and\ \citenamefont {Roushan}}]{Mi2022}%
  \BibitemOpen
  \bibfield  {author} {\bibinfo {author} {\bibfnamefont {X.}~\bibnamefont
  {Mi}}, \bibinfo {author} {\bibfnamefont {M.}~\bibnamefont {Sonner}}, \bibinfo
  {author} {\bibfnamefont {M.~Y.}\ \bibnamefont {Niu}}, \bibinfo {author}
  {\bibfnamefont {K.~W.}\ \bibnamefont {Lee}}, \bibinfo {author} {\bibfnamefont
  {B.}~\bibnamefont {Foxen}}, \bibinfo {author} {\bibfnamefont
  {R.}~\bibnamefont {Acharya}}, \bibinfo {author} {\bibfnamefont
  {I.}~\bibnamefont {Aleiner}}, \bibinfo {author} {\bibfnamefont {T.~I.}\
  \bibnamefont {Andersen}}, \bibinfo {author} {\bibfnamefont {F.}~\bibnamefont
  {Arute}}, \bibinfo {author} {\bibfnamefont {K.}~\bibnamefont {Arya}},
  \bibinfo {author} {\bibfnamefont {A.}~\bibnamefont {Asfaw}}, \bibinfo
  {author} {\bibfnamefont {J.}~\bibnamefont {Atalaya}}, \bibinfo {author}
  {\bibfnamefont {R.}~\bibnamefont {Babbush}}, \bibinfo {author} {\bibfnamefont
  {D.}~\bibnamefont {Bacon}}, \bibinfo {author} {\bibfnamefont {J.~C.}\
  \bibnamefont {Bardin}}, \bibinfo {author} {\bibfnamefont {J.}~\bibnamefont
  {Basso}}, \bibinfo {author} {\bibfnamefont {A.}~\bibnamefont {Bengtsson}},
  \bibinfo {author} {\bibfnamefont {G.}~\bibnamefont {Bortoli}}, \bibinfo
  {author} {\bibfnamefont {A.}~\bibnamefont {Bourassa}}, \bibinfo {author}
  {\bibfnamefont {L.}~\bibnamefont {Brill}}, \bibinfo {author} {\bibfnamefont
  {M.}~\bibnamefont {Broughton}}, \bibinfo {author} {\bibfnamefont {B.~B.}\
  \bibnamefont {Buckley}}, \bibinfo {author} {\bibfnamefont {D.~A.}\
  \bibnamefont {Buell}}, \bibinfo {author} {\bibfnamefont {B.}~\bibnamefont
  {Burkett}}, \bibinfo {author} {\bibfnamefont {N.}~\bibnamefont {Bushnell}},
  \bibinfo {author} {\bibfnamefont {Z.}~\bibnamefont {Chen}}, \bibinfo {author}
  {\bibfnamefont {B.}~\bibnamefont {Chiaro}}, \bibinfo {author} {\bibfnamefont
  {R.}~\bibnamefont {Collins}}, \bibinfo {author} {\bibfnamefont
  {P.}~\bibnamefont {Conner}}, \bibinfo {author} {\bibfnamefont
  {W.}~\bibnamefont {Courtney}}, \bibinfo {author} {\bibfnamefont {A.~L.}\
  \bibnamefont {Crook}}, \bibinfo {author} {\bibfnamefont {D.~M.}\ \bibnamefont
  {Debroy}}, \bibinfo {author} {\bibfnamefont {S.}~\bibnamefont {Demura}},
  \bibinfo {author} {\bibfnamefont {A.}~\bibnamefont {Dunsworth}}, \bibinfo
  {author} {\bibfnamefont {D.}~\bibnamefont {Eppens}}, \bibinfo {author}
  {\bibfnamefont {C.}~\bibnamefont {Erickson}}, \bibinfo {author}
  {\bibfnamefont {L.}~\bibnamefont {Faoro}}, \bibinfo {author} {\bibfnamefont
  {E.}~\bibnamefont {Farhi}}, \bibinfo {author} {\bibfnamefont
  {R.}~\bibnamefont {Fatemi}}, \bibinfo {author} {\bibfnamefont
  {L.}~\bibnamefont {Flores}}, \bibinfo {author} {\bibfnamefont
  {E.}~\bibnamefont {Forati}}, \bibinfo {author} {\bibfnamefont {A.~G.}\
  \bibnamefont {Fowler}}, \bibinfo {author} {\bibfnamefont {W.}~\bibnamefont
  {Giang}}, \bibinfo {author} {\bibfnamefont {C.}~\bibnamefont {Gidney}},
  \bibinfo {author} {\bibfnamefont {D.}~\bibnamefont {Gilboa}}, \bibinfo
  {author} {\bibfnamefont {M.}~\bibnamefont {Giustina}}, \bibinfo {author}
  {\bibfnamefont {A.~G.}\ \bibnamefont {Dau}}, \bibinfo {author} {\bibfnamefont
  {J.~A.}\ \bibnamefont {Gross}}, \bibinfo {author} {\bibfnamefont
  {S.}~\bibnamefont {Habegger}}, \bibinfo {author} {\bibfnamefont {M.~P.}\
  \bibnamefont {Harrigan}}, \bibinfo {author} {\bibfnamefont {J.}~\bibnamefont
  {Hilton}}, \bibinfo {author} {\bibfnamefont {M.}~\bibnamefont {Hoffmann}},
  \bibinfo {author} {\bibfnamefont {S.}~\bibnamefont {Hong}}, \bibinfo {author}
  {\bibfnamefont {T.}~\bibnamefont {Huang}}, \bibinfo {author} {\bibfnamefont
  {A.}~\bibnamefont {Huff}}, \bibinfo {author} {\bibfnamefont {W.~J.}\
  \bibnamefont {Huggins}}, \bibinfo {author} {\bibfnamefont {L.~B.}\
  \bibnamefont {Ioffe}}, \bibinfo {author} {\bibfnamefont {S.~V.}\ \bibnamefont
  {Isakov}}, \bibinfo {author} {\bibfnamefont {J.}~\bibnamefont {Iveland}},
  \bibinfo {author} {\bibfnamefont {E.}~\bibnamefont {Jeffrey}}, \bibinfo
  {author} {\bibfnamefont {Z.}~\bibnamefont {Jiang}}, \bibinfo {author}
  {\bibfnamefont {C.}~\bibnamefont {Jones}}, \bibinfo {author} {\bibfnamefont
  {D.}~\bibnamefont {Kafri}}, \bibinfo {author} {\bibfnamefont
  {K.}~\bibnamefont {Kechedzhi}}, \bibinfo {author} {\bibfnamefont
  {T.}~\bibnamefont {Khattar}}, \bibinfo {author} {\bibfnamefont
  {S.}~\bibnamefont {Kim}}, \bibinfo {author} {\bibfnamefont {A.}~\bibnamefont
  {Kitaev}}, \bibinfo {author} {\bibfnamefont {P.~V.}\ \bibnamefont {Klimov}},
  \bibinfo {author} {\bibfnamefont {A.~R.}\ \bibnamefont {Klots}}, \bibinfo
  {author} {\bibfnamefont {A.~N.}\ \bibnamefont {Korotkov}}, \bibinfo {author}
  {\bibfnamefont {F.}~\bibnamefont {Kostritsa}}, \bibinfo {author}
  {\bibfnamefont {J.~M.}\ \bibnamefont {Kreikebaum}}, \bibinfo {author}
  {\bibfnamefont {D.}~\bibnamefont {Landhuis}}, \bibinfo {author}
  {\bibfnamefont {P.}~\bibnamefont {Laptev}}, \bibinfo {author} {\bibfnamefont
  {K.-M.}\ \bibnamefont {Lau}}, \bibinfo {author} {\bibfnamefont
  {J.}~\bibnamefont {Lee}}, \bibinfo {author} {\bibfnamefont {L.}~\bibnamefont
  {Laws}}, \bibinfo {author} {\bibfnamefont {W.}~\bibnamefont {Liu}}, \bibinfo
  {author} {\bibfnamefont {A.}~\bibnamefont {Locharla}}, \bibinfo {author}
  {\bibfnamefont {E.}~\bibnamefont {Lucero}}, \bibinfo {author} {\bibfnamefont
  {O.}~\bibnamefont {Martin}}, \bibinfo {author} {\bibfnamefont {J.~R.}\
  \bibnamefont {McClean}}, \bibinfo {author} {\bibfnamefont {M.}~\bibnamefont
  {McEwen}}, \bibinfo {author} {\bibfnamefont {B.~M.}\ \bibnamefont {Costa}},
  \bibinfo {author} {\bibfnamefont {K.~C.}\ \bibnamefont {Miao}}, \bibinfo
  {author} {\bibfnamefont {M.}~\bibnamefont {Mohseni}}, \bibinfo {author}
  {\bibfnamefont {S.}~\bibnamefont {Montazeri}}, \bibinfo {author}
  {\bibfnamefont {A.}~\bibnamefont {Morvan}}, \bibinfo {author} {\bibfnamefont
  {E.}~\bibnamefont {Mount}}, \bibinfo {author} {\bibfnamefont
  {W.}~\bibnamefont {Mruczkiewicz}}, \bibinfo {author} {\bibfnamefont
  {O.}~\bibnamefont {Naaman}}, \bibinfo {author} {\bibfnamefont
  {M.}~\bibnamefont {Neeley}}, \bibinfo {author} {\bibfnamefont
  {C.}~\bibnamefont {Neill}}, \bibinfo {author} {\bibfnamefont
  {M.}~\bibnamefont {Newman}}, \bibinfo {author} {\bibfnamefont {T.~E.}\
  \bibnamefont {O'Brien}}, \bibinfo {author} {\bibfnamefont {A.}~\bibnamefont
  {Opremcak}}, \bibinfo {author} {\bibfnamefont {A.}~\bibnamefont {Petukhov}},
  \bibinfo {author} {\bibfnamefont {R.}~\bibnamefont {Potter}}, \bibinfo
  {author} {\bibfnamefont {C.}~\bibnamefont {Quintana}}, \bibinfo {author}
  {\bibfnamefont {N.~C.}\ \bibnamefont {Rubin}}, \bibinfo {author}
  {\bibfnamefont {N.}~\bibnamefont {Saei}}, \bibinfo {author} {\bibfnamefont
  {D.}~\bibnamefont {Sank}}, \bibinfo {author} {\bibfnamefont {K.}~\bibnamefont
  {Sankaragomathi}}, \bibinfo {author} {\bibfnamefont {K.~J.}\ \bibnamefont
  {Satzinger}}, \bibinfo {author} {\bibfnamefont {C.}~\bibnamefont {Schuster}},
  \bibinfo {author} {\bibfnamefont {M.~J.}\ \bibnamefont {Shearn}}, \bibinfo
  {author} {\bibfnamefont {V.}~\bibnamefont {Shvarts}}, \bibinfo {author}
  {\bibfnamefont {D.}~\bibnamefont {Strain}}, \bibinfo {author} {\bibfnamefont
  {Y.}~\bibnamefont {Su}}, \bibinfo {author} {\bibfnamefont {M.}~\bibnamefont
  {Szalay}}, \bibinfo {author} {\bibfnamefont {G.}~\bibnamefont {Vidal}},
  \bibinfo {author} {\bibfnamefont {B.}~\bibnamefont {Villalonga}}, \bibinfo
  {author} {\bibfnamefont {C.}~\bibnamefont {Vollgraff-Heidweiller}}, \bibinfo
  {author} {\bibfnamefont {T.}~\bibnamefont {White}}, \bibinfo {author}
  {\bibfnamefont {Z.~J.}\ \bibnamefont {Yao}}, \bibinfo {author} {\bibfnamefont
  {P.}~\bibnamefont {Yeh}}, \bibinfo {author} {\bibfnamefont {J.}~\bibnamefont
  {Yoo}}, \bibinfo {author} {\bibfnamefont {A.}~\bibnamefont {Zalcman}},
  \bibinfo {author} {\bibfnamefont {Y.}~\bibnamefont {Zhang}}, \bibinfo
  {author} {\bibfnamefont {N.}~\bibnamefont {Zhu}}, \bibinfo {author}
  {\bibfnamefont {H.}~\bibnamefont {Neven}}, \bibinfo {author} {\bibfnamefont
  {S.}~\bibnamefont {Boixo}}, \bibinfo {author} {\bibfnamefont
  {A.}~\bibnamefont {Megrant}}, \bibinfo {author} {\bibfnamefont
  {Y.}~\bibnamefont {Chen}}, \bibinfo {author} {\bibfnamefont {J.}~\bibnamefont
  {Kelly}}, \bibinfo {author} {\bibfnamefont {V.}~\bibnamefont {Smelyanskiy}},
  \bibinfo {author} {\bibfnamefont {D.~A.}\ \bibnamefont {Abanin}},\ and\
  \bibinfo {author} {\bibfnamefont {P.}~\bibnamefont {Roushan}},\ }\href
  {https://doi.org/10.48550/ARXIV.2204.11372} {\bibinfo {title}
  {Noise-resilient majorana edge modes on a chain of superconducting qubits}}
  (\bibinfo {year} {2022})\BibitemShut {NoStop}%
\bibitem [{\citenamefont {Partanen}\ \emph {et~al.}(2019)\citenamefont
  {Partanen}, \citenamefont {Goetz}, \citenamefont {Tan}, \citenamefont
  {Kohvakka}, \citenamefont {Sevriuk}, \citenamefont {Lake}, \citenamefont
  {Kokkoniemi}, \citenamefont {Ikonen}, \citenamefont {Hazra}, \citenamefont
  {M\"akinen}, \citenamefont {Hyypp\"a}, \citenamefont {Gr\"onberg},
  \citenamefont {Vesterinen}, \citenamefont {Silveri},\ and\ \citenamefont
  {M\"ott\"onen}}]{Silveri18}%
  \BibitemOpen
  \bibfield  {author} {\bibinfo {author} {\bibfnamefont {M.}~\bibnamefont
  {Partanen}}, \bibinfo {author} {\bibfnamefont {J.}~\bibnamefont {Goetz}},
  \bibinfo {author} {\bibfnamefont {K.~Y.}\ \bibnamefont {Tan}}, \bibinfo
  {author} {\bibfnamefont {K.}~\bibnamefont {Kohvakka}}, \bibinfo {author}
  {\bibfnamefont {V.}~\bibnamefont {Sevriuk}}, \bibinfo {author} {\bibfnamefont
  {R.~E.}\ \bibnamefont {Lake}}, \bibinfo {author} {\bibfnamefont
  {R.}~\bibnamefont {Kokkoniemi}}, \bibinfo {author} {\bibfnamefont
  {J.}~\bibnamefont {Ikonen}}, \bibinfo {author} {\bibfnamefont
  {D.}~\bibnamefont {Hazra}}, \bibinfo {author} {\bibfnamefont
  {A.}~\bibnamefont {M\"akinen}}, \bibinfo {author} {\bibfnamefont
  {E.}~\bibnamefont {Hyypp\"a}}, \bibinfo {author} {\bibfnamefont
  {L.}~\bibnamefont {Gr\"onberg}}, \bibinfo {author} {\bibfnamefont
  {V.}~\bibnamefont {Vesterinen}}, \bibinfo {author} {\bibfnamefont
  {M.}~\bibnamefont {Silveri}},\ and\ \bibinfo {author} {\bibfnamefont
  {M.}~\bibnamefont {M\"ott\"onen}},\ }\bibfield  {title} {\bibinfo {title}
  {{Exceptional points in tunable superconducting resonators}},\ }\href
  {https://doi.org/10.1103/PhysRevB.100.134505} {\bibfield  {journal} {\bibinfo
   {journal} {Phys. Rev. B}\ }\textbf {\bibinfo {volume} {100}},\ \bibinfo
  {pages} {134505} (\bibinfo {year} {2019})}\BibitemShut {NoStop}%
\bibitem [{\citenamefont {Silveri}\ \emph {et~al.}(2019)\citenamefont
  {Silveri}, \citenamefont {Masuda}, \citenamefont {Sevriuk}, \citenamefont
  {Tan}, \citenamefont {Jenei}, \citenamefont {Hyypp{\"a}}, \citenamefont
  {Hassler}, \citenamefont {Partanen}, \citenamefont {Goetz}, \citenamefont
  {Lake}, \citenamefont {Gr{\"o}nberg},\ and\ \citenamefont
  {M{\"o}tt{\"o}nen}}]{Silveri19a}%
  \BibitemOpen
  \bibfield  {author} {\bibinfo {author} {\bibfnamefont {M.}~\bibnamefont
  {Silveri}}, \bibinfo {author} {\bibfnamefont {S.}~\bibnamefont {Masuda}},
  \bibinfo {author} {\bibfnamefont {V.}~\bibnamefont {Sevriuk}}, \bibinfo
  {author} {\bibfnamefont {K.~Y.}\ \bibnamefont {Tan}}, \bibinfo {author}
  {\bibfnamefont {M.}~\bibnamefont {Jenei}}, \bibinfo {author} {\bibfnamefont
  {E.}~\bibnamefont {Hyypp{\"a}}}, \bibinfo {author} {\bibfnamefont
  {F.}~\bibnamefont {Hassler}}, \bibinfo {author} {\bibfnamefont
  {M.}~\bibnamefont {Partanen}}, \bibinfo {author} {\bibfnamefont
  {J.}~\bibnamefont {Goetz}}, \bibinfo {author} {\bibfnamefont {R.~E.}\
  \bibnamefont {Lake}}, \bibinfo {author} {\bibfnamefont {L.}~\bibnamefont
  {Gr{\"o}nberg}},\ and\ \bibinfo {author} {\bibfnamefont {M.}~\bibnamefont
  {M{\"o}tt{\"o}nen}},\ }\bibfield  {title} {\bibinfo {title} {{Broadband Lamb
  shift in an engineered quantum system}},\ }\href
  {https://doi.org/10.1038/s41567-019-0449-0} {\bibfield  {journal} {\bibinfo
  {journal} {Nature Physics}\ }\textbf {\bibinfo {volume} {15}},\ \bibinfo
  {pages} {533} (\bibinfo {year} {2019})}\BibitemShut {NoStop}%
\bibitem [{\citenamefont {Sevriuk}\ \emph {et~al.}(2019)\citenamefont
  {Sevriuk}, \citenamefont {Tan}, \citenamefont {Hyyppä}, \citenamefont
  {Silveri}, \citenamefont {Partanen}, \citenamefont {Jenei}, \citenamefont
  {Masuda}, \citenamefont {Goetz}, \citenamefont {Vesterinen}, \citenamefont
  {Grönberg},\ and\ \citenamefont {Möttönen}}]{Silveri19b}%
  \BibitemOpen
  \bibfield  {author} {\bibinfo {author} {\bibfnamefont {V.~A.}\ \bibnamefont
  {Sevriuk}}, \bibinfo {author} {\bibfnamefont {K.~Y.}\ \bibnamefont {Tan}},
  \bibinfo {author} {\bibfnamefont {E.}~\bibnamefont {Hyyppä}}, \bibinfo
  {author} {\bibfnamefont {M.}~\bibnamefont {Silveri}}, \bibinfo {author}
  {\bibfnamefont {M.}~\bibnamefont {Partanen}}, \bibinfo {author}
  {\bibfnamefont {M.}~\bibnamefont {Jenei}}, \bibinfo {author} {\bibfnamefont
  {S.}~\bibnamefont {Masuda}}, \bibinfo {author} {\bibfnamefont
  {J.}~\bibnamefont {Goetz}}, \bibinfo {author} {\bibfnamefont
  {V.}~\bibnamefont {Vesterinen}}, \bibinfo {author} {\bibfnamefont
  {L.}~\bibnamefont {Grönberg}},\ and\ \bibinfo {author} {\bibfnamefont
  {M.}~\bibnamefont {Möttönen}},\ }\bibfield  {title} {\bibinfo {title}
  {{Fast control of dissipation in a superconducting resonator}},\ }\href
  {https://doi.org/10.1063/1.5116659} {\bibfield  {journal} {\bibinfo
  {journal} {Applied Physics Letters}\ }\textbf {\bibinfo {volume} {115}},\
  \bibinfo {pages} {082601} (\bibinfo {year} {2019})}\BibitemShut {NoStop}%
\bibitem [{\citenamefont {Mörstedt}\ \emph {et~al.}(2022)\citenamefont
  {Mörstedt}, \citenamefont {Viitanen}, \citenamefont {Vadimov}, \citenamefont
  {Sevriuk}, \citenamefont {Partanen}, \citenamefont {Hyyppä}, \citenamefont
  {Catelani}, \citenamefont {Silveri}, \citenamefont {Tan},\ and\ \citenamefont
  {Möttönen}}]{Morstedt21}%
  \BibitemOpen
  \bibfield  {author} {\bibinfo {author} {\bibfnamefont {T.~F.}\ \bibnamefont
  {Mörstedt}}, \bibinfo {author} {\bibfnamefont {A.}~\bibnamefont {Viitanen}},
  \bibinfo {author} {\bibfnamefont {V.}~\bibnamefont {Vadimov}}, \bibinfo
  {author} {\bibfnamefont {V.}~\bibnamefont {Sevriuk}}, \bibinfo {author}
  {\bibfnamefont {M.}~\bibnamefont {Partanen}}, \bibinfo {author}
  {\bibfnamefont {E.}~\bibnamefont {Hyyppä}}, \bibinfo {author} {\bibfnamefont
  {G.}~\bibnamefont {Catelani}}, \bibinfo {author} {\bibfnamefont
  {M.}~\bibnamefont {Silveri}}, \bibinfo {author} {\bibfnamefont {K.~Y.}\
  \bibnamefont {Tan}},\ and\ \bibinfo {author} {\bibfnamefont {M.}~\bibnamefont
  {Möttönen}},\ }\bibfield  {title} {\bibinfo {title} {Recent {Developments}
  in {Quantum}-{Circuit} {Refrigeration}},\ }\href
  {https://doi.org/10.1002/andp.202100543} {\bibfield  {journal} {\bibinfo
  {journal} {Ann. Phys. (Berl.)}\ }\textbf {\bibinfo {volume} {534}},\ \bibinfo
  {pages} {2100543} (\bibinfo {year} {2022})}\BibitemShut {NoStop}%
\bibitem [{\citenamefont {Orell}\ \emph {et~al.}(2022)\citenamefont {Orell},
  \citenamefont {Zanner}, \citenamefont {Juan}, \citenamefont {Sharafiev},
  \citenamefont {Albert}, \citenamefont {Oleschko}, \citenamefont {Kirchmair},\
  and\ \citenamefont {Silveri}}]{Orell22}%
  \BibitemOpen
  \bibfield  {author} {\bibinfo {author} {\bibfnamefont {T.}~\bibnamefont
  {Orell}}, \bibinfo {author} {\bibfnamefont {M.}~\bibnamefont {Zanner}},
  \bibinfo {author} {\bibfnamefont {M.~L.}\ \bibnamefont {Juan}}, \bibinfo
  {author} {\bibfnamefont {A.}~\bibnamefont {Sharafiev}}, \bibinfo {author}
  {\bibfnamefont {R.}~\bibnamefont {Albert}}, \bibinfo {author} {\bibfnamefont
  {S.}~\bibnamefont {Oleschko}}, \bibinfo {author} {\bibfnamefont
  {G.}~\bibnamefont {Kirchmair}},\ and\ \bibinfo {author} {\bibfnamefont
  {M.}~\bibnamefont {Silveri}},\ }\bibfield  {title} {\bibinfo {title}
  {{Collective bosonic effects in an array of transmon devices}},\ }\href
  {https://doi.org/10.1103/PhysRevA.105.063701} {\bibfield  {journal} {\bibinfo
   {journal} {Phys. Rev. A}\ }\textbf {\bibinfo {volume} {105}},\ \bibinfo
  {pages} {063701} (\bibinfo {year} {2022})}\BibitemShut {NoStop}%
\bibitem [{\citenamefont {Takata}\ and\ \citenamefont
  {Notomi}(2018)}]{Takata2018}%
  \BibitemOpen
  \bibfield  {author} {\bibinfo {author} {\bibfnamefont {K.}~\bibnamefont
  {Takata}}\ and\ \bibinfo {author} {\bibfnamefont {M.}~\bibnamefont
  {Notomi}},\ }\bibfield  {title} {\bibinfo {title} {{Photonic Topological
  Insulating Phase Induced Solely by Gain and Loss}},\ }\href
  {https://doi.org/10.1103/PhysRevLett.121.213902} {\bibfield  {journal}
  {\bibinfo  {journal} {Phys. Rev. Lett.}\ }\textbf {\bibinfo {volume} {121}},\
  \bibinfo {pages} {213902} (\bibinfo {year} {2018})}\BibitemShut {NoStop}%
\bibitem [{\citenamefont {Orell}\ \emph {et~al.}(2019)\citenamefont {Orell},
  \citenamefont {Michailidis}, \citenamefont {Serbyn},\ and\ \citenamefont
  {Silveri}}]{Orell19}%
  \BibitemOpen
  \bibfield  {author} {\bibinfo {author} {\bibfnamefont {T.}~\bibnamefont
  {Orell}}, \bibinfo {author} {\bibfnamefont {A.~A.}\ \bibnamefont
  {Michailidis}}, \bibinfo {author} {\bibfnamefont {M.}~\bibnamefont
  {Serbyn}},\ and\ \bibinfo {author} {\bibfnamefont {M.}~\bibnamefont
  {Silveri}},\ }\bibfield  {title} {\bibinfo {title} {{Probing the many-body
  localization phase transition with superconducting circuits}},\ }\href
  {https://doi.org/10.1103/PhysRevB.100.134504} {\bibfield  {journal} {\bibinfo
   {journal} {Phys. Rev. B}\ }\textbf {\bibinfo {volume} {100}},\ \bibinfo
  {pages} {134504} (\bibinfo {year} {2019})}\BibitemShut {NoStop}%
\bibitem [{\citenamefont {Mansikkam\"aki}\ \emph {et~al.}(2021)\citenamefont
  {Mansikkam\"aki}, \citenamefont {Laine},\ and\ \citenamefont
  {Silveri}}]{Mansikkamaki21}%
  \BibitemOpen
  \bibfield  {author} {\bibinfo {author} {\bibfnamefont {O.}~\bibnamefont
  {Mansikkam\"aki}}, \bibinfo {author} {\bibfnamefont {S.}~\bibnamefont
  {Laine}},\ and\ \bibinfo {author} {\bibfnamefont {M.}~\bibnamefont
  {Silveri}},\ }\bibfield  {title} {\bibinfo {title} {{Phases of the disordered
  Bose-Hubbard model with attractive interactions}},\ }\href
  {https://doi.org/10.1103/PhysRevB.103.L220202} {\bibfield  {journal}
  {\bibinfo  {journal} {Phys. Rev. B}\ }\textbf {\bibinfo {volume} {103}},\
  \bibinfo {pages} {L220202} (\bibinfo {year} {2021})}\BibitemShut {NoStop}%
\bibitem [{\citenamefont {Prosen}\ and\ \citenamefont
  {Seligman}(2010)}]{Prosen_2010}%
  \BibitemOpen
  \bibfield  {author} {\bibinfo {author} {\bibfnamefont {T.}~\bibnamefont
  {Prosen}}\ and\ \bibinfo {author} {\bibfnamefont {T.~H.}\ \bibnamefont
  {Seligman}},\ }\bibfield  {title} {\bibinfo {title} {{Quantization over boson
  operator spaces}},\ }\href {https://doi.org/10.1088/1751-8113/43/39/392004}
  {\bibfield  {journal} {\bibinfo  {journal} {Journal of Physics A:
  Mathematical and Theoretical}\ }\textbf {\bibinfo {volume} {43}},\ \bibinfo
  {pages} {392004} (\bibinfo {year} {2010})}\BibitemShut {NoStop}%
\bibitem [{sup()}]{supplementary}%
  \BibitemOpen
  \bibinfo {note} {See Supplementary Material for more details.}\BibitemShut
  {Stop}%
\bibitem [{\citenamefont {Ma}\ \emph {et~al.}(2019)\citenamefont {Ma},
  \citenamefont {Saxberg}, \citenamefont {Owens}, \citenamefont {Leung},
  \citenamefont {Lu}, \citenamefont {Simon},\ and\ \citenamefont
  {Schuster}}]{Ma19}%
  \BibitemOpen
  \bibfield  {author} {\bibinfo {author} {\bibfnamefont {R.}~\bibnamefont
  {Ma}}, \bibinfo {author} {\bibfnamefont {B.}~\bibnamefont {Saxberg}},
  \bibinfo {author} {\bibfnamefont {C.}~\bibnamefont {Owens}}, \bibinfo
  {author} {\bibfnamefont {N.}~\bibnamefont {Leung}}, \bibinfo {author}
  {\bibfnamefont {Y.}~\bibnamefont {Lu}}, \bibinfo {author} {\bibfnamefont
  {J.}~\bibnamefont {Simon}},\ and\ \bibinfo {author} {\bibfnamefont {D.~I.}\
  \bibnamefont {Schuster}},\ }\bibfield  {title} {\bibinfo {title} {A
  dissipatively stabilized {Mott} insulator of photons},\ }\href
  {https://doi.org/10.1038/s41586-019-0897-9} {\bibfield  {journal} {\bibinfo
  {journal} {Nature}\ }\textbf {\bibinfo {volume} {566}},\ \bibinfo {pages}
  {51} (\bibinfo {year} {2019})}\BibitemShut {NoStop}%
\bibitem [{\citenamefont {Blais}\ \emph {et~al.}(2021)\citenamefont {Blais},
  \citenamefont {Grimsmo}, \citenamefont {Girvin},\ and\ \citenamefont
  {Wallraff}}]{Blais21}%
  \BibitemOpen
  \bibfield  {author} {\bibinfo {author} {\bibfnamefont {A.}~\bibnamefont
  {Blais}}, \bibinfo {author} {\bibfnamefont {A.~L.}\ \bibnamefont {Grimsmo}},
  \bibinfo {author} {\bibfnamefont {S.~M.}\ \bibnamefont {Girvin}},\ and\
  \bibinfo {author} {\bibfnamefont {A.}~\bibnamefont {Wallraff}},\ }\bibfield
  {title} {\bibinfo {title} {{Circuit quantum electrodynamics}},\ }\href
  {https://doi.org/10.1103/RevModPhys.93.025005} {\bibfield  {journal}
  {\bibinfo  {journal} {Rev. Mod. Phys.}\ }\textbf {\bibinfo {volume} {93}},\
  \bibinfo {pages} {025005} (\bibinfo {year} {2021})}\BibitemShut {NoStop}%
\bibitem [{\citenamefont {Elder}\ \emph {et~al.}(2020)\citenamefont {Elder},
  \citenamefont {Wang}, \citenamefont {Reinhold}, \citenamefont {Hann},
  \citenamefont {Chou}, \citenamefont {Lester}, \citenamefont {Rosenblum},
  \citenamefont {Frunzio}, \citenamefont {Jiang},\ and\ \citenamefont
  {Schoelkopf}}]{Elder20}%
  \BibitemOpen
  \bibfield  {author} {\bibinfo {author} {\bibfnamefont {S.~S.}\ \bibnamefont
  {Elder}}, \bibinfo {author} {\bibfnamefont {C.~S.}\ \bibnamefont {Wang}},
  \bibinfo {author} {\bibfnamefont {P.}~\bibnamefont {Reinhold}}, \bibinfo
  {author} {\bibfnamefont {C.~T.}\ \bibnamefont {Hann}}, \bibinfo {author}
  {\bibfnamefont {K.~S.}\ \bibnamefont {Chou}}, \bibinfo {author}
  {\bibfnamefont {B.~J.}\ \bibnamefont {Lester}}, \bibinfo {author}
  {\bibfnamefont {S.}~\bibnamefont {Rosenblum}}, \bibinfo {author}
  {\bibfnamefont {L.}~\bibnamefont {Frunzio}}, \bibinfo {author} {\bibfnamefont
  {L.}~\bibnamefont {Jiang}},\ and\ \bibinfo {author} {\bibfnamefont {R.~J.}\
  \bibnamefont {Schoelkopf}},\ }\bibfield  {title} {\bibinfo {title}
  {{High-Fidelity Measurement of Qubits Encoded in Multilevel Superconducting
  Circuits}},\ }\href {https://doi.org/10.1103/PhysRevX.10.011001} {\bibfield
  {journal} {\bibinfo  {journal} {Phys. Rev. X}\ }\textbf {\bibinfo {volume}
  {10}},\ \bibinfo {pages} {011001} (\bibinfo {year} {2020})}\BibitemShut
  {NoStop}%
\bibitem [{\citenamefont {Feiguin}(2011)}]{dmrg_lecture}%
  \BibitemOpen
  \bibfield  {author} {\bibinfo {author} {\bibfnamefont {A.~E.}\ \bibnamefont
  {Feiguin}},\ }\bibfield  {title} {\bibinfo {title} {The density matrix
  renormalization group and its time‐dependent variants},\ }\href
  {https://doi.org/10.1063/1.3667323} {\bibfield  {journal} {\bibinfo
  {journal} {AIP Conference Proceedings}\ }\textbf {\bibinfo {volume} {1419}},\
  \bibinfo {pages} {5} (\bibinfo {year} {2011})},\ \Eprint
  {https://arxiv.org/abs/https://aip.scitation.org/doi/pdf/10.1063/1.3667323}
  {https://aip.scitation.org/doi/pdf/10.1063/1.3667323} \BibitemShut {NoStop}%
\bibitem [{\citenamefont {Ollivier}\ and\ \citenamefont
  {Zurek}(2001)}]{QuantumDiscord}%
  \BibitemOpen
  \bibfield  {author} {\bibinfo {author} {\bibfnamefont {H.}~\bibnamefont
  {Ollivier}}\ and\ \bibinfo {author} {\bibfnamefont {W.~H.}\ \bibnamefont
  {Zurek}},\ }\bibfield  {title} {\bibinfo {title} {{Quantum Discord: A Measure
  of the Quantumness of Correlations}},\ }\href
  {https://doi.org/10.1103/PhysRevLett.88.017901} {\bibfield  {journal}
  {\bibinfo  {journal} {Phys. Rev. Lett.}\ }\textbf {\bibinfo {volume} {88}},\
  \bibinfo {pages} {017901} (\bibinfo {year} {2001})}\BibitemShut {NoStop}%
\bibitem [{\citenamefont {H{\"a}ffner}\ \emph {et~al.}(2005)\citenamefont
  {H{\"a}ffner}, \citenamefont {H{\"a}nsel}, \citenamefont {Roos},
  \citenamefont {Benhelm}, \citenamefont {{Chek-al-kar}}, \citenamefont
  {Chwalla}, \citenamefont {K{\"o}rber}, \citenamefont {Rapol}, \citenamefont
  {Riebe}, \citenamefont {Schmidt}, \citenamefont {Becher}, \citenamefont
  {G{\"u}hne}, \citenamefont {D{\"u}r},\ and\ \citenamefont
  {Blatt}}]{Haffner05}%
  \BibitemOpen
  \bibfield  {author} {\bibinfo {author} {\bibfnamefont {H.}~\bibnamefont
  {H{\"a}ffner}}, \bibinfo {author} {\bibfnamefont {W.}~\bibnamefont
  {H{\"a}nsel}}, \bibinfo {author} {\bibfnamefont {C.~F.}\ \bibnamefont
  {Roos}}, \bibinfo {author} {\bibfnamefont {J.}~\bibnamefont {Benhelm}},
  \bibinfo {author} {\bibfnamefont {D.}~\bibnamefont {{Chek-al-kar}}}, \bibinfo
  {author} {\bibfnamefont {M.}~\bibnamefont {Chwalla}}, \bibinfo {author}
  {\bibfnamefont {T.}~\bibnamefont {K{\"o}rber}}, \bibinfo {author}
  {\bibfnamefont {U.~D.}\ \bibnamefont {Rapol}}, \bibinfo {author}
  {\bibfnamefont {M.}~\bibnamefont {Riebe}}, \bibinfo {author} {\bibfnamefont
  {P.~O.}\ \bibnamefont {Schmidt}}, \bibinfo {author} {\bibfnamefont
  {C.}~\bibnamefont {Becher}}, \bibinfo {author} {\bibfnamefont
  {O.}~\bibnamefont {G{\"u}hne}}, \bibinfo {author} {\bibfnamefont
  {W.}~\bibnamefont {D{\"u}r}},\ and\ \bibinfo {author} {\bibfnamefont
  {R.}~\bibnamefont {Blatt}},\ }\bibfield  {title} {\bibinfo {title} {{Scalable
  multiparticle entanglement of trapped ions}},\ }\href
  {https://doi.org/10.1038/nature04279} {\bibfield  {journal} {\bibinfo
  {journal} {Nature}\ }\textbf {\bibinfo {volume} {438}},\ \bibinfo {pages}
  {643} (\bibinfo {year} {2005})}\BibitemShut {NoStop}%
\bibitem [{\citenamefont {Steffen}\ \emph {et~al.}(2006)\citenamefont
  {Steffen}, \citenamefont {Ansmann}, \citenamefont {Bialczak}, \citenamefont
  {Katz}, \citenamefont {Lucero}, \citenamefont {McDermott}, \citenamefont
  {Neeley}, \citenamefont {Weig}, \citenamefont {Cleland},\ and\ \citenamefont
  {Martinis}}]{Steffen06}%
  \BibitemOpen
  \bibfield  {author} {\bibinfo {author} {\bibfnamefont {M.}~\bibnamefont
  {Steffen}}, \bibinfo {author} {\bibfnamefont {M.}~\bibnamefont {Ansmann}},
  \bibinfo {author} {\bibfnamefont {R.~C.}\ \bibnamefont {Bialczak}}, \bibinfo
  {author} {\bibfnamefont {N.}~\bibnamefont {Katz}}, \bibinfo {author}
  {\bibfnamefont {E.}~\bibnamefont {Lucero}}, \bibinfo {author} {\bibfnamefont
  {R.}~\bibnamefont {McDermott}}, \bibinfo {author} {\bibfnamefont
  {M.}~\bibnamefont {Neeley}}, \bibinfo {author} {\bibfnamefont {E.~M.}\
  \bibnamefont {Weig}}, \bibinfo {author} {\bibfnamefont {A.~N.}\ \bibnamefont
  {Cleland}},\ and\ \bibinfo {author} {\bibfnamefont {J.~M.}\ \bibnamefont
  {Martinis}},\ }\bibfield  {title} {\bibinfo {title} {{Measurement of the
  Entanglement of Two Superconducting Qubits via State Tomography}},\ }\href
  {https://doi.org/10.1126/science.1130886} {\bibfield  {journal} {\bibinfo
  {journal} {Science}\ }\textbf {\bibinfo {volume} {313}},\ \bibinfo {pages}
  {1423} (\bibinfo {year} {2006})}\BibitemShut {NoStop}%
\bibitem [{\citenamefont {Filipp}\ \emph {et~al.}(2009)\citenamefont {Filipp},
  \citenamefont {Maurer}, \citenamefont {Leek}, \citenamefont {Baur},
  \citenamefont {Bianchetti}, \citenamefont {Fink}, \citenamefont {G\"oppl},
  \citenamefont {Steffen}, \citenamefont {Gambetta}, \citenamefont {Blais},\
  and\ \citenamefont {Wallraff}}]{Filipp09}%
  \BibitemOpen
  \bibfield  {author} {\bibinfo {author} {\bibfnamefont {S.}~\bibnamefont
  {Filipp}}, \bibinfo {author} {\bibfnamefont {P.}~\bibnamefont {Maurer}},
  \bibinfo {author} {\bibfnamefont {P.~J.}\ \bibnamefont {Leek}}, \bibinfo
  {author} {\bibfnamefont {M.}~\bibnamefont {Baur}}, \bibinfo {author}
  {\bibfnamefont {R.}~\bibnamefont {Bianchetti}}, \bibinfo {author}
  {\bibfnamefont {J.~M.}\ \bibnamefont {Fink}}, \bibinfo {author}
  {\bibfnamefont {M.}~\bibnamefont {G\"oppl}}, \bibinfo {author} {\bibfnamefont
  {L.}~\bibnamefont {Steffen}}, \bibinfo {author} {\bibfnamefont {J.~M.}\
  \bibnamefont {Gambetta}}, \bibinfo {author} {\bibfnamefont {A.}~\bibnamefont
  {Blais}},\ and\ \bibinfo {author} {\bibfnamefont {A.}~\bibnamefont
  {Wallraff}},\ }\bibfield  {title} {\bibinfo {title} {{Two-Qubit State
  Tomography Using a Joint Dispersive Readout}},\ }\href
  {https://doi.org/10.1103/PhysRevLett.102.200402} {\bibfield  {journal}
  {\bibinfo  {journal} {Phys. Rev. Lett.}\ }\textbf {\bibinfo {volume} {102}},\
  \bibinfo {pages} {200402} (\bibinfo {year} {2009})}\BibitemShut {NoStop}%
\end{thebibliography}%


\end{document}